\documentclass[aps,nofootinbib,preprintnumbers]{revtex4}
\usepackage{lipsum}
\usepackage{CJK}
\usepackage{amsfonts}
\usepackage{amsmath}
\usepackage{amssymb}
\usepackage[english]{babel}
\usepackage{graphicx}
\usepackage{epsfig}
\usepackage{bm}
\usepackage{verbatim}
\usepackage[utf8]{inputenc}
\usepackage{booktabs}
\usepackage{multirow}
\usepackage{subfig}
\usepackage{slashed}
\usepackage{xcolor}
\usepackage[colorlinks=true,urlcolor=red,citecolor=red]{hyperref}
\usepackage[font=small]{caption}
\usepackage{float}
\usepackage{blindtext}
\usepackage{placeins}
\usepackage[utf8]{inputenc}
\usepackage{bbm}
\usepackage[colorinlistoftodos]{todonotes}

\newenvironment{Eqnarray}{\arraycolsep 0.14em\begin{eqnarray}}{\end{eqnarray}}
\newcommand{\ba}{\begin{Eqnarray}}
\newcommand{\ea}{\end{Eqnarray}}
\newcommand{\be}{\begin{equation}}
\newcommand{\ee}{\end{equation}}
\newcommand{\bal}{\begin{aligned}}
\newcommand{\eal}{\end{aligned}}
\newcommand{\bea}{\begin{eqnarray}}
\newcommand{\eea}{\end{eqnarray}}
\newcommand{\ben}{\begin{enumerate}}
\newcommand{\een}{\end{enumerate}}
\newcommand{\bit}{\begin{itemize}}
\newcommand{\eit}{\end{itemize}}
\newcommand{\bde}{\begin{widetext}}
\newcommand{\ede}{\end{widetext}}

\renewcommand{\[}{\left[}

\def\lsim{\mathrel{\rlap{\lower4pt\hbox{\hskip1pt$\sim$}}
    \raise1pt\hbox{$<$}}}
\def\gsim{\mathrel{\rlap{\lower4pt\hbox{\hskip1pt$\sim$}}
    \raise1pt\hbox{$>$}}}

\def\3211{$\mathrm{SU(3) \otimes SU(2)_L \otimes U(1)_R \otimes U(1)_{B-L}}$ }
\def\321{$\mathrm{SU(3) \otimes SU(2) \otimes U(1)}$ }
\def\422{$\mathrm{SU(4) \otimes SU(2) \otimes SU(2)_R}$ }

\newcommand{\mathsym}[1]{{}}

\definecolor{bostonuniversityred}{rgb}{0.8, 0.0, 0.0}

\newcommand{\aech}[1]{{#1}}
\newcommand{\ac}[1]{{#1}}

\input{tcilatex}
\begin{document}

\title{A renormalizable left-right symmetric model with low scale seesaw
mechanisms}
\author{A. E. C\'arcamo Hern\'andez}
\email{antonio.carcamo@usm.cl}
\affiliation{{Departamento de F\'{\i}sica, Universidad T\'ecnica Federico Santa Mar\'{\i}%
a, Casilla 110-V, Valpara\'{\i}so, Chile}}
\affiliation{{Centro Cient\'{\i}fico-Tecnol\'ogico de Valpara\'{\i}so, Casilla 110-V,
Valpara\'{\i}so, Chile}}
\affiliation{{Millennium Institute for Subatomic Physics at High-Energy Frontier
(SAPHIR), Fern\'andez Concha 700, Santiago, Chile}}
\author{Ivan Schmidt}
\email{ivan.schmidt@usm.cl}
\affiliation{{Departamento de F\'{\i}sica, Universidad T\'ecnica Federico Santa Mar\'{\i}%
a, Casilla 110-V, Valpara\'{\i}so, Chile}}
\affiliation{{Centro Cient\'{\i}fico-Tecnol\'ogico de Valpara\'{\i}so, Casilla 110-V,
Valpara\'{\i}so, Chile}}
\affiliation{{Millennium Institute for Subatomic Physics at High-Energy Frontier
(SAPHIR), Fern\'andez Concha 700, Santiago, Chile}}
\date{\today }

\begin{abstract}
We propose a low scale renormalizable left-right symmetric theory that
successfully explains the observed SM fermion mass hierarchy, the tiny
values for the light active neutrino masses and is consistent with the
lepton and baryon asymmetries of the Universe, 
the muon and electron anomalous magnetic moments as well as the with the
constraints arising from the meson oscillations. In the proposed model the
top and exotic quarks obtain masses at tree level, whereas the masses of the
bottom, charm and strange quarks, tau and muon leptons are generated from a
tree level Universal Seesaw mechanism, thanks to their mixings with the
charged exotic vector like fermions. The masses for the first generation SM
charged fermions arise from a radiative seesaw mechanism at one loop level,
mediated by charged vector like fermions and electrically neutral scalars.
The light active neutrino masses are produced from a one-loop level inverse
seesaw mechanism mediated by electrically neutral scalar singlets and right
handed Majorana neutrinos. Our model is also consistent with the
experimental constraints arising from the Higgs diphoton decay rate as well as with the constraints arising from charged lepton flavor violation. We also
discuss the $Z^{\prime }$ and heavy scalar production at a proton-proton
collider.

\footnotesize
DOI:\href{https://doi.org/10.1016/j.nuclphysb.2022.115696}{10.1016/j.nuclphysb.2022.115696}
\normalsize
\end{abstract}

\maketitle

\section{Introduction}

Despite the great success of the Standard Model (SM) as a theory of
fundamental interactions, it features drawbacks such as, for example, the
lack of explanation of the SM flavor structure; in particular, the observed
pattern of SM fermion masses and mixings, the origin of Dark Matter (DM),
the source of parity violation in electroweak (EW) interactions, the lepton
and baryon asymmetries of the Universe and the anomalous magnetic moments of
the muon and electron. In order to address these issues, it is necessary to
propose a possible more general higher energy theory. In this sense,
left-right symmetric electroweak extensions of the Weinberg-Salam theory
have many appealing features, foremost of which is to address the origin of
parity violation as a low energy effect, a remanent of its breaking at a
certain high energy scale. We are therefore proposing, as a possible
explanation of the problems listed before, a minimal renormalizable
Left-right symmetric theory ~\cite{Pati:1974yy,Mohapatra:1974gc} based on
the gauge symmetry $SU(3)_{C}\times SU\left( 2\right) _{L}\times SU\left(
2\right) _{R}\times U\left( 1\right) _{B-L}$, supplemented by the $%
Z_{4}^{\left( 1\right) }\times Z_{4}^{\left( 2\right) }$ discrete group,
where the $Z_{4}^{\left( 1\right) }$ symmetry is completely broken, whereas
the $Z_{4}^{\left( 2\right) }$ symmetry is broken down to the preserved $%
Z_{2}$, thus allowing the implementation of a radiative inverse seesaw
mechanism to generate the tiny masses of the light active neutrinos. In the
proposed model, the top and exotic quarks obtain masses at tree level from
the Yukawa interactions, whereas the masses of the bottom, charm and strange
quarks, tau and muon leptons arise from a tree level Universal Seesaw
mechanism \cite{Davidson:1987mh,Davidson:1987mi}. The masses for the first
generation SM charged fermions are generated from a one loop level radiative
seesaw mechanism mediated by charged vector like fermions and electrically
neutral scalars. Unlike \cite{Davidson:1987mh}, where the tree level
Universal Seesaw mechanism was first implemented to generate the masses of
all SM charged fermions and light active neutrinos, in our model we use the
tree level Universal Seesaw mechanism only for the charm, bottom, strange
quarks, tau and muon leptons. Furthermore, whereas in the model of \cite%
{Davidson:1987mh} the light active neutrino masses are generated from a type
I seesaw mechanism, in our model we implement the one loop level inverse
seesaw mechanism mediated by electrically neutral scalar singlets and right
handed Majorana neutrinos, in order to produce the tiny masses of the light
active neutrinos. Some recent left-right symmetric models have been
considered in Refs. \cite%
{CarcamoHernandez:2018hst,Dekens:2014ina,Nomura:2016run,Brdar:2018sbk,Ma:2020lnm,Babu:2020bgz}%
. Unlike the model of Ref \cite{CarcamoHernandez:2018hst}, where non
renormalizable Yukawa interactions are employed for the implementation of a
Froggatt Nielsen mechanism to produce the current SM fermion mass and mixing
pattern, our proposed model is a fully renormalizable theory, with minimal
particle content and symmetries, where tree level Universal as well as
one-loop level radiative seesaw and inverse seesaw mechanisms are combined
to explain the observed hierarchy of SM fermion masses and fermionic mixing
parameters. Furthermore, unlike Ref. \cite{CarcamoHernandez:2018hst} our
model successfully explains the electron and muon anomalous magnetic moments
and includes a discussion about leptogenesis and collider signatures of
heavy scalar and $Z^{\prime }$ gauge bosons, which is not presented in \cite%
{CarcamoHernandez:2018hst}.

In our current model, the charged vector-like leptons responsible for the
tree level Universal and one-loop level radiative seesaw mechanism that
produces the SM charged fermion mass hierarchy, allows to reproduce the
measured values of the muon and electron anomalous magnetic moments, thus
linking the fermion mass generation mechanism and the $g-2$ anomalies, which
is not given in the left-right symmetric model of Ref. \cite%
{CarcamoHernandez:2018hst}. Moreover, unlike the left-right symmetric theory
of Ref. \cite{Ma:2020lnm}, our model does not rely on the inclusion of
scalar leptoquarks to generate one loop level masses for the SM charged
fermions and light active neutrinos. Besides that, whereas in the left-right
symmetric model of \cite{Babu:2020bgz} the light active neutrino masses are
generated from a combination of type I and type II seesaw mechanisms, in our
model the tiny masses of the light active neutrinos are produced from an
inverse seesaw mechanism at one loop level. Another difference of our model
with the one proposed of \cite{Babu:2020bgz} is that in the former a
mechanism for explaining the SM charged fermion mass hierarchy is presented,
whereas in the latter such mechanism is not given. Furthermore, whereas in
the models of Refs. \cite{Brdar:2018sbk} and \cite{Nomura:2016run}, the
masses of the light active neutrinos are generated from a tree level inverse
and radiative type I sessaw mechanisms, respectively, in our model we use
the inverse seesaw mechanism at one loop level to produce the tiny masses of
the light active neutrinos. In addition, our model includes a dynamical
mechanism to generate the SM charged fermion mass pattern, which is not
presented in the model of Ref. \cite{Brdar:2018sbk}.

On the other hand, the renormalizable left-right symmetric theory proposed
in this paper has similar amount of particle content compared to the
left-right symmetric model considered in \cite{Dekens:2014ina}. For
instance, whereas the scalar sector of left-right symmetric model of Ref. 
\cite{Dekens:2014ina} has one scalar bidoublet (having 8 degrees of
freedom), one $SU(2)_{L}$ scalar triplet (transforming as a $SU(2)_{R}$
singlet) (having 6 degrees of freedom) and one $SU(2)_{R}$ scalar triplet
(transforming as a $SU(2)_{L}$ singlet) (having 6 degrees of freedom), thus
amounting to 14 physical scalar degrees of freedom (after substracting the
number of Goldstone bosons), our current left-right model has one scalar
bidoublet (8 degrees of freedom), two $SU(2)_{L}$ scalar doublets (8 degrees
of freedom), two $SU(2)_{R}$ scalar doublets (8 degrees of freedom), two
electrically neutral gauge singlet real scalars (2 degrees of freedom) and
two electrically neutral gauge singlet complex scalars (4 degrees of
freedom), which corresponds to 24 physical scalar degrees of freedom.
Despite our model has more scalar degrees of freedom than the one proposed
in \cite{Dekens:2014ina}, the advantage of our proposal with respect to the
ones presented in \cite{Dekens:2014ina,Brdar:2018sbk} is that in the former
a mechanism that naturally explains the SM fermion mass hierarchy is
presented, whereas the latter does not include such mechanism.

The paper is organized as follows. In section \ref{model} we outline the
proposed model. The implications of our model in the SM fermion hierarchy is
discussed in section \ref{fermionmasses}. The implications of our model in
charged lepton flavor violation are described in section \ref{LFV}. The consequences of our model in
leptogenesis are described in section \ref{leptogenesis}, while the model
scalar potential is analyzed in section \ref{scalarpotential}. The
implications of our model in the Higgs diphoton decay are discussed in
section \ref{sec.Higgsdiphoton}, and in section \ref{sec.gminus2} we analyze
its application to the muon and electron anomalous magnetic moments. The $%
Z^\prime$ and heavy scalar production at a proton-proton collider are
discussed in sections \ref{HeavyScalar} and \ref{Zprime}, respectively. The
implications of our model in meson oscillations are discussed in section \ref%
{FCNC}. We conclude in section \ref{conclusions}. An analytical argument of
the minimal number of fermionic seesaw mediators required to generate the
masses of SM fermions via a seesaw-like mechanism is presented in Appendix %
\ref{M}.


\section{An extended Left-Right symmetric model}

\label{model} Before providing a detailed explanation of our left-right
symmetric model, we will explain the reasoning behind introducing extra
scalars, fermions and symmetries, needed for implementing an interplay of
tree level universal and radiative seesaw mechanism to explain the SM
charged fermion mass hierarchy and one loop level inverse seesaw mechanism
to generate the tiny neutrino masses. It is worth mentioning that in our
proposed model, the mass of the top quark will be generated from a
renormalizable Yukawa operator, with an order one Yukawa coupling, i.e. 
\begin{equation}
\overline{Q}_{3L}\Phi Q_{iR},\hspace{1.5cm}i=1,2,3
\end{equation}%
where $Q_{3L}$ and $Q_{iR}$ are $SU\left( 2\right) _{L}$ and $SU\left(
2\right) _{R}$ quark doublets, respectively: 
\begin{equation}
Q_{iL}=\left( 
\begin{array}{c}
u_{iL} \\ 
d_{iL}%
\end{array}%
\right) ,\hspace{1.5cm}Q_{iR}=\left( 
\begin{array}{c}
\overline{u}_{iR} \\ 
\overline{d}_{iR}%
\end{array}%
\right) ,\hspace{1.5cm}i=1,2,3,
\end{equation}%
whereas $\Phi $ is a scalar bidoblet, with the VEV pattern 
\begin{equation}
\left\langle \Phi \right\rangle =\left( 
\begin{array}{cc}
v_{1} & 0 \\ 
0 & v_{2}%
\end{array}%
\right) ,
\end{equation}%
where we have set $v_{2}=0$ to prevent a bottom quark mass arising from the
above given Yukawa interaction. Now, to generate tree level masses via a
Universal Seesaw mechanism for the bottom, strange and charm quarks, as well
as for the tau and muon leptons, one loop level masses for the first
generation SM charged fermions and the tiny masses for the light active
neutrinos via a one loop level inverse seesaw mechanism, we need to forbid
the operators: 
\begin{eqnarray}
&&\overline{Q}_{nL}\Phi Q_{iR},\hspace{1.5cm}\overline{Q}_{nL}\widetilde{%
\Phi }Q_{iR},\hspace{1.5cm}n=1,2,\hspace{1.5cm}i=1,2,3,  \notag \\
&&\overline{L}_{iL}\widetilde{\Phi }L_{jR},\hspace{1.5cm}\overline{L}_{iL}%
\widetilde{\chi }_{L}N_{jR},\hspace{1.5cm}\left( m_{N}\right) _{ij}\overline{%
N}_{iR}N_{jR}^{C},\hspace{1.5cm}i,j=1,2,3.
\end{eqnarray}%
where $\chi_L$ ($\chi_R$) is a $SU\left ( 2\right ) _{L}$ ($SU\left ( 2\right ) _{R}$) scalar doublet. Furthermore, $L_{iL}$ and $L_{iR}$ are $SU\left( 2\right) _{L}$ and $SU\left(
2\right) _{R}$ lepton doublets, respectively:%
\begin{equation}
L_{iL}=\left( 
\begin{array}{c}
\nu _{iL} \\ 
e_{iL}%
\end{array}%
\right) ,\hspace{1.5cm}L_{iR}=\left( 
\begin{array}{c}
\nu _{iR} \\ 
e_{iR}%
\end{array}%
\right) ,\hspace{1.5cm}i=1,2,3,
\end{equation}%
while $N_{iR}$ ($i=1,2,3$) are gauge singlet
neutral leptons. As it will be shown in the following, the aforementioned
gauge singlet neutral leptons are necessary for the implementation of the
one loop level inverse seesaw mechanism that produces the tiny masses of the
light active neutrinos.

Furthermore, the successfull implentation of the tree level universal and
radiative seesaw mechanism to explain the SM charged fermion mass hierarchy
and of the one loop level inverse seesaw mechanism to generate the tiny
neutrino masses, requires the following operators: 
\begin{eqnarray}
&&\overline{Q}_{3L}\chi _{L}B_{1R},\hspace{1.5cm}\overline{Q}_{nL}\chi
_{L}B_{2R},\hspace{1.5cm}\overline{B}_{nL}\chi _{R}^{\dagger }Q_{iR},\hspace{%
1.5cm}\overline{B}_{1L}\rho B_{1R},\hspace{1.5cm}\overline{B}_{2L}\sigma
B_{2R},  \notag \\
&&\overline{Q}_{nL}\widetilde{\chi }_{L}T_{R},\hspace{1.5cm}\overline{T}_{L}%
\widetilde{\chi }_{R}^{\dagger }Q_{iR},\hspace{1.5cm}\overline{T}_{L}\sigma
T_{R},\hspace{1.5cm}n=1,2,\hspace{1.5cm}i=1,2,3,  \notag \\
&&\overline{Q}_{nL}\phi _{L}B_{R}^{\prime },\hspace{1cm}\bar{B}_{L}^{\prime
}\phi _{R}^{\dagger }Q_{iR},\hspace{1cm}\overline{Q}_{nL}\widetilde{\phi }%
_{L}T_{R}^{\prime },\hspace{1cm}\bar{T}_{L}^{\prime }\widetilde{\phi }%
_{R}^{\dagger }Q_{iR},\hspace{1cm}\bar{B}_{L}^{\prime }\sigma B_{R}^{\prime
},\hspace{1cm}\bar{T}_{L}^{\prime }\sigma T_{R}^{\prime },  \notag \\
&&\overline{L}_{iL}\chi _{L}E_{nR},\hspace{1cm}\overline{E}_{nL}\chi
_{R}^{\dagger }L_{jR},\hspace{1cm}\overline{L}_{iL}\phi _{L}E_{R}^{\prime },%
\hspace{1cm}\bar{E}_{L}^{\prime }\phi _{R}^{\dagger }L_{iR},\hspace{1cm}%
\overline{E}_{nL}\rho E_{nR},\hspace{1cm}\bar{E}_{L}^{\prime }\rho
E_{R}^{\prime },  \notag \\
&&\overline{L}_{iL}\Phi L_{jR},\hspace{1cm}\overline{N_{iR}^{C}}\widetilde{%
\chi }_{R}^{\dagger }L_{jR},\hspace{1cm}\overline{\Omega }_{nR}\Omega
_{nR}^{C}\eta ,\hspace{1cm}\overline{N}_{nR}\Omega _{kR}^{C}\varphi ,\hspace{%
1cm}n,k=1,2.
\end{eqnarray}%
This requires to add $Z_{4}^{\left( 1\right) }$ and $Z_{4}^{\left( 2\right)
} $ discrete symmetries, which are spontaneously broken, where the former is
completely broken, and the latter is broken down to the preserved $Z_{2}$
symmetry. Such remaining conserved $Z_{2}$ symmetry allows to implement an
inverse seesaw mechanism at one loop level to produce the tiny neutrino
masses. Let us note that the gauge singlet neutral leptons $\Omega _{nR}$ ($%
n=1,2$) are crucial for generating the term $\left( m_{N}\right) _{ij}%
\overline{N}_{iR}N_{jR}^{C}$ ($i,j=1,2,3$) at one loop level, thus allowing
the implementation of the one-loop level inverse seesaw mechanism.
Additionally, the above mentioned exotic neutral lepton content is the
minimal one required to generate the masses for two light active neutrinos,
as required from the neutrino oscillation experimental data. Besides that,
the SM charged fermion sector has to be extended to include the following
heavy fermions: up type quarks $T$, $T^{\prime }$, down type quarks $B_{n}$, 
$B^{\prime }$ and charged leptons $E_{n}$ ($n=1,2$), $E^{\prime }$ in
singlet representations under $SU\left( 2\right) _{L}\times SU\left(
2\right) _{R}$. As a consequence of the above mentioned exotic charged
fermion spectrum, the rows and columns of the tree level SM charged fermion
mass matrices will be linearly dependent, thus implying that the first
generation SM charged fermions will be massless at tree level. The one loop
level corrections to these matrices mediated by the $T^{\prime }$, $%
B^{\prime }$ and $E^{\prime }$ fermionic fields will make thus rows and
columns linearly indenpendent, thus yielding one-loop level masses for the
up and down quarks as well as for the electron. Consequently, the
aforementioned exotic charged fermion spectrum is the minimal necessary so
that no massless charged SM-fermions would appear in the model, provided
that one loop level corrections are taken into account. For a more detailed
explanation of the analytical argument of the minimal number of fermionic
seesaw mediators required to generate the masses of SM fermions via a
seesaw-like mechanism the reader is referred to Appendix \ref{M}.

On the other hand, in what regards the scalar sector, it is worth mentioning
that $\chi _{L}$, $\phi _{L}$ and $\chi _{R}$, $\phi _{R}$ are $SU\left(
2\right) _{L}$ and $SU\left( 2\right) _{R}$ scalar doublets, respectively,
whereas $\eta $, $\sigma $, $\rho $ and $\varphi $ are gauge singlet
scalars. Furthermore, the $\chi _{L}$ ($\chi _{R}$) scalar is crucial for
generating mass mixing terms between left (right) handed SM charged fermions
and right (left) handed exotic charged fermions. Furthermore, the $SU\left(
2\right) _{R}$ scalar doublet $\chi _{R}$ is crucial for triggering the
spontaneous breaking of the $SU\left( 2\right) _{L}\times SU\left( 2\right)
_{R}\times U\left( 1\right) _{B-L}$ symmetry down to the SM electroweak
gauge group. Besides that, the $\sigma $ and $\rho $ are gauge singlet
scalars, whose inclusion is necessary for generating the masses of the
charged exotic fermions. On the other hand, the gauge singlet scalars $\eta $
and $\varphi $ are required to generate tree and one loop level masses for
the Majorana neutrinos $\Omega _{nR}$ ($n=1,2$) and $N_{iR}$ ($i=1,2,3$),
which is crucial for a radiative generation of the $\mu $ parameter of the
inverse seesaw mechanism. Moreover, the inclusion of the scalar bidoublet $%
\Phi $ is crucial to generate a tree level top quark mass, as well as the
Dirac neutrino submatrix, as will be shown below. The aforementioned scalar
content is the minimal required for a successful implementation of the tree
level universal and one loop level radiative seesaw mechanisms to explain
the SM charged fermion mass hierarchy, as well as of the one loop level
inverse seesaw mechanism to produce the tiny neutrino masses. By suitable
charge assignments to be specified below, we can implement the
aforementioned seesaw mechanisms, useful for explaining the SM fermion mass
hierarchy.

Our proposed model is based on the gauge symmetry $SU(3)_{C}\times SU\left(
2\right) _{L}\times SU\left( 2\right) _{R}\times U\left( 1\right) _{B-L}$,
supplemented by the $Z_{4}^{\left( 1\right) }\times Z_{4}^{\left( 2\right) }$
discrete group, where the full symmetry $\mathcal{G}$ exhibites the
following breaking scheme: 
\begin{eqnarray}
&&\mathcal{G}=SU(3)_{C}\times SU\left( 2\right) _{L}\times SU\left( 2\right)
_{R}\times U\left( 1\right) _{B-L}\times Z_{4}^{\left( 1\right) }\times
Z_{4}^{\left( 2\right) }  \notag \\
&&\hspace{35mm}\Downarrow v_{\sigma },v_{\eta },v_{\rho }  \notag \\[0.12in]
&&\hspace{15mm}SU(3)_{C}\times SU\left( 2\right) _{L}\times SU\left(
2\right) _{R}\times U\left( 1\right) _{B-L}  \notag \\
&&\hspace{35mm}\Downarrow v_{R}  \notag \\[0.12in]
&&\hspace{15mm}SU(3)_{C}\times SU\left( 2\right) _{L}\times U\left( 1\right)
_{Y}\times Z_{2}  \notag \\[0.12in]
&&\hspace{35mm}\Downarrow v_{1},v_{L}  \notag \\[0.12in]
&&\hspace{15mm}SU(3)_{C}\otimes U\left( 1\right) _{Q}\times Z_{2}
\end{eqnarray}%
Both $Z_{4}^{\left( 1\right) }$ and $Z_{4}^{\left( 2\right) }$ discrete
groups are spontaneously broken, and are crucial for avoiding a tree level
inverse seesaw mechanism. The $Z_{4}^{\left( 1\right) }$ symmetry is
completely broken, whereas the $Z_{4}^{\left( 2\right) }$ symmetry is broken
down to the preserved $Z_{2}$ symmetry. It is assumed that such discrete
symmetries are broken at the scale much larger than the scale of breaking of
the left-right symmetry. We further assume that the left-right symmetry
breaking scale is about $v_{R}\sim \mathcal{O}(10)$ TeV. In addition, the $%
Z_{4}^{\left( 2\right) }$ symmetry, which is spontaneously broken to the
preserved $Z_{2}$, is crucial in order to forbid the appearance of the term $%
\left( m_{N}\right) _{ij}\overline{N}_{iR}N_{jR}^{C}$ at tree level, thus
allowing the implementation of the one loop level inverse seesaw mechanism
that generates the light active neutrino masses. Besides that, the
spontaneously broken $Z_{4}^{\left( 1\right) }$ symmetry is crucial to
prevent tree level Yukawa mass terms involving the scalar bidoublet and SM
charged fermions lighter than the top quark. As we will see in the
following, in the SM fermion sector only the top quark will acquire its mass
from a renormalizable Yukawa interaction with the scalar bidoublet, whereas
the SM charged fermions lighter than the top quark will get their masses
from tree level Universal seesaw and radiative seesaw mechanisms.

The fermion assignments under the $SU(3)_{C}\times SU\left( 2\right)
_{L}\times SU\left( 2\right) _{R}\times U\left( 1\right) _{B-L}$ group are: 
\begin{eqnarray}
Q_{iL} &=&\left( 
\begin{array}{c}
u_{iL} \\ 
d_{iL}%
\end{array}%
\right) \sim \left( \mathbf{3},\mathbf{2,1},\frac{1}{3}\right) ,\hspace{1.5cm%
}Q_{iR}=\left( 
\begin{array}{c}
\overline{u}_{iR} \\ 
\overline{d}_{iR}%
\end{array}%
\right) \sim \left( \mathbf{3},\mathbf{1,2},\frac{1}{3}\right) ,\hspace{1.5cm%
}i=1,2,3,  \notag \\
L_{iL} &=&\left( 
\begin{array}{c}
\nu _{iL} \\ 
e_{iL}%
\end{array}%
\right) \sim \left( \mathbf{1},\mathbf{2,1},-1\right) ,\hspace{1.5cm}%
L_{iR}=\left( 
\begin{array}{c}
\nu _{iR} \\ 
e_{iR}%
\end{array}%
\right) \sim \left( \mathbf{1},\mathbf{1,2},-1\right) ,\hspace{1.5cm}i=1,2,3,
\notag \\
T_{R} &\sim &\left( \mathbf{3},\mathbf{1,1},\frac{4}{3}\right) ,\hspace{1.5cm%
}T_{L}\sim \left( \mathbf{3},\mathbf{1,1},\frac{4}{3}\right) ,\hspace{1.5cm}%
T_{R}^{\prime }\sim \left( \mathbf{3},\mathbf{1,1},\frac{4}{3}\right) ,%
\hspace{1.5cm}T_{L}^{\prime }\sim \left( \mathbf{3},\mathbf{1,1},\frac{4}{3}%
\right) ,  \notag \\
B_{nR} &\sim &\left( \mathbf{3},\mathbf{1,1},-\frac{2}{3}\right) ,\hspace{%
1.5cm}B_{nL}\sim \left( \mathbf{3},\mathbf{1,1},-\frac{2}{3}\right) ,\hspace{%
1.5cm}B_{R}^{\prime }\sim \left( \mathbf{3},\mathbf{1,1},-\frac{2}{3}\right)
,\hspace{1.5cm}B_{L}^{\prime }\sim \left( \mathbf{3},\mathbf{1,1},-\frac{2}{3%
}\right) ,  \notag \\
E_{nR} &\sim &\left( \mathbf{1},\mathbf{1,1},-2\right) ,\hspace{1.5cm}%
E_{nL}\sim \left( \mathbf{1},\mathbf{1,1},-2\right) ,\hspace{1.5cm}%
E_{R}^{\prime }\sim \left( \mathbf{1},\mathbf{1,1},-2\right) ,\hspace{1.5cm}%
E_{L}^{\prime }\sim \left( \mathbf{1},\mathbf{1,1},-2\right) ,  \notag \\
N_{iR} &\sim &\left( \mathbf{1},\mathbf{1,1},0\right) ,\hspace{1.5cm}\Omega
_{nR}\sim \left( \mathbf{1},\mathbf{1,1},0\right) ,\hspace{1.5cm}n=1,2.
\end{eqnarray}%
Let us note that we have extended the fermion sector of the original
left-right symmetric model model by introducing two exotic up type quarks $T$%
, $T^{\prime }$, three exotic down type quarks $B_{n}$ ($n=1,2$), $B^{\prime
}$, three charged leptons $E_{n}$, $E^{\prime }$ and five Majorana
neutrinos, i.e., $N_{iR}$ ($i=1,2,3$)\ and $\Omega _{nR}$ ($n=1,2$). Such
exotic fermions are assigned as singlet representations of the $SU\left(
2\right) _{L}\times SU\left( 2\right) _{R}$ group. The above mentioned
exotic fermion content is the minimal one required to generate tree level
masses via a Universal seesaw mechanism for the bottom, charm and strange
quarks, as well as the tau and muon, and one loop level masses for the first
generation SM charged fermions, i.e., the up, down quarks, and the electron.

The scalar assignments under the $SU(3)_{C}\times SU\left( 2\right)
_{L}\times SU\left( 2\right) _{R}\times U\left( 1\right) _{B-L}$ group are: 
\begin{eqnarray}
\Phi &=&\left( 
\begin{array}{cc}
\frac{1}{\sqrt{2}}\left( v_{1}+\phi _{1R}^{0}+i\phi _{1I}^{0}\right) & \phi
_{2}^{+} \\ 
\phi _{1}^{-} & \frac{1}{\sqrt{2}}\left( v_{2}+\phi _{2R}^{0}+i\phi
_{2I}^{0}\right)%
\end{array}%
\right) \sim \left( \mathbf{1},\mathbf{2,2},0\right) ,  \notag \\
\chi _{L} &=&\left( 
\begin{array}{c}
\chi _{L}^{+} \\ 
\frac{1}{\sqrt{2}}\left( v_{L}+\func{Re}\chi _{L}^{0}+i\func{Im}\chi
_{L}^{0}\right)%
\end{array}%
\right) \sim \left( \mathbf{1},\mathbf{2,1},1\right) ,\hspace{1cm}\chi
_{R}=\left( 
\begin{array}{c}
\chi _{R}^{+} \\ 
\frac{1}{\sqrt{2}}\left( v_{R}+\func{Re}\chi _{R}^{0}+i\func{Im}\chi
_{R}^{0}\right)%
\end{array}%
\right) \sim \left( \mathbf{1},\mathbf{1,2},1\right) ,  \notag \\
\phi _{L} &=&\left( 
\begin{array}{c}
\phi _{L}^{+} \\ 
\frac{1}{\sqrt{2}}\left( \func{Re}\phi _{L}^{0}+i\func{Im}\phi
_{L}^{0}\right)%
\end{array}%
\right) \sim \left( \mathbf{1},\mathbf{2,1},1\right) ,\hspace{1cm}\phi
_{R}=\left( 
\begin{array}{c}
\phi _{R}^{+} \\ 
\frac{1}{\sqrt{2}}\left( \func{Re}\phi _{R}^{0}+i\func{Im}\phi
_{R}^{0}\right)%
\end{array}%
\right) \sim \left( \mathbf{1},\mathbf{1,2},1\right) ,  \notag \\
\sigma &\sim &\left( \mathbf{1},\mathbf{1,1},0\right) ,\hspace{1cm}\varphi
\sim \left( \mathbf{1},\mathbf{1,1},0\right) ,\hspace{1cm}\eta \sim \left( 
\mathbf{1},\mathbf{1,1},0\right) ,\hspace{1cm}\rho \sim \left( \mathbf{1},%
\mathbf{1,1},0\right) .
\end{eqnarray}%
To implement the tree level Universal mechanism we have introduced the
scalars $\chi _{L}$, $\chi _{R}$ which are responsible for generating tree
level mixings between the exotic and SM fermions. Besides that, the scalar
fields $\phi _{L}$, $\phi _{R}$ are required for the implementation of the
radiative seesaw mechanism that produces the masses for the first generation
SM charged fermions.

We have further introduced the gauge singlet scalars $\eta $ and $\varphi $
which are crucial for the implementation of the radiative inverse seesaw
mechanism necessary to produce the light active neutrino masses.
Furthermore, the gauge singlet scalar $\sigma $ provides tree level masses
for the exotic $T$ , $T^{\prime }$, $B_{2}$ and $B^{\prime }$ quarks.
Besides that, the gauge singlet scalars $\rho $ and $\eta $ are included in
the scalar spectrum in order to provide tree level masses for the exotic
down type quark $B_{1}$, for the exotic leptons $E_{n}$, $E^{\prime }$\ and $%
\Omega _{nR}$ ($n=1,2$), without the need of invoking soft-breaking mass
terms. Furthermore, we have also included the scalar bidoublet $\Phi $,
which is responsible for generating the top quark mass from the
renormalizable Yukawa operator $\overline{Q}_{3L}\Phi Q_{iR}$ ($i=1,2,3$).%

The vacuum expectation values (VEVs) of the scalars $\Phi $, $\chi _{L}$ and 
$\chi _{R}$ are: 
\begin{equation}
\left\langle \Phi \right\rangle =\left( 
\begin{array}{cc}
v_{1} & 0 \\ 
0 & v_{2}%
\end{array}%
\right) ,\hspace{1.5cm}\left\langle \chi _{L}\right\rangle =\left( 
\begin{array}{c}
0 \\ 
v_{L}%
\end{array}%
\right) ,\hspace{1.5cm}\left\langle \chi _{R}\right\rangle =\left( 
\begin{array}{c}
0 \\ 
v_{R}%
\end{array}%
\right) ,
\end{equation}
where for the sake of simplicity we will set $v_{2}=0$.

The fermion assignments under $Z_{4}^{\left( 1\right) }\times Z_{4}^{\left(
2\right) }$ are: 
\begin{eqnarray*}
Q_{nL} &\sim &\left( -1,-1\right) ,\hspace{1cm}Q_{3L}\sim \left( i,-1\right)
,\hspace{1cm}Q_{jR}\sim \left( 1,1\right) ,\hspace{1cm} \\
T_{L} &\sim &\left( 1,1\right) ,\hspace{1cm}T_{R}\sim \left( 1,-1\right) ,%
\hspace{1cm}T_{L}^{\prime }\sim \left( 1,-i\right) ,\hspace{1cm}%
T_{R}^{\prime }\sim \left( 1,i\right) , \\
B_{nL} &\sim &\left( 1,1\right) ,\hspace{1cm}B_{1R}\sim \left( -i,-1\right) ,%
\hspace{1cm}B_{2R}\sim \left( 1,-1\right) ,\hspace{1cm}B_{L}^{\prime }\sim
\left( 1,i\right) ,\hspace{1cm}B_{R}^{\prime }\sim \left( 1,-i\right) , \\
L_{jL} &\sim &\left( 1,i\right) ,\hspace{1cm}L_{jR}\sim \left( -i,-i\right) ,%
\hspace{1cm}N_{jR}\sim \left( i,i\right) ,\hspace{1cm}\Omega _{nR}\sim
\left( -i,1\right) ,\hspace{1cm}j=1,2,3, \\
E_{nL} &\sim &\left( -i,-i\right) ,\hspace{1cm}E_{nR}\sim \left( -1,i\right)
,\hspace{1cm}E_{L}^{\prime }\sim \left( -i,1\right) ,\hspace{1cm}%
E_{R}^{\prime }\sim \left( -1,-1\right) ,\hspace{1cm}n=1,2.
\end{eqnarray*}

The scalar fields have the following $Z_{4}^{\left( 1\right) }\times
Z_{4}^{\left( 2\right) }$ assignments: 
\begin{eqnarray}
\Phi &\sim &\left( i,-1\right) ,\hspace{1cm}\chi _{L}\sim \left( -1,1\right)
,\hspace{1cm}\chi _{R}\sim \left( 1,1\right) ,\hspace{1cm}\phi _{L}\sim
\left( -1,-i\right) ,\hspace{1cm}\phi _{R}\sim \left( 1,-i\right)  \notag \\
\varphi &\sim &\left( 1,i\right) ,\hspace{1cm}\sigma \sim \left( 1,-1\right)
,\hspace{1cm}\eta \sim \left( -1,1\right) ,\hspace{1cm}\rho \sim \left(
i,-1\right) .
\end{eqnarray}%
The fermion and scalar assignments under the $SU(3)_{C}\times
SU(2)_{L}\times SU(2)_{R}\times U(1)_{B-L}\times Z_{4}^{\left( 1\right)
}\times Z_{4}^{\left( 2\right) }$ symmetry are shown in Tables \ref{fermions}
and \ref{scalars}, respectively.

\begin{table}[tbp]
\begin{equation*}
\begin{array}{|c|c|c|c|c|c|c|c|c|c|c|c|c|c|c|c|c|c|c|c|c|}
\hline
& Q_{nL} & Q_{3L} & Q_{iR} & L_{iL} & L_{iR} & T_{L} & T_{R} & T_{L}^{\prime
} & T_{R}^{\prime } & B_{nL} & B_{1R} & B_{2R} & B_{L}^{\prime } & 
B_{R}^{\prime } & E_{nL} & E_{nR} & E_{L}^{\prime } & E_{R}^{\prime } & 
N_{iR} & \Omega _{nR} \\ \hline
SU(3)_{C} & \mathbf{3} & \mathbf{3} & \mathbf{3} & \mathbf{1} & \mathbf{1} & 
\mathbf{3} & \mathbf{3} & \mathbf{3} & \mathbf{3} & \mathbf{3} & \mathbf{3}
& \mathbf{3} & \mathbf{3} & \mathbf{3} & \mathbf{1} & \mathbf{1} & \mathbf{1}
& \mathbf{1} & \mathbf{1} & \mathbf{1} \\ \hline
SU\left( 2\right) _{L} & \mathbf{2} & \mathbf{2} & \mathbf{1} & \mathbf{2} & 
\mathbf{1} & \mathbf{1} & \mathbf{1} & \mathbf{1} & \mathbf{1} & \mathbf{1}
& \mathbf{1} & \mathbf{1} & \mathbf{1} & \mathbf{1} & \mathbf{1} & \mathbf{1}
& \mathbf{1} & \mathbf{1} & \mathbf{1} & \mathbf{1} \\ \hline
SU\left( 2\right) _{R} & \mathbf{1} & \mathbf{1} & \mathbf{2} & \mathbf{1} & 
\mathbf{2} & \mathbf{1} & \mathbf{1} & \mathbf{1} & \mathbf{1} & \mathbf{1}
& \mathbf{1} & \mathbf{1} & \mathbf{1} & \mathbf{1} & \mathbf{1} & \mathbf{1}
& \mathbf{1} & \mathbf{1} & \mathbf{1} & \mathbf{1} \\ \hline
U\left( 1\right) _{B-L} & \frac{1}{3} & \frac{1}{3} & \frac{1}{3} & -1 & -1
& \frac{4}{3} & \frac{4}{3} & \frac{4}{3} & \frac{4}{3} & -\frac{2}{3} & -%
\frac{2}{3} & -\frac{2}{3} & -\frac{2}{3} & -\frac{2}{3} & -2 & -2 & -2 & -2
& 0 & 0 \\ \hline
Z_{4}^{\left( 1\right) } & -1 & i & 1 & 1 & -i & 1 & 1 & 1 & 1 & 1 & -i & 1
& 1 & 1 & -i & -1 & -i & -1 & i & -i \\ \hline
Z_{4}^{\left( 2\right) } & -1 & -1 & 1 & i & -i & 1 & -1 & -i & i & 1 & -1 & 
-1 & i & -i & -i & i & 1 & -1 & i & 1 \\ \hline
\end{array}%
\end{equation*}%
\caption{Fermion assignments under $SU(3)_{C}\times SU(2)_{L}\times
SU(2)_{R}\times U(1)_{B-L}\times Z_{4}^{\left( 1\right) }\times
Z_{4}^{\left( 2\right) }$. Here $i=1,2,3$ and $n=1,2$}
\label{fermions}
\end{table}
\begin{table}[tbp]
\begin{equation*}
\begin{array}{|c|c|c|c|c|c|c|c|c|c|}
\hline
& \Phi & \chi _{L} & \chi _{R} & \phi _{L} & \phi _{R} & \varphi & \sigma & 
\eta & \rho \\ \hline
SU(3)_{C} & \mathbf{1} & \mathbf{1} & \mathbf{1} & \mathbf{1} & \mathbf{1} & 
\mathbf{1} & \mathbf{1} & \mathbf{1} & \mathbf{1} \\ \hline
SU\left( 2\right) _{L} & \mathbf{2} & \mathbf{2} & \mathbf{1} & \mathbf{2} & 
\mathbf{1} & \mathbf{1} & \mathbf{1} & \mathbf{1} & \mathbf{1} \\ \hline
SU\left( 2\right) _{R} & \mathbf{2} & \mathbf{1} & \mathbf{2} & \mathbf{1} & 
\mathbf{2} & \mathbf{1} & \mathbf{1} & \mathbf{1} & \mathbf{1} \\ \hline
U\left( 1\right) _{B-L} & 0 & 1 & 1 & 1 & 1 & 0 & 0 & 0 & 0 \\ \hline
Z_{4}^{\left( 1\right) } & i & -1 & 1 & -1 & 1 & 1 & 1 & -1 & i \\ \hline
Z_{4}^{\left( 2\right) } & -1 & 1 & 1 & -i & -i & i & -1 & 1 & -1 \\ \hline
\end{array}%
\end{equation*}%
\caption{Scalar assignments under $SU(3)_{C}\times SU(2)_{L}\times
SU(2)_{R}\times U(1)_{B-L}\times Z_{4}^{\left( 1\right) }\times
Z_{4}^{\left( 2\right) }$.}
\label{scalars}
\end{table}

Let us note that all scalar fields acquire nonvanishing vacuum expectation
values, excepting the scalar singlet $\varphi $, as well as the $\phi _{L}$
and $\phi _{R}$ fields whose $Z_{4}^{\left( 2\right) }$ charges correspond
to nontrivial charges under the preserved remnant $Z_{2}$ symmetry.
Furthermore, due to such remnant $Z_{2}$ symmetry, the real and imaginary
parts of the scalar singlet $\varphi $ and of the neutral components of the $%
\phi _{L}$ and $\phi _{R}$ fields will not have mixings with the remaining
CP even and CP odd neutral scalar fields of the model.

It is worth mentioning that the preserved $Z_{2}$ symmetry allows for stable
scalar and fermionic dark matter candidates. The scalar dark matter
candidate is the lightest among the $\func{Re}\varphi $, $\func{Im}\varphi $%
, $\func{Re}\phi _{L}^{0}$, $\func{Re}\phi _{R}^{0}$, $\func{Im}\phi
_{L}^{0} $\ and $\func{Im}\phi _{R}^{0}$\ fields. The fermionic dark matter
candidate is the lightest among the right handed Majorana neutrinos $N_{iR}$
($i=1,2,3$). In the scenario of a scalar DM candidate, it annihilates mainly
into $WW$, $ZZ$, $t\overline{t}$, $b\overline{b}$ and $h_{SM}h_{SM}$ via a
Higgs portal scalar interaction. These annihilation channels will contribute
to the DM relic density, which can be accommodated for appropriate values of
the scalar DM mass and of the coupling of the Higgs portal scalar
interaction. Some studies of the dark matter constraints for the scenario of
scalar singlet dark matter candidate are provided in ~\cite%
{Escudero:2016gzx,Bernal:2017xat,CarcamoHernandez:2020ehn}. Thus, for the DM
direct detection prospects, the scalar DM candidate would scatter off a
nuclear target in a detector via Higgs boson exchange in the $t$-channel,
giving rise to a constraint on the Higgs portal scalar interaction coupling.
Regarding the scenario of fermionic DM candidate, the Dark matter relic
abundance can be obtained through freeze-in, as shown in \cite%
{Bernal:2017xat}. The resulting constraints can therefore be fulfilled for
an appropriate region of parameter space, along similar lines of Refs.~\cite%
{Bernal:2017xat,Han:2019lux,Cabrera:2020lmg,CarcamoHernandez:2021iat,Abada:2021yot}%
. A detailed study of the implications of our model in dark matter is beyond
the scope of this work and will be done elsewhere.

With the above particle content, the following relevant Yukawa terms arise: 
\begin{eqnarray}
-\mathcal{L}_{Y} &=&\dsum\limits_{i=1}^{3}\alpha _{i}\overline{Q}_{3L}\Phi
Q_{iR}+\dsum\limits_{n=1}^{2}x_{n}^{\left( T\right) }\overline{Q}_{nL}%
\widetilde{\chi }_{L}T_{R}+\dsum\limits_{i=1}^{3}z_{i}^{\left( T\right) }%
\overline{T}_{L}\widetilde{\chi }_{R}^{\dagger
}Q_{iR}+\dsum\limits_{n=1}^{2}w_{n}^{\left( T^{\prime }\right) }\overline{Q}%
_{nL}\widetilde{\phi }_{L}T_{R}^{\prime
}+\dsum\limits_{i=1}^{3}r_{i}^{\left( T^{\prime }\right) }\bar{T}%
_{L}^{\prime }\widetilde{\phi }_{R}^{\dagger }Q_{iR}  \notag \\
&&+x_{3}^{\left( B\right) }\overline{Q}_{3L}\chi
_{L}B_{1R}+\dsum\limits_{n=1}^{2}x_{n2}^{\left( B\right) }\overline{Q}%
_{nL}\chi
_{L}B_{2R}+\dsum\limits_{n=1}^{2}\dsum\limits_{i=1}^{3}z_{ni}^{\left(
B\right) }\overline{B}_{nL}\chi _{R}^{\dagger
}Q_{iR}+\dsum\limits_{n=1}^{2}w_{n}^{\left( B^{\prime }\right) }\overline{Q}%
_{nL}\phi _{L}B_{R}^{\prime }+\dsum\limits_{i=1}^{3}r_{i}^{\left( B^{\prime
}\right) }\bar{B}_{L}^{\prime }\phi _{R}^{\dagger }Q_{iR}  \notag \\
&&+y_{T}\overline{T}_{L}\sigma T_{R}+y_{T^{\prime }}\bar{T}_{L}^{\prime
}\sigma T_{R}^{\prime }+y_{B_{1}}\overline{B}_{1L}\rho B_{1R}+y_{B_{2}}%
\overline{B}_{2L}\sigma B_{2R}+y_{B^{\prime }}\bar{B}_{L}^{\prime }\sigma
B_{R}^{\prime }+\dsum\limits_{n=1}^{2}y_{E_{n}}\overline{E}_{nL}\rho
E_{nR}+y_{E^{\prime }}\bar{E}_{L}^{\prime }\rho E_{R}^{\prime }  \notag \\
&&+\dsum\limits_{i=1}^{3}\dsum\limits_{n=1}^{2}x_{in}^{\left( E\right) }%
\overline{L}_{iL}\chi
_{L}E_{nR}+\dsum\limits_{n=1}^{2}\dsum\limits_{i=1}^{3}z_{nj}^{\left(
E\right) }\overline{E}_{nL}\chi _{R}^{\dagger
}L_{jR}+\dsum\limits_{i=1}^{3}w_{i}^{\left( E^{\prime }\right) }\overline{L}%
_{iL}\phi _{L}E_{R}^{\prime }+\dsum\limits_{i=1}^{3}r_{i}^{\left( E^{\prime
}\right) }\bar{E}_{L}^{\prime }\phi _{R}^{\dagger }L_{iR}  \notag \\
&&+\dsum\limits_{i=1}^{3}\dsum\limits_{j=1}^{3}y_{ij}^{\left( L\right) }%
\overline{L}_{iL}\Phi
L_{jR}+\dsum\limits_{i=1}^{3}\dsum\limits_{j=1}^{3}x_{ij}^{\left( N\right) }%
\overline{N_{iR}^{C}}\widetilde{\chi }_{R}^{\dagger
}L_{jR}+\dsum\limits_{n=1}^{2}\left( y_{\Omega }\right) _{n}\overline{\Omega 
}_{nR}\Omega _{nR}^{C}\eta
+\dsum\limits_{i=1}^{3}\dsum\limits_{k=1}^{2}x_{ik}^{\left( S\right) }%
\overline{N}_{iR}\Omega _{kR}^{C}\varphi +H.c.  \label{Ly}
\end{eqnarray}

To close this section, in the following we discuss the implications of our
model for flavor changing neutral currents (FCNC). The FCNC in the down type
quark sector are expected to be very suppressed since at energies below the
scale $v_{R}$ of breaking of the left-right symmetry, only the $SU\left(
2\right) _{L}$ scalar doublet $\chi _{L}$ will appear in the down type quark
Yukawa terms. In what regards the up type quark sector, there would be FCNC
at tree level, since at low energies (below $v_{R}$), the bidoblet scalar $%
\Phi $ and the $SU\left( 2\right) _{L}$ scalar doublet $\chi _{L}$
participate in the up type quark Yukawa interactions. However, such FCNC
which can give rise to meson oscillations, can be suppressed by appropiate
values of the Yukawa couplings and heavy non SM neutral scalar masses.
Furthermore, concerning the charged lepton sector, the corresponding FCNC
can be suppressed by making the matrix $y_{ij}^{\left( L\right) }$ diagonal. 
%
\newpage

\section{Fermion mass matrices.}

\label{fermionmasses} From the Yukawa interactions, we find that the mass
matrices for SM charged fermions are given by:

\begin{eqnarray}
M_{U} &=&\left( 
\begin{array}{ccc}
\Delta _{U} & 0_{2\times 1} & A_{U} \\ 
0_{1\times 2} & m_{t} & 0 \\ 
B_{U} & 0 & m_{T}%
\end{array}%
\right) ,\hspace{1cm}\hspace{1cm}A_{U}=\left( 
\begin{array}{c}
x_{1}^{\left( T\right) } \\ 
x_{2}^{\left( T\right) }%
\end{array}%
\right) \frac{v_{_{L}}}{\sqrt{2}},  \notag \\
B_{U} &=&\left( 
\begin{array}{cc}
z_{1}^{\left( T\right) }, & z_{2}^{\left( T\right) }%
\end{array}%
\right) \frac{v_{R}}{\sqrt{2}},\hspace{1cm}m_{t}=\alpha _{3}\frac{v_{1}}{%
\sqrt{2}},  \label{MU}
\end{eqnarray}

\begin{eqnarray}
M_{D} &=&\left( 
\begin{array}{cc}
\Delta _{D} & A_{D} \\ 
B_{D} & M_{B}%
\end{array}%
\right) ,\hspace{1cm}\hspace{1cm}A_{D}=\left( 
\begin{array}{cc}
0 & x_{12}^{\left( B\right) } \\ 
0 & x_{22}^{\left( B\right) } \\ 
x_{3}^{\left( B\right) } & 0%
\end{array}%
\right) \frac{v_{L}}{\sqrt{2}},  \notag \\
B_{D} &=&\left( 
\begin{array}{ccc}
z_{11}^{\left( B\right) }, & z_{12}^{\left( B\right) }, & z_{13}^{\left(
B\right) } \\ 
z_{21}^{\left( B\right) }, & z_{22}^{\left( B\right) }, & z_{23}^{\left(
B\right) }%
\end{array}%
\right) \frac{v_{R}}{\sqrt{2}},\hspace{1cm}\hspace{1cm}M_{B}=\left( 
\begin{array}{cc}
m_{B_{1}} & 0 \\ 
0 & m_{B_{2}}%
\end{array}%
\right) ,  \label{MD}
\end{eqnarray}

\begin{eqnarray}
M_{E} &=&\left( 
\begin{array}{cc}
\Delta _{E} & A_{E} \\ 
B_{E} & C_{E}%
\end{array}%
\right) ,\hspace{1cm}\hspace{1cm}A_{E}=\left( 
\begin{array}{cc}
x_{11}^{\left( E\right) } & x_{12}^{\left( E\right) } \\ 
x_{21}^{\left( E\right) } & x_{22}^{\left( E\right) } \\ 
x_{31}^{\left( E\right) } & x_{32}^{\left( E\right) }%
\end{array}%
\right) \frac{v_{L}}{\sqrt{2}},  \notag \\
B_{D} &=&\left( 
\begin{array}{ccc}
z_{11}^{\left( E\right) }, & z_{12}^{\left( E\right) }, & z_{13}^{\left(
E\right) } \\ 
z_{21}^{\left( E\right) }, & z_{22}^{\left( E\right) }, & z_{23}^{\left(
E\right) }%
\end{array}%
\right) \frac{v_{R}}{\sqrt{2}},\hspace{1cm}\hspace{1cm}C_{E}=\left( 
\begin{array}{cc}
m_{E_{1}} & 0 \\ 
0 & m_{E_{2}}%
\end{array}%
\right) ,  \label{ME}
\end{eqnarray}

\begin{figure}[tbp]
\centering
\includegraphics[width = 0.9\textwidth]{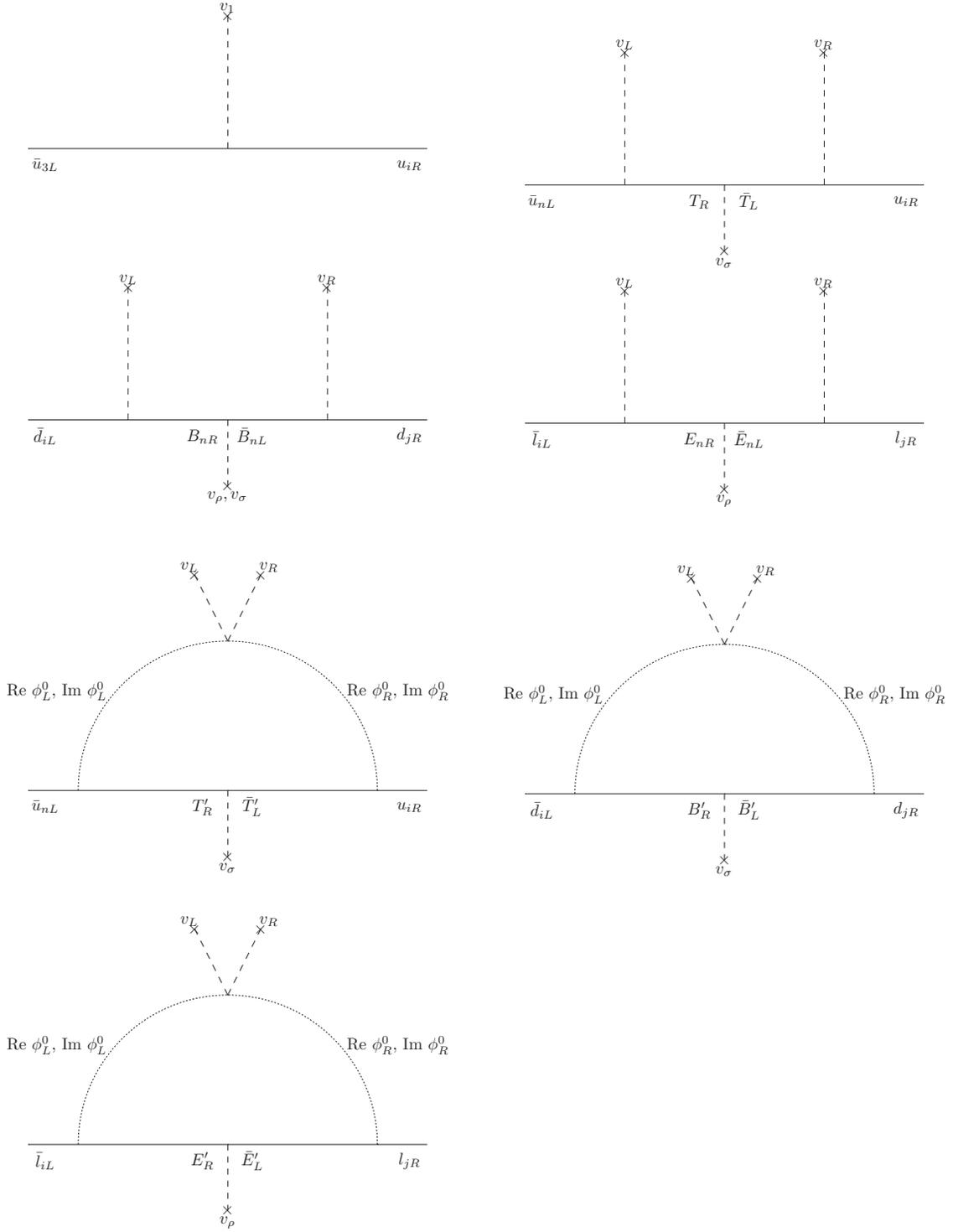}
\caption{Feynman diagrams contributing to the entries of the SM charged
fermion mass matrices. Here, $n=1,2$ and $i,j=1,2,3$.}
\label{Diagramschargedfermions}
\end{figure}

\begin{figure}[tbp]
\centering
\includegraphics[width = 0.9\textwidth]{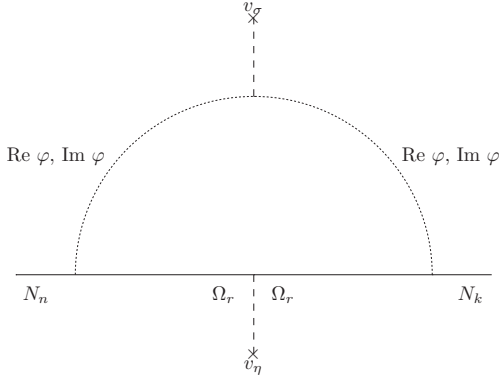}\vspace{-15cm}
\caption{One-loop Feynman diagram contributing to the Majorana neutrino mass
submatrix $\protect\mu $. Here, $n,k=1,2,3$ and $r=1,2$.}
\label{Loopdiagrammu}
\end{figure}

\begin{figure}[tbp]
\centering
\includegraphics[width = 0.9\textwidth]{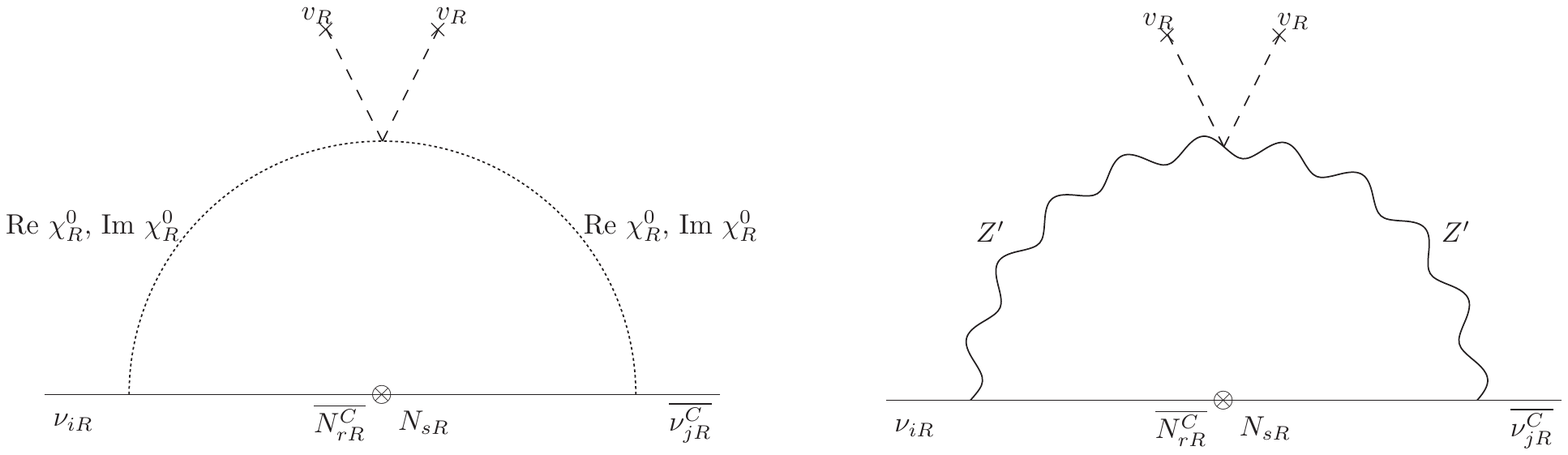}\vspace{-15cm}
\caption{Feynman diagram contributing to the Majorana neutrino mass
submatrix $\tilde{\protect\mu}$. Here, $i,j,r,s=1,2,3$ and the cross mark $%
\otimes$ in the internal lines corresponds to the one loop level induced
Majorana mass term.}
\label{Loopdiagrammutilde}
\end{figure}
\begin{figure}[tbp]
\centering
\includegraphics[width=8.3cm, height=7.5cm]{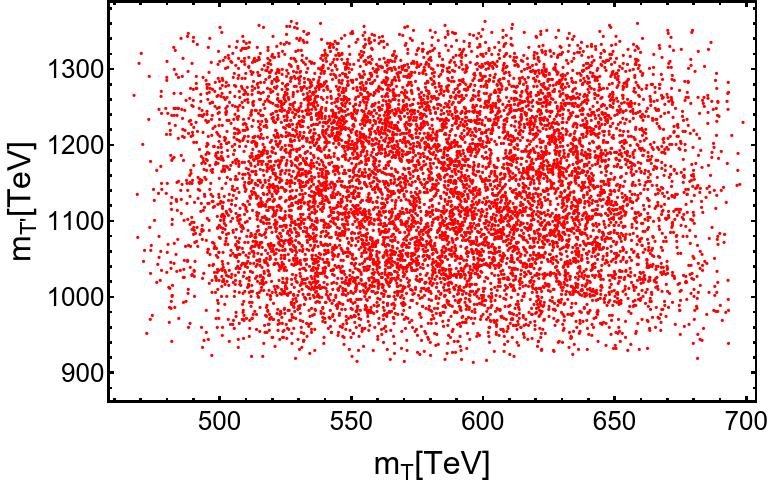}%
\includegraphics[width=8.3cm, height=7.5cm]{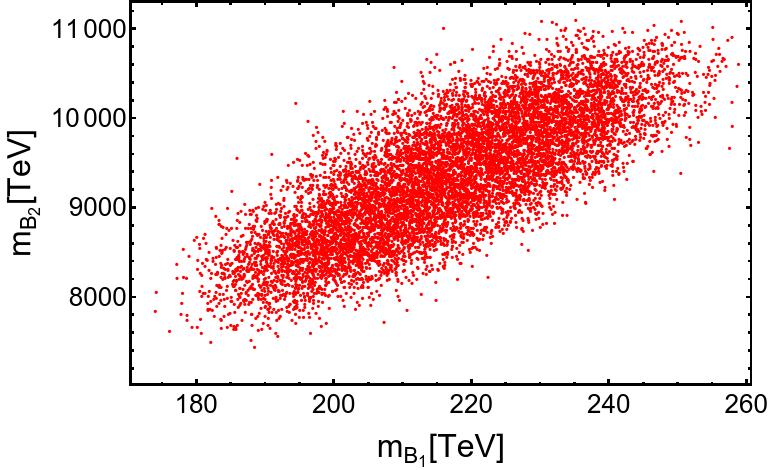}\newline
\includegraphics[width=8.3cm, height=7.5cm]{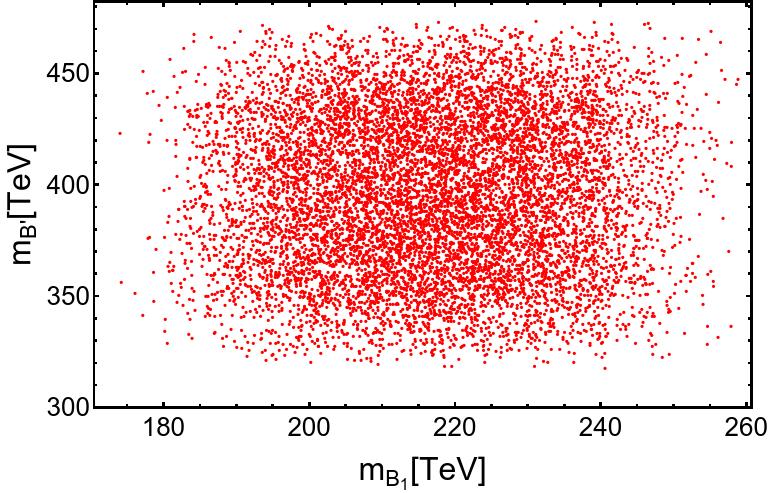}%
\includegraphics[width=8.3cm, height=7.5cm]{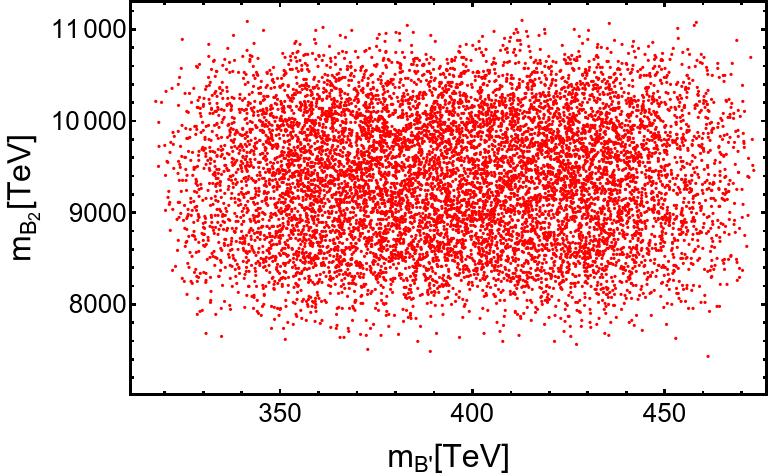}\newline
\includegraphics[width=8.3cm, height=7.5cm]{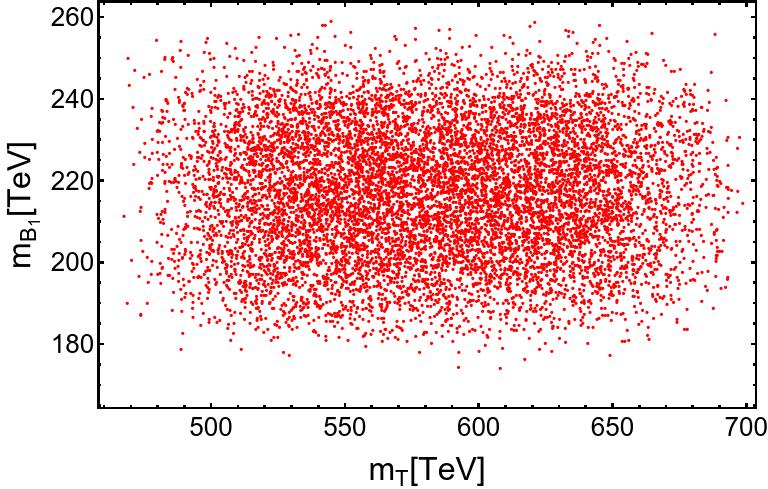}%
\includegraphics[width=8.3cm, height=7.5cm]{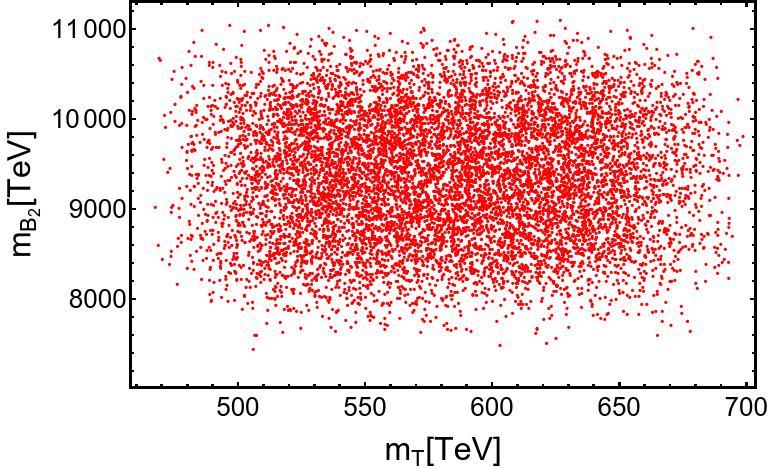}\newline
\caption{Correlations between the heavy exotic quark masses.}
\label{nonSMquarkmasses}
\end{figure}
\begin{figure}[tbp]
\centering
\includegraphics[width=0.5\textwidth]{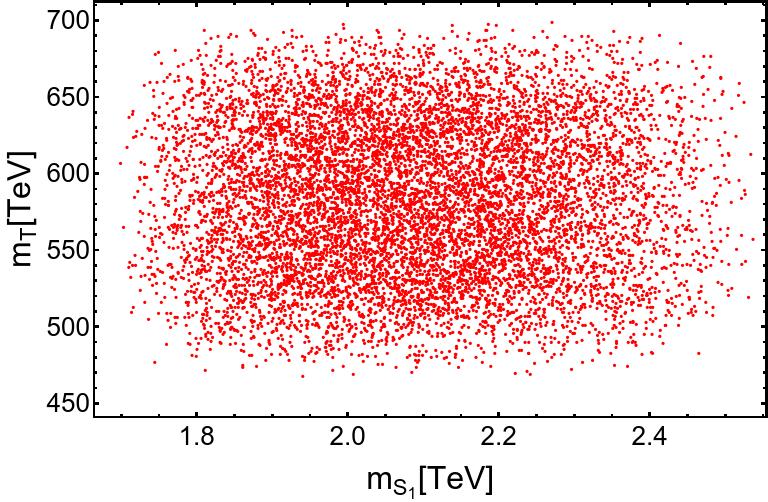}%
\includegraphics[width=0.5\textwidth]{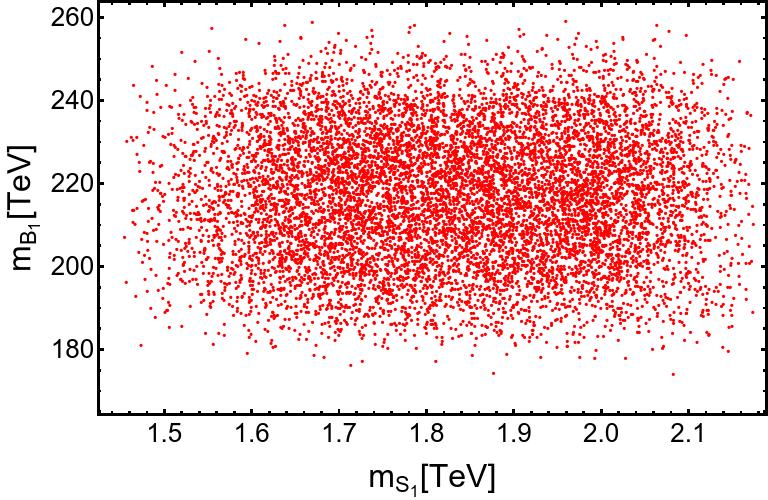}\newline
\includegraphics[width=0.5\textwidth]{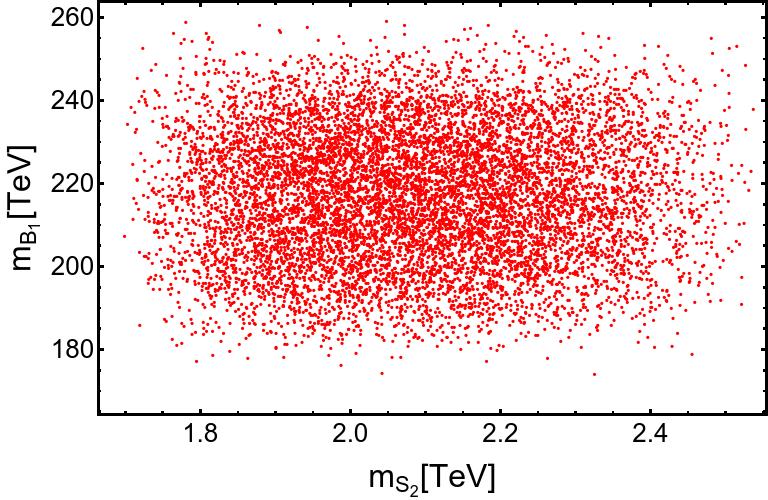}%
\includegraphics[width=0.5\textwidth]{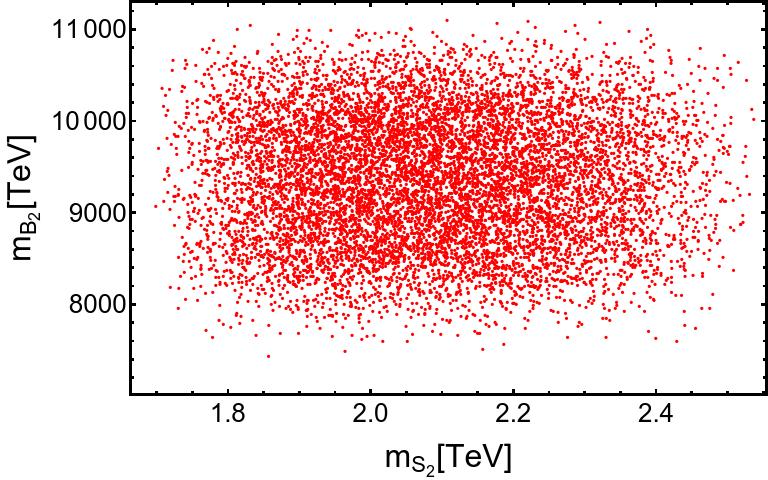}\newline
\caption{Correlations between the exotic quark masses and the masses $m_{S_1}$ and $m_{S_2}$ of the inert scalars $S_1$ and $S_2$, respectively. }
\label{heavyquarkvsscalars}
\end{figure}

where we have set $\alpha _{1}=\alpha _{2}=0$ to strongly suppress the tree
level FCNC in the quark sector. As seen from Eqs. (\ref{MU}), (\ref{MD}) and
(\ref{ME}), the exotic heavy vector-like fermions mix with the SM fermions
lighter than top quark. The masses of these vector-like fermions are much
larger than the scale of breaking of the left-right symmetry $v_{R}\sim 
\mathcal{O}(10)$ TeV, since the gauge singlet scalars $\eta $, $\sigma $ and 
$\rho $ are assumed to acquire vacuum expectation values much larger than
this scale. Therefore, the charm, bottom and strange quarks, as well as the
tau and muon leptons, acquire their masses from the tree-level Universal
seesaw mechanism, whereas the first generation SM charged fermions, i.e.,
the up, down quarks and the electron get one-loop level masses from a
radiative seesaw mechanism. Thus, the SM charged fermion mass matrices take
the form: 
\begin{eqnarray}
\widetilde{M}_{U} &=&\left( 
\begin{array}{cc}
\Delta _{U}-A_{U}M_{\widetilde{T}}^{-1}B_{U} & 0_{2\times 1} \\ 
0_{1\times 2} & m_{t}%
\end{array}%
\right) , \\
\widetilde{M}_{D} &=&\Delta _{D}-A_{D}M_{B}^{-1}B_{D}, \\
\widetilde{M}_{E} &=&\Delta _{E}-A_{E}M_{E}^{-1}B_{E},
\end{eqnarray}%
where $\Delta _{U}$, $\Delta _{D}$ and $\Delta _{E}$ are the one loop level
contributions to the SM charged fermion mass matrices arising from the
one-loop Feynman diagrams of Figure \ref{Diagramschargedfermions}. It is
worth mentioning that the first and second Feynman diagrams of the first row
of Figure \ref{Diagramschargedfermions} contribute to the $\left( 3,i\right) 
$ and $\left( n,i\right) $ ($i=1,2,3$ and $n=1,2$) entries of the SM up type
quark mass matrix, respectively. The first and the second diagrams from the
second row of Figure \ref{Diagramschargedfermions} contribute to the $\left(
i,j\right) $ ($i,j=1,2,3$) entries of the SM down type quark and SM charged
lepton mass matrices, respectively. Furthermore, the one loop level
contributions to the $\left( n,i\right) $ entries of the SM up type quark
mass matrix arise from the first diagram of the third row of Figure \ref%
{Diagramschargedfermions}. On the other hand, the second diagram of the
third row of Figure \ref{Diagramschargedfermions} generates the one loop
level contribution to the $\left( i,j\right) $ entries of the SM down type
quark mass matrix. Finally, the last diagram of Figure \ref%
{Diagramschargedfermions} yields the one loop level contribution to the $%
\left( i,j\right) $ entries of the SM charged lepton mass matrix. The one
loop level contributions to the SM charged fermion mass matrices are given
by: 
\begin{eqnarray}
\Delta _{U} &=&\frac{m_{T^{\prime }}}{16\pi ^{2}}\left( 
\begin{array}{ccc}
w_{1}^{\left( T^{\prime }\right) }r_{1}^{\left( T^{\prime }\right) } & 
w_{1}^{\left( T^{\prime }\right) }r_{2}^{\left( T^{\prime }\right) } & 
w_{1}^{\left( T^{\prime }\right) }r_{3}^{\left( T^{\prime }\right) } \\ 
w_{2}^{\left( T^{\prime }\right) }r_{1}^{\left( T^{\prime }\right) } & 
w_{2}^{\left( T^{\prime }\right) }r_{2}^{\left( T^{\prime }\right) } & 
w_{2}^{\left( T^{\prime }\right) }r_{3}^{\left( T^{\prime }\right) } \\ 
0 & 0 & 0%
\end{array}%
\right)  \label{DeltaU} \\
&&\times \left\{ \left[ f\left( m_{S_{1}}^{2},m_{T^{\prime }}^{2}\right)
-f\left( m_{S_{2}}^{2},m_{T^{\prime }}^{2}\right) \right] \sin 2\theta _{S}-%
\left[ f\left( m_{P_{1}}^{2},m_{T^{\prime }}^{2}\right) -f\left(
m_{P_{2}}^{2},m_{T^{\prime }}^{2}\right) \right] \sin 2\theta _{P}\right\} ,
\notag \\
\Delta _{D} &=&\frac{2m_{B^{\prime }}}{16\pi ^{2}}\left( 
\begin{array}{ccc}
w_{1}^{\left( B^{\prime }\right) }r_{1}^{\left( B^{\prime }\right) } & 
w_{1}^{\left( B^{\prime }\right) }r_{2}^{\left( B^{\prime }\right) } & 
w_{1}^{\left( B^{\prime }\right) }r_{3}^{\left( B^{\prime }\right) } \\ 
w_{2}^{\left( B^{\prime }\right) }r_{1}^{\left( B^{\prime }\right) } & 
w_{2}^{\left( B^{\prime }\right) }r_{2}^{\left( B^{\prime }\right) } & 
w_{2}^{\left( B^{\prime }\right) }r_{3}^{\left( B^{\prime }\right) } \\ 
0 & 0 & 0%
\end{array}%
\right)  \label{DeltaD} \\
&&\times \left\{ \left[ f\left( m_{S_{1}}^{2},m_{B^{\prime }}^{2}\right)
-f\left( m_{S_{2}}^{2},m_{B^{\prime }}^{2}\right) \right] \sin 2\theta _{S}-%
\left[ f\left( m_{P_{1}}^{2},m_{B^{\prime }}^{2}\right) -f\left(
m_{P_{2}}^{2},m_{B^{\prime }}^{2}\right) \right] \sin 2\theta _{P}\right\} ,
\notag \\
\Delta _{E} &=&\frac{2m_{E^{\prime }}}{16\pi ^{2}}\left( 
\begin{array}{ccc}
w_{1}^{\left( E^{\prime }\right) }r_{1}^{\left( E^{\prime }\right) } & 
w_{1}^{\left( E^{\prime }\right) }r_{2}^{\left( E^{\prime }\right) } & 
w_{1}^{\left( E^{\prime }\right) }r_{3}^{\left( E^{\prime }\right) } \\ 
w_{2}^{\left( E^{\prime }\right) }r_{1}^{\left( E^{\prime }\right) } & 
w_{2}^{\left( E^{\prime }\right) }r_{2}^{\left( E^{\prime }\right) } & 
w_{2}^{\left( E^{\prime }\right) }r_{3}^{\left( E^{\prime }\right) } \\ 
w_{3}^{\left( E^{\prime }\right) }r_{1}^{\left( E^{\prime }\right) } & 
w_{3}^{\left( E^{\prime }\right) }r_{2}^{\left( E^{\prime }\right) } & 
w_{3}^{\left( E^{\prime }\right) }r_{3}^{\left( E^{\prime }\right) }%
\end{array}%
\right)  \label{DeltaE} \\
&&\times \left\{ \left[ f\left( m_{S_{1}}^{2},m_{E^{\prime }}^{2}\right)
-f\left( m_{S_{2}}^{2},m_{E^{\prime }}^{2}\right) \right] \sin 2\theta _{S}-%
\left[ f\left( m_{P_{1}}^{2},m_{E^{\prime }}^{2}\right) -f\left(
m_{P_{2}}^{2},m_{E^{\prime }}^{2}\right) \right] \sin 2\theta _{P}\right\} ,
\notag
\end{eqnarray}%
where $f\left( m_{1},m_{2}\right) $ is given by: 
\begin{equation}
f\left( m_{1},m_{2}\right) =\frac{m_{1}^{2}}{m_{1}^{2}-m_{2}^{2}}\ln \left( 
\frac{m_{1}^{2}}{m_{2}^{2}}\right) ,
\end{equation}%
and the physical scalars $S_{1}$, $S_{2}$ and pseudoscalars $P_{1}$ and $%
P_{2}$ are given by: 
\begin{equation}
\left( 
\begin{array}{c}
S_{1} \\ 
S_{2}%
\end{array}%
\right) =\left( 
\begin{array}{cc}
\cos \theta _{S} & \sin \theta _{S} \\ 
-\sin \theta _{S} & \cos \theta _{S}%
\end{array}%
\right) \left( 
\begin{array}{c}
\func{Re}\phi _{L}^{0} \\ 
\func{Re}\phi _{R}^{0}%
\end{array}%
\right) ,\hspace{1cm}\left( 
\begin{array}{c}
\ P_{1} \\ 
P_{2}%
\end{array}%
\right) =\left( 
\begin{array}{cc}
\cos \theta _{P} & \sin \theta _{P} \\ 
-\sin \theta _{P} & \cos \theta _{P}%
\end{array}%
\right) \left( 
\begin{array}{c}
\func{Im}\phi _{L}^{0} \\ 
\func{Im}\phi _{R}^{0}%
\end{array}%
\right) .
\end{equation}%
It is worth mentioning that the SM charged fermion mass hierarchy can be
successfully reproduced by having appropriate values for the exotic fermion
masses. For instance, to successfully explain the GeV scale value of the
bottom quark and tau lepton masses, we have that such masses can be
estimated as: 
\begin{equation}
m_{b}\sim m_{\tau }\sim \frac{y^{2}v_{L}v_{R}}{m_{F}}  \label{estimate}
\end{equation}%
where $m_{\mathrm{F}}$ is the mass scale of the exotic fermions, $y$ the SM
fermion-exotic fermion Yukawa coupling and $\lambda $ the quartic scalar
coupling. Taking $v_{L}\sim \mathcal{O}\left( 100\right) $ GeV, $v_{R}\sim 
\mathcal{O}\left( 10\right) $ TeV, $m_{F}\sim \mathcal{O}\left( 100\right) $
TeV and $y\sim \mathcal{O}\left( 0.4\right) $, Eq. (\ref{estimate}) takes
the form $m_{b}\sim m_{\tau }\sim \mathcal{O}\left( 1\right) $ GeV, thus
showing that our model naturally explains the smallness of the bottom and
tau masses with respect to the top quark mass. Furthermore, the hierarchy
between the masses of the remaining SM charged fermions lighter than the top
quark can be accommodated by having some deviation from the scenario of
universality of the Yukawa couplings in both quark and lepton sectors. This
would imply some moderate tuning among the Yukawa couplings. However, such a
situation is considerably better compared to that of the minimal Left-Right
symmetric model, where a significant tuning of the Yukawa couplings is
required. In order to find the best fit point that successfully
reproduces the SM quark masses and CKM parameters, we proceed to minimize
the following $\chi ^{2}$ function: 
\begin{equation}
\chi ^{2}=\sum_{f}\frac{(m_{f}^{\text{th}}-m_{f}^{\text{exp}})^{2}}{\sigma
_{f}^{2}}+\frac{(|\mathbf{V}_{12}^{\text{th}}|-|\mathbf{V}_{12}^{\text{exp}%
}|)^{2}}{\sigma _{12}^{2}}+\frac{(|\mathbf{V}_{23}^{\text{th}}|-|\mathbf{V}%
_{23}^{\text{exp}}|)^{2}}{\sigma _{23}^{2}}+\frac{(|\mathbf{V}_{13}^{\text{th%
}}|-|\mathbf{V}_{13}^{\text{exp}}|)^{2}}{\sigma _{13}^{2}}+\frac{(J_{q}^{%
\text{th}}-J_{q}^{\text{exp}})^{2}}{\sigma _{J}^{2}}\,\;,
\end{equation}%
where $f=u,c,t,d,s,b$ and $J_{q}$ is the Jarlskog parameter. The
experimental values for the quark masses are given by~\cite{Xing:2020ijf}, 
\begin{equation*}
\begin{split}
m_{u}^{\text{exp}}(M_{Z})& =1.24\pm 0.22\text{MeV}\;, \\
m_{c}^{\text{exp}}(M_{Z})& =0.626\pm 0.020\text{GeV}\;, \\
m_{t}^{\text{exp}}(M_{Z})& =172.9\pm 0.04\text{GeV}\;, \\
m_{d}^{\text{exp}}(M_{Z})& =2.69\pm 0.19\text{MeV}\;, \\
m_{s}^{\text{exp}}(M_{Z})& =53.5\pm 4.6\text{MeV}\;, \\
m_{b}^{\text{exp}}(M_{Z})& =2.86\pm 0.03\text{GeV}\;,
\end{split}%
\end{equation*}%
and the CKM parameters are~\cite{Zyla:2020zbs} 
\begin{equation*}
\begin{split}
|\mathbf{V}_{12}^{\text{exp}}|& =0.22452\pm 0.00044\;, \\
|\mathbf{V}_{23}^{\text{exp}}|& =0.04214\pm 0.00076\;, \\
|\mathbf{V}_{13}^{\text{exp}}|& =0.00365\pm 0.00012\;, \\
J_{q}^{\text{exp}}& =(3.18\pm 0.15)\times 10^{-5}\;.
\end{split}%
\end{equation*}%
The magnitudes of the quark Yukawa couplings are randomly varied in the
range $[0.1,1.5]$, whereas their complex phases are ranged between $0$ and $2\pi $. Furthermore, we have fixed $v_{L}=$ $100$\mbox{GeV} and $v_{R}=10$\mbox{TeV} and randomly varied $\theta =\theta _{S}=-\theta _{P}$ in a
small range around $\frac{\pi }{3}$. The masses of the vector like quarks
and inert scalar mediators are varied in the ranges: 
\begin{eqnarray}
0.5\mbox{TeV} &\leq &m_{S_{1}}=m_{P_{1}}\leq 10\mbox{TeV},\hspace{1cm}%
1.01m_{S_{1}}\leq m_{S_{2}}=m_{P_{2}}\leq 1.03m_{S_{1}},\hspace{1cm}1%
\mbox{TeV}\leq m_{T^{\prime }},m_{B^{\prime }}\leq 10^{3}\mbox{TeV},  \notag
\\
10^{2}\mbox{TeV} &\leq &m_{B_{1}}\leq 2\times 10^{2}\mbox{TeV},\hspace{0.9cm}%
10^{2}\frac{m_{b}}{m_{c}}\mbox{TeV}\leq m_{T}\leq 2\times 10^{2}\frac{m_{b}}{%
m_{c}}\mbox{TeV},\hspace{0.9cm}10^{2}\frac{m_{b}}{m_{s}}\mbox{TeV}\leq
m_{B_{2}}\leq 2\times 10^{2}\frac{m_{b}}{m_{s}}\mbox{TeV},\hspace{0.9cm} 
\notag
\end{eqnarray}%
In the above described range of parameters, we find that the minimization of
the $\chi ^{2}$ function yields the following benchmark point, consistent
with the experimental values of the SM quark masses and CKM parameters:
\begin{eqnarray}
\theta &\simeq &85.9^{\circ },\hspace{1cm}m_{S_{1}}=m_{P_{1}}\simeq 1.9%
\mbox{TeV},\hspace{1cm}m_{S_{2}}=m_{P_{2}}\simeq 2.1\mbox{TeV},\hspace{1cm}%
v_{L}\simeq 100\mbox{GeV},\hspace{1cm}v_{R}\simeq 10\mbox{TeV},  \notag \\
m_{T} &\simeq &583\mbox{TeV},\hspace{0.9cm}m_{T^{\prime }}\simeq 1.1\times
10^{3}\mbox{TeV},\hspace{0.9cm}m_{B_{1}}\simeq 216\mbox{TeV},\hspace{0.9cm}%
m_{B_{2}}\simeq 9.3\times 10^{3}\mbox{TeV},\hspace{0.9cm}m_{B^{\prime
}}\simeq 396\mbox{TeV},  \notag \\
x_{1}^{\left( T\right) } &\simeq &0.24-0.02i,\hspace{1cm}x_{2}^{\left(
T\right) }\simeq 0.96-0.06i,\hspace{1cm}z_{1}^{\left( T\right)
}=z_{2}^{\left( T\right) }\simeq -0.16+0.08i,\hspace{1cm}x_{12}^{\left(
B\right) }\simeq -0.05-0.03i,  \notag \\
x_{22}^{\left( B\right) } &\simeq &-0.62-0.06i,\hspace{1cm}x_{3}^{\left(
B\right) }\simeq 0.07-0.62i,\hspace{1cm}z_{11}^{\left( B\right) }\simeq 0.25,%
\hspace{1cm}z_{12}^{\left( B\right) }\simeq 0.49,\hspace{1cm}z_{13}^{\left(
B\right) }\simeq -0.44,  \notag \\
z_{21}^{\left( B\right) } &\simeq &-1.17i,\hspace{1cm}z_{22}^{\left(
B\right) }\simeq 0.95i,\hspace{1cm}z_{23}^{\left( B\right) }\simeq 0.80i,%
\hspace{1cm}w_{1}^{\left( T^{\prime }\right) }\simeq -0.39+0.197i,\hspace{1cm%
}w_{2}^{\left( T^{\prime }\right) }\simeq -0.58+0.29i,  \notag \\
r_{1}^{\left( T^{\prime }\right) } &\simeq &-0.154+0.084i,\hspace{1cm}%
r_{2}^{\left( T^{\prime }\right) }\simeq -0.875+0.48i,\hspace{1cm}%
r_{3}^{\left( T^{\prime }\right) }\simeq 0.30-0.16i,\hspace{1cm}%
w_{1}^{\left( B^{\prime }\right) }\simeq 0.12-0.087i,  \notag \\
w_{2}^{\left( B^{\prime }\right) } &\simeq &0.44+0.74i,\hspace{1cm}%
r_{1}^{\left( B^{\prime }\right) }\simeq -0.17-0.98i,\hspace{1cm}%
r_{2}^{\left( B^{\prime }\right) }\simeq -1.22,\hspace{1cm}r_{3}^{\left(
B^{\prime }\right) }\simeq 1.299+0.698i,
\label{bfpquarks}
\end{eqnarray}
As we can see, the dimensionless quark Yukawa couplings are of order unity
with moderate deviations. This shows that the proposed model is able to
explain the existing pattern of the observed quark spectrum. The resulting correlations among the heavy exotic quark masses and between the exotic quark masses and the masses $m_{S_1}$ and $m_{S_2}$ of the inert scalars $S_1$ and $S_2$ are shown in Figures \ref{nonSMquarkmasses} and \ref{heavyquarkvsscalars}, respectively, which present the allowed region of parameter space for the seesaw mediator masses, consistent with a successful description of the observed pattern of SM quark masses and CKM parameters. As shown in Figures \ref{nonSMquarkmasses} and \ref{heavyquarkvsscalars}, the observed SM quark mass and mixing hierarchy can be successfully accounted for, provided that the heavy vector like quark have masses in the ranges $450$ TeV $\lesssim m_T\lesssim 700$ TeV, $900$ TeV $\lesssim m_{T^{\prime}}\lesssim 1400$ TeV, $7.5\times 10^3$ TeV $\lesssim m_{B_2}\lesssim 11 \times 10^3$ TeV, $300$ TeV $\lesssim m_{T^{\prime}}\lesssim 500$ TeV, wheras the masses of the inert scalars are constrained to be in the ranges $1.4$ TeV $\lesssim m_{S_1}\lesssim 2.2$ TeV and $1.7$ TeV $\lesssim m_{S_2}\lesssim 2.5$ TeV  for $m_{S_1}=m_{P_1}$ and $m_{S_2}=m_{P_2}$.

It may seem that the problem of the hierarchies of SM fermions is not solved
but simply reparameterized in terms of unknown vector-like fermion masses.
However, there are six advantages to this approach. Firstly, the approach is
dynamical, since the vector-like masses, which are dynamically generated
from Yukawa interactions involving the gauge singlet scalars neutral under
the remnant $Z_{2}$ symmetry, are new physical quantities, which could in
principle be determined by a future theory. Secondly, it has experimental
consequences, since the new vector-like charged exotic fermions and right
handed neutrinos can be discovered directly at proton-proton colliders via
their production by gluon fusion (for the exotic quarks only) and Drell Yan
mechanisms, or indirectly from their loop contributions to certain
observables. For instance, the charged exotic vector like leptons, which
mediate the Universal seesaw mechanism that produces the SM charged lepton
masses, are also crucial for accommodating the experimental values of the
muon and electron anomalous magnetic moments, whose magnitudes do not find
an explanation within the context of the Standard Model. Thirdly, this
approach can also account for the small quark mixing angles, as well as the
large lepton mixing angles arising from the neutrino sector. Fourthly, the
effective Yukawa couplings are proportional to a product of two other
dimensionless couplings, so a small hierarchy in those couplings can yield a
quadratically larger hierarchy in the effective couplings. Fifthly, the
masses of the light active neutrinos are dynamically generated via a
radiative inverse seesaw mechanism at one loop level, thanks to the remnant $%
Z_{2}$ symmetry arisen from the spontaneous breaking of the $Z_{4}$
symmetry. Sixthly, the remnant $Z_{2}$ symmetry allows for stable scalar and
fermionic dark matter candidates. For all these reasons, the approach we
follow in this paper is both well motivated and interesting.

Concerning the neutrino sector, we find that the neutrino Yukawa
interactions give rise to the following neutrino mass terms: 
\begin{equation}
-\mathcal{L}_{mass}^{\left( \nu \right) }=\frac{1}{2}\left( 
\begin{array}{ccc}
\overline{\nu _{L}^{C}} & \overline{\nu _{R}} & \overline{N_{R}}%
\end{array}%
\right) M_{\nu }\left( 
\begin{array}{c}
\nu _{L} \\ 
\nu _{R}^{C} \\ 
N_{R}^{C}%
\end{array}%
\right) +\dsum\limits_{n=1}^{2}\left( m_{\Omega }\right) _{n}\overline{%
\Omega }_{nR}\Omega _{nR}^{C}+H.c,  \label{Lnu}
\end{equation}%
where the neutrino mass matrix reads: 
\begin{equation}
M_{\nu }=\left( 
\begin{array}{ccc}
0_{3\times 3} & m_{\nu D} & 0_{3\times 3} \\ 
m_{\nu D}^{T} & \widetilde{\mu } & M \\ 
0_{3\times 3} & M^{T} & \mu%
\end{array}%
\right) ,  \label{Mnu}
\end{equation}%
and the submatrices are given by: 
\begin{eqnarray}
\left( m_{\nu D}\right) _{ij} &=&y_{ij}^{\left( L\right) }\frac{v_{1}}{\sqrt{%
2}},\hspace{0.7cm}\hspace{0.7cm}M_{ij}=x_{ij}^{\left( N\right) }\frac{v_{R}}{%
\sqrt{2}},\hspace{0.7cm}\hspace{0.7cm}i,j,n,k=1,2,3,\hspace{0.7cm}\hspace{%
0.7cm}r=1,2,  \notag \\
\mu _{nk} &=&\dsum\limits_{r=1}^{2}\frac{x_{nr}^{\left( S\right)
}x_{kr}^{\left( S\right) }m_{\Omega _{r}}}{16\pi ^{2}}\left[ \frac{%
m_{\varphi _{R}}^{2}}{m_{\varphi _{R}}^{2}-m_{\Omega _{r}}^{2}}\ln \left( 
\frac{m_{\varphi _{R}}^{2}}{m_{\Omega _{r}}^{2}}\right) -\frac{m_{\varphi
_{I}}^{2}}{m_{\varphi _{I}}^{2}-m_{\Omega _{r}}^{2}}\ln \left( \frac{%
m_{\varphi _{I}}^{2}}{m_{\Omega _{r}}^{2}}\right) \right] ,\hspace{0.7cm}%
\hspace{0.7cm}.
\end{eqnarray}%
The $\mu $ block is generated at one loop level due to the exchange of $%
\Omega _{rR}$ ($r=1,2$) and $\varphi $ in the internal lines, as shown in
Figure \ref{Loopdiagrammu}. To close the corresponding one loop diagram, the
following trilinear scalar interaction is needed: 
\begin{equation}
V_{\mu }=A\left( \varphi ^{\ast }\right) ^{2}\sigma ,
\end{equation}

Furthermore, the $\widetilde{\mu }$ submatrix is generated from the Feynman
diagram of Figure \ref{Loopdiagrammutilde}, which involves the virtual
exchange of $\func{Re}\chi _{R}^{0}$, $\func{Im}\chi _{R}^{0}$, $Z^{\prime }$
as well as the one loop level induced Majorana mass term in the internal
lines of the loop, in analogy with \cite{Pilaftsis:1991ug}. The entries of
the submatrix $\widetilde{\mu }$ are given by:
\begin{eqnarray}
\widetilde{\mu }_{ij} &=&\frac{g_{R}^{2}}{16\pi ^{2}}\mu _{ij}\frac{%
m_{Z^{\prime }}^{2}}{m_{Z^{\prime }}^{2}-\mu _{ij}^{2}}\ln \left( \frac{%
m_{Z^{\prime }}^{2}}{\mu _{ij}^{2}}\right)  \label{mutilde} \\
&&+\dsum\limits_{r=1}^{3}\dsum\limits_{s=1}^{3}\frac{x_{ri}^{\left( N\right)
}x_{sj}^{\left( N\right) }}{16\pi ^{2}}\mu _{rs}\left[ \frac{m_{\chi
_{R}^{0}}^{2}}{m_{\chi _{R}^{0}}^{2}-\left\vert \mu _{rs}\right\vert ^{2}}%
\ln \left( \frac{m_{\chi _{R}^{0}}^{2}}{\left\vert \mu _{rs}\right\vert ^{2}}%
\right) -\frac{m_{\chi _{I}^{0}}^{2}}{m_{\chi _{I}^{0}}^{2}-\left\vert \mu
_{rs}\right\vert ^{2}}\ln \left( \frac{m_{\chi _{I}^{0}}^{2}}{\left\vert \mu
_{rs}\right\vert ^{2}}\right) \right] ,  \notag
\end{eqnarray}
Then, as follows from Eq. (\ref{mutilde}), we have $\left\vert\widetilde{\mu }_{ij}\right\vert<<\left\vert \mu_{ij}\right\vert$ ($i,j=1,2,3$), since the entries of the $\widetilde{\mu }$ submatrix are much smaller than the entries  of the $\mu$ submatrix, by at least two orders of magnitude.

The light active masses arise from an inverse seesaw mechanism and the
physical neutrino mass matrices are: 
\begin{eqnarray}
\widetilde{\mathbf{M}}_{\nu } &=&m_{\nu D}\left( M^{T}\right) ^{-1}\mu
M^{-1}m_{\nu D}^{T},\hspace{0.7cm}  \label{M1nu} \\
\mathbf{M}_{\nu }^{\left( 1\right) } &=&-\frac{1}{2}\left( M+M^{T}\right) +%
\frac{1}{2}\left( \mu +\widetilde{\mu }\right) ,\hspace{0.7cm} \\
\mathbf{M}_{\nu }^{\left( 2\right) } &=&\frac{1}{2}\left( M+M^{T}\right) +%
\frac{1}{2}\left( \mu +\widetilde{\mu }\right) .
\end{eqnarray}%
where $M_{\nu }^{(1)}$ corresponds to the mass matrix for light active
neutrinos $\nu _{a}$ ($a=1,2,3$), whereas $M_{\nu }^{(2)}$ and $M_{\nu
}^{(3)}$ are the mass matrices for sterile neutrinos ($N_{a}^{-},N_{a}^{+}$)
which are superpositions of mostly $\nu _{aR}$ and $N_{aR}$ as $N_{a}^{\pm
}\sim \frac{1}{\sqrt{2}}\left( \nu _{aR}\mp N_{aR}\right) $. In the limit $%
\mu \rightarrow 0$, which corresponds to unbroken lepton number, the light
active neutrinos become massless. The smallness of the $\mu $ and $%
\widetilde{\mu }$ parameters is responsible for a small mass splitting
between the three pairs of sterile neutrinos, thus implying that the sterile
neutrinos form pseudo-Dirac pairs. 
The full neutrino mass matrix given by Eq. (\ref{Mnu}) can be diagonalized
by the following rotation matrix \cite{Catano:2012kw}: 
\begin{equation}
\mathbb{R}=%
\begin{pmatrix}
\mathbf{R}_{\nu } & \mathbf{R}_{1}\mathbf{R}_{M}^{\left( 1\right) } & 
\mathbf{R}_{2}\mathbf{R}_{M}^{\left( 2\right) } \\ 
-\frac{(\mathbf{R}_{1}^{\dagger }+\mathbf{R}_{2}^{\dagger })}{\sqrt{2}}%
\mathbf{R}_{\nu } & \frac{(1-\mathbf{S})}{\sqrt{2}}\mathbf{R}_{M}^{\left(
1\right) } & \frac{(1+\mathbf{S})}{\sqrt{2}}\mathbf{R}_{M}^{\left( 2\right) }
\\ 
-\frac{(\mathbf{R}_{1}^{\dagger }-\mathbf{R}_{2}^{\dagger })}{\sqrt{2}}%
\mathbf{R}_{\nu } & \frac{(-1-\mathbf{S})}{\sqrt{2}}\mathbf{R}_{M}^{\left(
1\right) } & \frac{(1-\mathbf{S})}{\sqrt{2}}\mathbf{R}_{M}^{\left( 2\right) }%
\end{pmatrix}%
,  \label{U}
\end{equation}%
where 
\begin{equation}
\mathbf{S}=-\frac{1}{4}M^{-1}\mu ,\hspace{1cm}\hspace{1cm}\mathbf{R}%
_{1}\simeq \mathbf{R}_{2}\simeq \frac{1}{\sqrt{2}}m_{\nu D}^{\ast }M^{-1}.
\end{equation}%
Notice that the physical neutrino spectrum is composed of three light active
neutrinos and six exotic neutrinos. The exotic neutrinos are pseudo-Dirac,
with masses $\sim \pm \frac{1}{2}\left( M+M^{T}\right) $ and a small
splitting $\mu $. Furthermore, $\mathbf{R}_{\nu }$, $\mathbf{R}_{M}^{\left(
1\right) }$ and $\mathbf{R}_{M}^{\left( 2\right) }$ are the rotation
matrices which diagonalize $\widetilde{\mathbf{M}}_{\nu }$, $\mathbf{M}_{\nu
}^{\left( 1\right) }$ and $\mathbf{M}_{\nu }^{\left( 2\right) }$,
respectively.

On the other hand, using Eq. (\ref{U}) we find that the neutrino fields $\nu
_{L}=\left( \nu _{1L},\nu _{2L},\nu _{3L}\right) ^{T}$, $\nu _{R}^{C}=\left(
\nu _{1R}^{C},\nu _{2R}^{C},\nu _{3R}^{C}\right) $ and $N_{R}^{C}=\left(
N_{1R}^{C},N_{2R}^{C},N_{3R}^{C}\right) $ are related with the physical
neutrino fields by the following relations: 
\begin{equation}
\left( 
\begin{array}{c}
\nu _{L} \\ 
\nu _{R}^{C} \\ 
N_{R}^{C}%
\end{array}%
\right) =\mathbb{R}\Psi _{L}\simeq 
\begin{pmatrix}
\mathbf{R}_{\nu } & \mathbf{R}_{1}\mathbf{R}_{M}^{\left( 1\right) } & 
\mathbf{R}_{2}\mathbf{R}_{M}^{\left( 2\right) } \\ 
-\frac{(\mathbf{R}_{1}^{\dagger }+\mathbf{R}_{2}^{\dagger })}{\sqrt{2}}%
\mathbf{R}_{\nu } & \frac{(1-\mathbf{S})}{\sqrt{2}}\mathbf{R}_{M}^{\left(
1\right) } & \frac{(1+\mathbf{S})}{\sqrt{2}}\mathbf{R}_{M}^{\left( 2\right) }
\\ 
-\frac{(\mathbf{R}_{1}^{\dagger }-\mathbf{R}_{2}^{\dagger })}{\sqrt{2}}%
\mathbf{R}_{\nu } & \frac{(-1-\mathbf{S})}{\sqrt{2}}\mathbf{R}_{M}^{\left(
1\right) } & \frac{(1-\mathbf{S})}{\sqrt{2}}\mathbf{R}_{M}^{\left( 2\right) }%
\end{pmatrix}%
\left( 
\begin{array}{c}
\Psi _{L}^{\left( 1\right) } \\ 
\Psi _{L}^{\left( 2\right) } \\ 
\Psi _{L}^{\left( 3\right) }%
\end{array}%
\right) ,\hspace{0.5cm}\hspace{0.5cm}\hspace{0.5cm}\hspace{0.5cm}\Psi
_{L}=\left( 
\begin{array}{c}
\Psi _{L}^{\left( 1\right) } \\ 
\Psi _{L}^{\left( 2\right) } \\ 
\Psi _{L}^{\left( 3\right) }%
\end{array}%
\right) ,
\end{equation}%
where $\Psi _{jL}^{\left( 1\right) }$, $\Psi _{jL}^{\left( 2\right)
}=N_{j}^{+}$ and $\Psi _{jL}^{\left( 3\right) }=N_{j}^{-}$ ($j=1,2,3$) are
the three active neutrinos and six exotic neutrinos, respectively.

Finally to close this section we provide a discussion about collider
signatures of exotic fermions of our model. From the Yukawa interactions it
follows that the charged exotic fermions have mixing mass terms with the SM
charged fermions, which allows the former to decay into any of the scalars
of the model and SM charged fermions. These heavy charged exotic fermions
can be produced in association with the charged fermions and can be pair
produced as well at the LHC via gluon fusion (for the exotic quarks only)
and Drell Yan mechanism. Consequently, observing an excess of events in the
multijet and multilepton final state can be a signal of support of this
model at the LHC. Regarding the sterile neutrino sector, it is worth
mentioning that the sterile neutrinos can be produced at the LHC in
association with a SM charged lepton, via quark-antiquark annihilation
mediated by a $W^{\prime }$ gauge boson. The corresponding total cross
section for the process $pp\rightarrow W^{\prime }\rightarrow lN_{a}^{\pm }$ 
$(a=1,2,3)$ will be sizeable provided that $m_{W^{\prime }}>m_{N_{a}^{\pm }}$%
, which implies that in the $s$-channel the $W^{\prime }$ gauge boson is on
its mass shell. Furthermore, in our model the sterile neutrinos have the
following two body decay modes: $N_{a}^{\pm }\rightarrow l_{i}^{\pm }W^{\mp
} $, $N_{a}^{\pm }\rightarrow \nu _{i}Z$ and $N_{a}^{\pm }\rightarrow \nu
_{i}S $ (where $i=1,2,3$ is a flavor index and $S$ correponds to any of the
scalars of our model lighter than the sterile neutrinos), which are
suppressed by the small active-sterile neutrino mixing angle, taken to
fullfill $\theta \sim \mathcal{O}(10^{-3})$, in order to keep charged have
lepton flavor violating decays well below their current experimental upper
limit and at the same time successfully comply with the constraints arising
from the unitarity \cite{Abada:2018nio,Fernandez-Martinez:2016lgt}.
Furthermore the heavy sterile neutrinos $N_{a}^{\pm }$ can decay via
off-shell gauge bosons via the following modes: $N_{a}^{\pm }\rightarrow
l_{i}^{+}l_{j}^{-}\nu _{k}$, $N_{a}^{\pm }\rightarrow l_{i}^{-}u_{j}\bar{d}%
_{k}$, $N_{a}^{\pm }\rightarrow b\bar{b}\nu _{k}$ (where $i,j,k=1,2,3$ are
flavor indices). Consequently, the heavy sterile neutrino can be detected
the LHC via the observation of an excess of events with respect to the SM
background in a final state composed of a pair of opposite sign charged
leptons plus two jets. This signal of a pair of opposite sign charged
leptons plus two jets arising from the decay of sterile neutrinos via an
offshell $W^{\prime }$ gauge boson features a much lower SM background than
the ones arising from the pair production and decays of sterile neutrinos,
thus making the sterile neutrino much easier to detect at the LHC in
left-right symmetric models than in models having only an extra $U^{\prime
}(1)$ symmetry \cite{AguilarSaavedra:2012fu,Das:2012ii}. Studies of inverse
seesaw neutrino signatures at colliders as well as the production of heavy
neutrinos at the LHC are carried out in Refs. \cite%
{Dev:2009aw,BhupalDev:2012zg,Das:2012ze,AguilarSaavedra:2012fu,Das:2012ii,Dev:2013oxa,Das:2014jxa,Das:2016hof,Das:2017gke,Das:2017nvm,Das:2017zjc,Das:2017rsu,Das:2018usr,Das:2018hph,Bhardwaj:2018lma,Helo:2018rll,Pascoli:2018heg}%
. A comprehensive study of the exotic fermion production at the LHC and the
exotic fermion decay modes is beyond the scope of this work and is left for
future studies. 

\section{Charged lepton flavor violation}
\label{LFV}\ac{In this section we will discuss the implications of the model in charged lepton flavor violation. 
As mentioned in the previous section, the sterile neutrino
spectrum of the model is composed of six nearly degenerate heavy neutrinos.
These sterile neutrinos, together with the heavy $W^{\prime }$ gauge boson,
induce the $l_{i}\rightarrow l_{j}\gamma $ decay at one loop level, whose
Branching ratio is given by: \cite%
{Ilakovac:1994kj,Deppisch:2004fa,Lindner:2016bgg}: 
\begin{eqnarray}
Br\left( l_{i}\rightarrow l_{j}\gamma \right) &=&\frac{\alpha
_{W}^{3}s_{W}^{2}m_{l_{i}}^{5}\kappa ^{2}}{256\pi ^{2}m_{W^{\prime
}}^{4}\Gamma _{i}}\dsum\limits_{r=1}^{3}\left\vert G\left( \frac{%
m_{N_{r}}^{2}}{m_{W^{\prime }}^{2}}\right) \right\vert ^{2},\hspace{0.5cm}%
\hspace{0.5cm}\hspace{0.5cm}G\left( x\right) =-\frac{2x^{3}+5x^{2}-x}{%
4\left( 1-x\right) ^{2}}-\frac{3x^{3}}{2\left( 1-x\right) ^{4}}\ln x,  \notag
\\
\kappa &=&\left\vert \dsum\limits_{k=1}^{3}\left( V_{lL}^{\dagger }\right)
_{ik}\left( V_{lL}^{\dagger }\right) _{jk}\right\vert ,
\end{eqnarray}%
where the one loop level contribution arising from the $W$ gauge boson
exchange has been neglected, because it is suppressed by the quartic power of
the active-sterile neutrino mixing angle $\theta $, assumed to be of the
order of $10^{-3}$, for sterile neutrino masses of about $1$ TeV. It has
been shown in Ref. \cite{Deppisch:2013cya} that for such mixing angle the
contribution of the $W$ gauge boson to the branching ratio for the $\mu
\rightarrow e\gamma $ decay rate takes values of the order of $10^{-16}$,
which corresponds to three orders of magnitude below its experimental upper
limit of $4.2\times 10^{-13}$. Thus, in this work we only consider the
dominant $W^{\prime }$ contribution to the $\mu \rightarrow e\gamma $ decay
rate. Furthermore, there will be additional contributions to the $\mu
\rightarrow e\gamma $ decay, arising from the virtual exchange of
electrically neutral scalars and charged exotic leptons. We have numerically
checked that this contribution is close to $10^{-13}$ for charged exotic
leptons with masses $m_{E}\sim \mathcal{O}\left( 100\right) $ TeV\ and
flavor violating Yukawa couplings around $10^{-3}$. In order to simplify our
analysis, we will consider a benchmark scenario where the couplings of the
lepton violating scalar interactions are much lower than $10^{-3}$, thus
allowing us to consider the $\mu \rightarrow e\gamma $ decay as mainly
arising from the $W^{\prime }$ and heavy neutrino virtual exchange.
Furthermore, in our analysis we consider the simplified scenario of
degenerate heavy neutrinos with a common mass $m_{N}$ and we also set $%
g_{R}=g$ and $\kappa =10^{-2}$, which corresponds to off-diagonal elements
of $V_{lL}$ left handed leptonic rotation matrix of the order of $0.1$. 
\begin{figure}[tbp]
\includegraphics[width=0.6\textwidth]{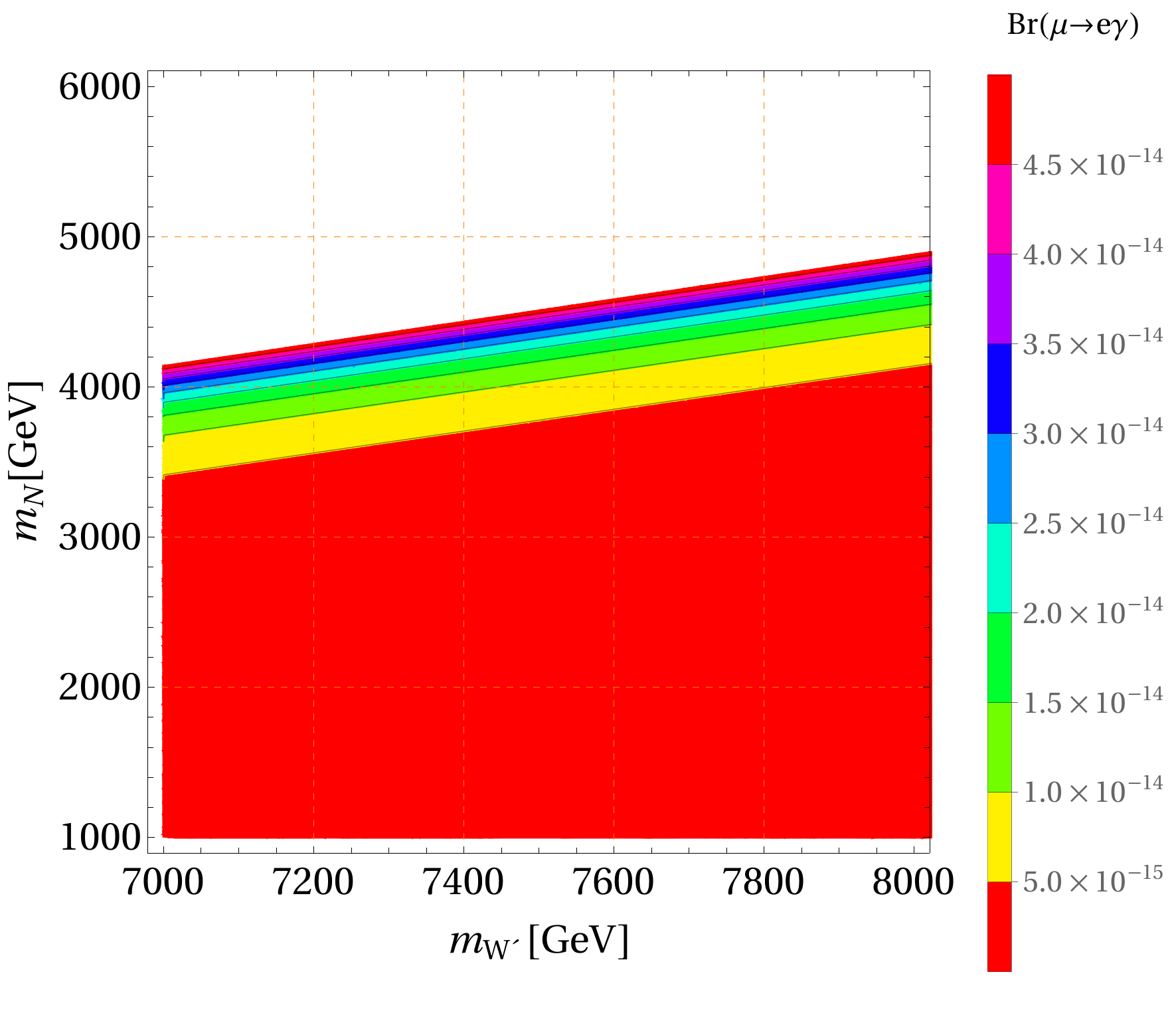}
\caption{Allowed parameter space in the $m_{W^{\prime }}-m_{N}$ plane
consistent with the LFV constraints.}
\label{LFVplot}
\end{figure}
Figure \ref{LFVplot} shows the allowed parameter space in the $m_{W^{\prime
}}-m_{N}$ plane, consistent with the constraints arising from charged lepton
flavor violating decays. The $W^{\prime }$ gauge boson and the sterile
neutrino masses have been taken to be in the ranges $7$ TeV$\lesssim
m_{W^{\prime }}\lesssim 10$ TeV and $1$ TeV$\lesssim m_{N}\lesssim 10$ TeV,
respectively. As seen from Figure \ref{LFVplot}, the $\mu \rightarrow
e\gamma $ decay branching ratio reach values of the order of $10^{-14}$ and
lower, which are below its experimental upper limit of $4.2\times 10^{-13}$
and are within the reach of future experimental sensitivity, in the allowed
model parameter space. In the region of parameter space consistent with $\mu
\rightarrow e\gamma $ decay rate constraints, the maximum obtained branching
ratios for the $\tau \rightarrow \mu \gamma $ and $\tau \rightarrow e\gamma $
decays can reach values below their corresponding upper experimental bounds
of $4.4\times 10^{-8}$ and $3.3\times 10^{-8}$, respectively. Consequently,
our model is compatible with the current charged lepton flavor violating
decay constraints.

On the other hand, the Effective Lagrangian approach for describing LFV
processes, used in \cite{Kuno:1999jp}, in the regime of low momentum limit,
where the off-shell contributions from photon exchange are negligible with
respect to the contributions arising from real photon emission, imply that
the dipole operators shown in Ref.~\cite{Kuno:1999jp} will dominate the
Lepton Flavor Violating (LFV) transitions $\mu \rightarrow 3e$, $\mu {\text{%
Al}}\rightarrow e{\text{Al and }}\mu {\text{Ti}}\rightarrow e{\text{T}}$,
yielding the following relations \cite{Kuno:1999jp,Lindner:2016bgg}: 
\begin{equation}
{\text{Br}}\left( \mu \rightarrow 3e\right) \simeq \frac{1}{160}{\text{Br}}%
\left( \mu \rightarrow e\gamma \right) ,\hspace{1cm}{\text{CR}}\left( \mu {%
\text{Ti}}\rightarrow e{\text{Ti}}\right) \simeq \frac{1}{200}{\text{Br}}%
\left( \mu \rightarrow e\gamma \right) ,\hspace{1cm}{\text{CR}}\left( \mu {%
\text{Al}}\rightarrow e{\text{Al}}\right) \simeq \frac{1}{350}{\text{Br}}%
\left( \mu \rightarrow e\gamma \right)  \label{eq:CR-BR}
\end{equation}
where the $\mu^{-}-e^{-}$ conversion ratio is defined~\cite{Lindner:2016bgg}
as follows: 
\begin{equation}  \label{eq:Conversion-Rate}
{\text{CR}}\left(\mu-e\right)=\frac{\Gamma\left(\mu^{-}+{\text{Nucleus}}%
\left(A,Z\right)\rightarrow e^{-}+{\text{Nucleus}}\left(A,Z\right)\right)}{%
\Gamma\left(\mu^{-}+{\text{Nucleus}}\left(A,Z\right)\rightarrow\nu_{\mu}+{%
\text{Nucleus}}\left(A,Z-1\right)\right)}
\end{equation}
Consequently, for our model we expect that the resulting rates for the LFV
transitions $\mu \rightarrow 3e$, $\mu {\text{Al}}\rightarrow e{\text{Al and 
}}\mu {\text{Ti}}\rightarrow e{\text{T}}$ will be of the order of $10^{-16}$%
, i.e, two orders of magnitude lower than the obtained rate for the $\mu
\rightarrow e\gamma$ decay, implying that in our model the
corresponding values are below their current experimental bounds of about $%
10^{-12}$ 
for these LFV transitions.

\section{Leptogenesis}
\label{leptogenesis} 
In this section we will analyze the implications of our model in leptogenesis. In our analysis of leptogenesis we follow the approach of Ref. \cite{Blanchet:2009kk}. To simplify our analysis, we work in the basis where the SM charged lepton mass matrix is diagonal, assume that $y^{(L)}$ and $x^{(N)}$ are diagonal matrices and consider the scenario where $\left\vert y_{11}^{(L)}\right\vert \ll \left\vert y_{22}^{(L)}\right\vert ,\left\vert y_{33}^{(L)}\right\vert $ and $\left\vert x_{11}^{(N)}\right\vert \ll \left\vert x_{22}^{(N)}\right\vert,\left\vert x_{33}^{(N)}\right\vert $. In that scenario only the first generation of pseudo-Dirac fermions $N_{a}^{\pm }$, i.e, $N_{1}^{\pm }$ will be much lighter than the second and third generation ones. This implies that the decay of {$N_{1}^{\pm }$ provides the dominant contribution to the Baryon asymmetry of the Universe (BAU), whereas the decays of the heavier pseudo-Dirac fermions $N_{2}^{\pm }$ and $N_{3}^{\pm }$ will give subleading contributions to the $B-L$ asymmetry. This is due to the fact that the lepton asymmetry generated by the decays of the heavier pseudo-Dirac pairs $N_{2}^{\pm }$ and $N_{3}^{\pm }$} gets washed out very quickly, yielding a very small impact on the lepton asymmetry produced by the decay of the lightest pair $N_{1}^{\pm }$, as discussed in Ref. \cite{Blanchet:2009kk}. We are considering the scenario of diagonal $y^{(L)}$ matrix in order to suppress tree level FCNCs in the charged lepton sector. We also take the initial temperature larger than the mass $m_{{N^{\pm }}}$ of the lightest pair of pseudo-Dirac fermions $N_{1}^{\pm }=N^{\pm }$. Within this minimal scenario, the Boltzmann equations take the form \cite{Buchmuller:2004nz}: 
\begin{eqnarray}
\frac{dN_{{N_{1}^{\pm }}}\left( z\right) }{dz} &=&-\left[ D\left( z\right)
+S\left( z\right) \right] \left[ N_{{N_{1}^{\pm }}}\left( z\right) -N_{{%
N_{1}^{\pm }}}^{eq}\left( z\right) \right] ,  \notag \\
\frac{dN_{N_{B-L}}\left( z\right) }{dz} &=&-\varepsilon _{\pm }D\left(
z\right) \left[ N_{{N_{1}^{\pm }}}\left( z\right) -N_{{N_{1}^{\pm }}%
}^{eq}\left( z\right) \right] -W\left( z\right) N_{N_{B-L}}\left( z\right) ,
\end{eqnarray}%
where $z=\frac{m_{{N_{1}^{\pm }}}}{T}$, whereas $N_{{N_{1}^{\pm }}}$ and $N_{N_{B-L}}$ are the number density and the amount of $B-L$ asymmetry, respectively. Here $\varepsilon _{\pm }$ are the lepton asymmetry parameters, which are induced by the $N^{\pm }$ decay processes and have the following form \cite{Covi:1996wh,Rangarajan:1999kt,Gu:2010xc,Pilaftsis:1997jf}: 
\begin{eqnarray}
\varepsilon _{\pm } &=&\dsum\limits_{i=1}^{3}\dsum\limits_{r=1}^{2}\frac{%
\left[ \Gamma \left( N_{\pm }\rightarrow l_{i}H_{r}^{+}\right) -\Gamma
\left( N_{\pm }\rightarrow \bar{l}_{i}H_{r}^{-}\right) \right] }{\left[
\Gamma \left( N_{\pm }\rightarrow l_{i}H_{r}^{+}\right) +\Gamma \left(
N_{\pm }\rightarrow \bar{l}_{i}H_{r}^{-}\right) \right] }+\dsum%
\limits_{i=1}^{3}\frac{\left[ \Gamma \left( N_{\pm }\rightarrow h\nu
_{i}\right) -\Gamma \left( N_{\pm }\rightarrow h\nu _{i}\right) \right] }{%
\left[ \Gamma \left( N_{\pm }\rightarrow h\nu _{i}\right) +\Gamma \left(
N_{\pm }\rightarrow h\nu _{i}\right) \right] }  \notag \\
&\simeq &\frac{\func{Im}\left\{ \left( \left[ \left( y_{N_{+}}\right)
^{\dagger }\left( y_{N_{-}}\right) \right] ^{2}\right) _{11}\right\} }{8\pi
A_{\pm }}\frac{r}{r^{2}+\frac{\Gamma _{\pm }^{2}}{m_{N_{\pm }}^{2}}},
\label{ep}
\end{eqnarray}%
with: 
\begin{eqnarray}
r &=&\frac{m_{N_{+}}^{2}-m_{N_{-}}^{2}}{m_{N_{+}}m_{N_{-}}},\hspace{0.7cm}%
\hspace{0.7cm}A_{\pm }=\left[ \left( y_{N_{\pm }}\right) ^{\dagger
}y_{N_{\pm }}\right] _{11},\hspace{0.7cm}\hspace{0.7cm}\Gamma _{\pm }=\frac{%
A_{\pm }m_{N_{\pm }}}{8\pi },  \notag \\
y_{N_{\pm }} &=&\frac{y^{\left( L\right) }}{\sqrt{2}}\left( 1\mp S\right) =%
\frac{y^{\left( L\right) }}{\sqrt{2}}\left[ 1\pm \frac{1}{4}M^{-1}\left( \mu
+\widetilde{\mu }\right) \right] ,  \label{yN}
\end{eqnarray}
where we have assumed that the exotic leptonic fields $E_{nR}$, $E^{\prime }$
and $\Omega _{nR}$ ($n=1,2$)\ are heavier than the lightest pseudo-Dirac
fermions $N_{1}^{\pm }=N^{\pm }$.}

\ac{On the other hand, it is worth mentioning that $N_{N_{1}^{\pm }}$ and $N_{N_{B-L}}$ are computed in a portion of comoving volume that contains one
photon at temperatures much larger than $m_{{N_{1}^{\pm }}}$, thus implying that $N_{{N_{1}^{\pm }}}^{eq}\left( z<<1\right) =\frac{3}{4}$ \cite%
{Buchmuller:2004nz}. Besides that, $D\left( z\right) $, $S\left(z\right)$ and $W\left( z\right) $, are the thermally averaged rates corresponding to the decays of ${N_{1}^{\pm }}$, to the scattering processes and to the inverse
decays, respectively. These thermally averaged rates are given by: 
\begin{eqnarray}
D\left( z\right) &=&D\left( z\right) _{N_{1}}+D_{N_{1}}^{\left( W^{\prime
}\right) }\left( z\right) ,\hspace{0.7cm}\hspace{0.7cm}{S\left( z\right)
=S_{Z^{\prime }}\left( z\right) +}S_{W^{\prime }}\left( z\right) , \\
W\left( z\right) &=&W_{N_{1}}^{ID}\left( z\right) +W_{N_{1}}^{ID\left(
W^{\prime }\right) }\left( z\right) ,
\end{eqnarray}%
 where $D\left( z\right) _{N_{1}}$ is the thermally averaged rate
associated with the two body decays $N_{1}^{\pm }\rightarrow l_{i}H{_{r}^{+}}$ ($r=1,2$), $N_{1}^{\pm }\rightarrow \nu _{i}h$ ($i=1,2,3$%
) whereas $D_{N_{1}}^{\left( W^{\prime }\right) }\left( z\right) $ corresponds to the thermally averaged rate arising from the $W^{\prime }$
mediated three body decay $N_{1}^{\pm }\rightarrow l_{i}^{-}u_{j}\bar{d}_{k} $ ($i,j,k=1,2,3$). Furthermore, $S_{Z^{\prime }}\left( z\right) $ is the thermally \aech{averaged} rate arising from the $Z^{\prime }$ mediated scattering processes $N_{1}^{\pm }N_{1}^{\pm }\longleftrightarrow l_{i}\overline{l}_{j}$ ($i,j=1,2,3$), $N_{1}^{\pm }N_{1}^{\pm }\longleftrightarrow u_{i}\overline{u}_{j}$ and $N_{1}^{\pm }N_{1}^{\pm }\longleftrightarrow d_{i}\overline{d}_{j}$, whereas the thermally averaged rate $S_{W^{\prime }}\left( z\right) $ is caused by the $W^{\prime }$ mediated processes {$N_{1}^{\pm }$}$l_{iR}\longleftrightarrow \overline{u}_{jR}d_{kR}$, $N_{1}^{\pm }\overline{u}_{iR}\longleftrightarrow l_{jR}\overline{d}_{kR}$, {$N_{1}^{\pm }d$}$_{iR}\longleftrightarrow l_{jR}u_{kR}$ ($i,j,k=1,2,3$). In addition, $W_{N_{1}}^{ID}\left( z\right) $ and $W_{N_{1}}^{ID\left( W^{\prime }\right) }\left( z\right) $ are the thermally
averaged rates arising from the inverse two and three body decays of $N_{1}^{\pm }$, respectively. The above mentioned thermally averaged rates
are given by \cite{Plumacher:1996kc,Buchmuller:2004nz,Cosme:2004xs,Frere:2008ct,Blanchet:2010kw,Dolan:2018qpy}:
\begin{eqnarray}
D\left( z\right) _{N_{1}} &=&\frac{\Gamma _{D}}{H\left( z=1\right) z}=Kz%
\frac{\mathcal{K}_{1}\left( z\right) }{\mathcal{K}_{2}\left( z\right) },%
\hspace{0.7cm}\hspace{0.7cm}W_{N_{1}}^{ID}\left( z\right) =\frac{1}{2}\frac{%
\Gamma _{ID}\left( z\right) }{H\left( z\right) z}=\frac{1}{2}\frac{\Gamma
_{D}\left( z\right) }{H\left( z\right) z}\frac{N^{eq}\left( z\right) }{N_{l}}%
=\frac{1}{4}K\mathcal{K}_{1}\left( z\right) z^{3},  \notag \\
\Gamma _{ID}\left( z\right) &=&\Gamma _{D}\left( z\right) \frac{N^{eq}\left(
z\right) }{N_{l}^{eq}},\hspace{0.7cm}N_{{N_{1}^{\pm }}}^{eq}\left( z\right) =%
\frac{3}{8}z^{2}\mathcal{K}_{2}\left( z\right) ,\hspace{0.7cm}N_{l}^{eq}=%
\frac{3}{4},\hspace{0.7cm}K=\frac{\left[ y^{\left( L\right) }\left(
y^{\left( L\right) }\right) ^{\dagger }\right] _{11}v_{1}^{2}}{2m_{\ast }m_{{%
{N_{1}^{\pm }}}}},  \notag \\
D_{N_{1}}^{\left( W^{\prime }\right) }\left( z\right) &=&\frac{\gamma _{{%
N_{1}^{\pm }}}^{\left( W_{R}\right) }}{n_{{N_{1}^{\pm }}}^{eq}\left(
z\right) H\left( z=1\right) z},\hspace{0.7cm}n_{{N_{1}^{\pm }}}^{eq}\left(
z\right) =\frac{3}{4}n_{\gamma }\left( z\right) N_{{N_{1}^{\pm }}%
}^{eq}\left( z\right) =\frac{9}{64}z^{2}\mathcal{K}_{2}\left( z\right)
n_{\gamma }\left( z\right) ,\hspace{0.7cm}n_{\gamma }=\frac{2\zeta \left(
3\right) }{\pi ^{2}}T^{3}, \\
W_{N_{1}}^{ID\left( W^{\prime }\right) }\left( z\right) &=&\frac{1}{2}%
D_{W^{\prime }}\left( z\right) \frac{N^{eq}\left( z\right) }{N_{l}},\hspace{%
0.7cm}\hspace{0.7cm}S\left( z\right) _{Z^{\prime },W^{\prime }}=\frac{\Gamma
_{S}^{\left( Z^{\prime },W^{\prime }\right) }}{H\left( z=1\right) z},\hspace{%
0.7cm}\hspace{0.7cm}\Gamma _{S}=\frac{\gamma _{{S}}^{\left( Z^{\prime
},W^{\prime }\right) }}{n_{{N_{1}^{\pm }}}^{eq}\left( z\right) H\left(
z=1\right) z},  \notag \\
\gamma _{{N_{1}^{\pm }}}^{\left( W^{\prime }\right) } &=&n_{{N_{1}^{\pm }}%
}^{eq}\left( z\right) \frac{\mathcal{K}_{1}\left( z\right) }{\mathcal{K}%
_{2}\left( z\right) }\Gamma _{N}^{\left( W^{\prime }\right) },\hspace{0.7cm}%
\Gamma _{{N_{1}^{\pm }}}^{\left( W^{\prime }\right) }=\frac{3g_{R}^{4}}{%
2^{9}\pi ^{3}m_{{{N_{1}^{\pm }}}}^{3}}\int_{0}^{m_{{{N_{1}^{\pm }}}}^{2}}ds%
\frac{m_{{{N_{1}^{\pm }}}}^{6}-3m_{{{N_{1}^{\pm }}}}^{2}s^{2}+2s^{3}}{\left(
s-m_{W^{\prime }}^{2}\right) ^{2}+m_{W^{\prime }}^{2}\Gamma _{W^{\prime
}}^{2}},\hspace{0.7cm}\Gamma _{W^{\prime }}=\frac{g_{R}^{2}}{4\pi }%
m_{W^{\prime }}^{2},  \notag
\end{eqnarray}%
where $\mathcal{K}_{r}\left( z\right) $ ($r=1,2$) is the modified Besel
function of the $r$th type, $\Gamma _{{N_{1}^{\pm }}}^{\left( W^{\prime
}\right) }$ the total three body decay of ${N_{1}^{\pm }}$, $\Gamma
_{W^{\prime }}$ the $W^{\prime }$ total decay width, $n_{{N_{1}^{\pm }}%
}^{eq}\left( z\right) $ is the right handed equilibrium distribution
density, $n_{\gamma }\left( z\right) $ is the number density of photons, $K$
the washout parameter, $m_{\ast }$ the equilibrium neutrino mass, $H\left(
z\right) $ the Hubble expansion rate, whereas $\gamma_{S}^{\left(
Z^{\prime },W^{\prime }\right) }$, $\gamma _{N_{1}^{\pm }}^{\left(
W_{R}\right) }$ are the scattering and decay reaction densities,
respectively. The equilibrium neutrino mass $m_{\ast }$ and the Hubble
expansion rate $H$ are given by \cite{Buchmuller:2004nz}: 
\begin{equation}
m_{\ast }=\frac{16\pi ^{\frac{5}{2}}\sqrt{g_{\ast }}v_{1}^{2}}{3\sqrt{5}M_{P}%
}=1.08\times 10^{-3}eV,\hspace{0.7cm}\hspace{0.7cm}H=\sqrt{\frac{4\pi
^{3}g_{\ast }}{45}}\frac{T^{2}}{M_{P}}=\sqrt{\frac{4\pi ^{3}g_{\ast }}{45}}%
\frac{m_{{{N_{1}^{\pm }}}}^{2}}{z^{2}M_{P}}\simeq 1.66\sqrt{g_{\ast }}\frac{%
m_{{{N_{1}^{\pm }}}}^{2}}{z^{2}M_{P}},  \notag
\end{equation}%
where $g_{\ast }=118$ is the number of effective relativistic degrees of
freedom, $M_{Pl}=1.2\times 10^{9}$ GeV is the Planck constant.

Furthermore, the scattering reaction densities $\gamma _{{S}%
}^{\left(W^{\prime }\right) }$ and $\gamma _{{S}}^{\left( Z^{\prime }\right)
}$ are given by: 
\begin{eqnarray}
\gamma _{{S}}^{\left( W^{\prime }\right) } &=&\gamma _{{{N_{1}^{\pm }}%
l_{iR}\longleftrightarrow \overline{u}_{jR}d_{kR}}}+\gamma _{{{N_{1}^{\pm }}%
\overline{u}_{iR}\longleftrightarrow l_{jR}\overline{d}_{kR}}}+\gamma _{{{%
N_{1}^{\pm }d}_{iR}\longleftrightarrow l_{jR}u_{kR}}},  \notag \\
\gamma _{{S}}^{\left( Z^{\prime }\right) } &=&\gamma _{{{N_{1}^{\pm
}N_{1}^{\pm }}\longleftrightarrow l}_{i}{\overline{l}}_{j}}+\gamma _{{{%
N_{1}^{\pm }N_{1}^{\pm }}\longleftrightarrow u}_{i}{\overline{u}}%
_{j}}+\gamma _{{{N_{1}^{\pm }N_{1}^{\pm }}\longleftrightarrow d}_{i}{%
\overline{d}}_{j}}
\end{eqnarray}
where the scattering reaction density for the process $ab\longleftrightarrow cd$ is defined as follows \cite%
{Luty:1992un,Plumacher:1996kc,Buchmuller:2004nz,Cosme:2004xs,Frere:2008ct,Blanchet:2010kw,Dolan:2018qpy}: 
\begin{eqnarray}
\gamma _{ab\longleftrightarrow cd} &=&\frac{T}{64\pi ^{4}}\int_{s_{\min
}}^{\infty }ds\sqrt{s}\widehat{\sigma }_{ab\longleftrightarrow cd}\left(
s\right) \mathcal{K}_{1}\left( \frac{\sqrt{s}}{T}\right) =\frac{m_{{{%
N_{1}^{\pm }}}}^{4}}{64\pi ^{4}z}\int_{x_{0}}^{\infty }dx\sqrt{x}\widehat{%
\sigma }\left( x\right) \mathcal{K}_{1}\left( z\sqrt{x}\right) ,\hspace{0.7cm%
}\hspace{0.7cm}  \notag \\
x &=&\frac{s}{m_{{{N_{1}^{\pm }}}}^{2}},\hspace{0.7cm}\hspace{0.7cm}x_{0}=%
\frac{1}{m_{{{N_{1}^{\pm }}}}^{2}}\max \left[ \left( m_{a}+m_{b}\right)
^{2},\left( m_{c}+m_{d}\right) ^{2}\right] ,\hspace{0.7cm}\hspace{0.7cm}z=%
\frac{m_{{{N_{1}^{\pm }}}}}{T},
\end{eqnarray}
and $\widehat{\sigma }_{ab\longleftrightarrow cd}\left( s\right) $ is the reduced cross section corresponding to the scattering process $ab\longleftrightarrow cd$. In the left-right model under consideration, the relevant reduced cross sections are given by \cite{Plumacher:1996kc,Plumacher:1996kc,Cosme:2004xs,Frere:2008ct,Blanchet:2010kw,Dolan:2018qpy}
\begin{eqnarray}
\widehat{\sigma }_{{{N_{1}^{\pm }}l_{iR}\longleftrightarrow \overline{u}%
_{jR}d_{kR}}}\left( x\right) &=&\frac{9g_{R}^{4}}{48\pi x}\frac{%
1-3x^{2}+2x^{3}}{\left[ \left( x-\frac{m_{W^{\prime }}^{2}}{m_{{{N_{1}^{\pm }%
}}}^{2}}\right) ^{2}+\frac{m_{W^{\prime }}^{2}\Gamma _{W^{\prime }}^{2}}{m_{{%
{N_{1}^{\pm }}}}^{2}}\right] }, \\
\widehat{\sigma }_{{{N_{1}^{\pm }}\overline{u}_{iR}\longleftrightarrow l_{jR}%
\overline{d}_{kR}}}\left( x\right) &=&\frac{9g_{R}^{4}}{8\pi x}%
\int_{1-x}^{0}du\frac{\left( x+u\right) \left( x+u-1\right) }{\left( u-\frac{%
m_{W^{\prime }}^{2}}{m_{{{N_{1}^{\pm }}}}^{2}}\right) ^{2}}, \\
\widehat{\sigma }_{{{N_{1}^{\pm }d}_{iR}\longleftrightarrow l_{jR}u_{kR}}%
}\left( x\right) &=&\frac{9g_{R}^{4}}{8\pi }\frac{m_{{{N_{1}^{\pm }}}}^{2}}{%
m_{W^{\prime }}^{2}}\frac{\left( 1-x\right) ^{2}}{x+\frac{m_{W^{\prime }}^{2}%
}{m_{{{N_{1}^{\pm }}}}^{2}}-1}, \\
\widehat{\sigma }_{{{N_{1}^{\pm }N_{1}^{\pm }}\longleftrightarrow l\overline{%
l}}}\left( x\right) +\gamma _{{{N_{1}^{\pm }N_{1}^{\pm }}\longleftrightarrow
u}_{i}{\overline{u}}_{j}}+\gamma _{{{N_{1}^{\pm }N_{1}^{\pm }}%
\longleftrightarrow d}_{i}{\overline{d}}_{j}} &=&\frac{13g_{B-L}^{2}}{6\pi }%
\frac{\sqrt{x\left( x-4\right) ^{3}}}{\left( x-\frac{m_{Z^{\prime }}^{2}}{m_{%
{{N_{1}^{\pm }}}}^{2}}\right) ^{2}+\frac{m_{Z^{\prime }}^{2}\Gamma
_{Z^{\prime }}^{2}}{m_{{{N_{1}^{\pm }}}}^{2}}},
\end{eqnarray}
where $\Gamma _{Z^{\prime }}$ is the total $Z^{\prime }$ decay given by:
\begin{equation}
\Gamma _{Z^{\prime }}=\frac{g_{B-L}^{2}}{24\pi }m_{Z^{\prime }}\left[
13+3\left( 1-\frac{4m_{{{N_{1}^{\pm }}}}^{2}}{m_{Z^{\prime }}^{2}}\right) ^{%
\frac{3}{2}}\right], 
\end{equation}
It is worth mentioning that we are not considering the contributions arising
from the $t$-channel scattering processes $N_{1}^{\pm }N_{1}^{\pm }$$\longleftrightarrow l_{i}\overline{l}_{j}$ ($i,j=1,2,3$) since its
corresponding rates have a very fast decrease for $z${$=\frac{m_{{N_{1}^{\pm }}}}{T}$}$>1$, as discussed in Ref. \cite{Blanchet:2009kk}. Furthermore, we are also not considering contributions arising from $\Delta L=1$ scatterings involving scalars, since they are subleading, as discussed in Ref. \cite{Blanchet:2009kk}. Moreover, we are not considering scattering processes involving heavy charged exotic fermions, since they are very heavy with masses larger than $100$ TeV (see Eq. (\ref{bfpquarks})) in order to naturally reproduce the SM fermion mass hierarchy.

The numerical solution of the Boltzmann equations allows to determine the amount of $B-L$ asymmetry $N_{B-L}$, and then the baryon to photon ratio, by using the following relation \cite{Buchmuller:2004nz,Frere:2008ct}:
\begin{equation}
\eta _{B}=\frac{n_{B}}{n_{\gamma }}=\frac{3}{4}a_{sph}N_{B-L},\hspace{0.7cm}%
\hspace{0.7cm}a_{sph}=\frac{8n_{f}+4n_{H}}{22n_{f}+13n_{H}},
\end{equation}
where $a_{sph}$ is the $L$ to $B$ sphaleron conversion rate. Furthermore, $n_{f}$ is the number of fermion families and $n_{H}$ is the number of Higgs doublets.  

As shown in Ref. \cite{Blanchet:2010kw}, the contributions arising from the aforementioned scattering processes, as well as from the inverse decays, are subdominant for temperatures sufficiently lower than the mass $m_N$, i.e, $z>>1$, thus implying that the lepton asymmetry mainly arises from
the decay of the the lightest pair of pseudo-Dirac fermions $N_{1}^{\pm }$. This is confirmed in Figure \ref{scattrate}, which shows the thermally averaged rates corresponding to the decays, scattering and washouts, as a function of $z=\frac{m_{N}}{T}$, where $m_{N}$ is the mass of the lightest pair of pseudo-Dirac fermions $N_{1}^{\pm }=N^{\pm }$ and $T$ the temperature. Here we have set $v_R=14$ TeV, $m_{W^{\prime }}=7$ TeV and $m_{Z^{\prime }}=7.2$ TeV. As shown in Figure \ref{scattrate}, for $z>\mathcal{O}(10)$ the thermally averaged scattering rate corresponding to the decays is much larger by several orders of magnitudes than the ones associated with the scattering and inverse decays (washouts). Furthermore, it has been shown in Ref. \cite{Blanchet:2010kw} that the contribution arising from the $W_{R}$
mediated three body decay $N_{1}^{\pm }\rightarrow l_{i}^{-}u_{j}\bar{d}_{k}$
is much smaller than the ones arising from the {$N_{1}^{\pm }$}$\rightarrow
l_{i}H{_{r}^{+}}$, {$N_{1}^{\pm }$}$\rightarrow \nu _{i}h$ decays. On the
other hand, if the temperature of the Universe
drops below the scale of breaking of the left-right symmetry, the inverse
decays producing ${N_{1}^{\pm }}$ fall out of thermal equilibrium, and
thermal leptogenesis can take place.
\begin{figure}[tbp]
\centering
\includegraphics[width=14cm,height=10cm]{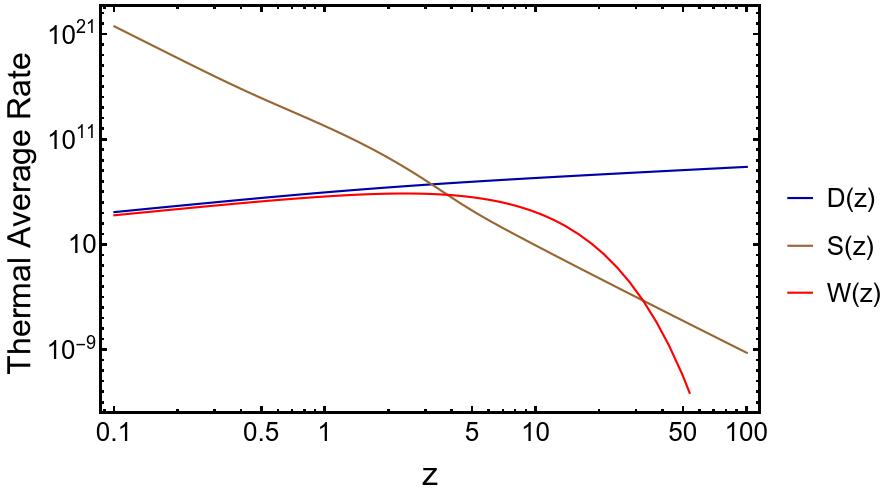}
\caption{Thermally averaged scattering rates $D\left(z\right)$, $S\left(z\right)$ and $W\left(z\right)$ as functions of $z=\frac{m_{N}}{T}$, with $m_{N}$ the mass of the lightest pair of pseudo-Dirac fermions $N_{1}^{\pm }=N^{\pm }$ and $T$ the temperature. Here we have set $v_R=14$ TeV, $m_{W^{\prime }}=7$ TeV and $m_{Z^{\prime }}=7.2$ TeV.}
\label{scattrate}
\end{figure}
It is worth mentioning that CP violation in the lepton sector, necessary to
generate the lepton asymmetry parameter, can arise from complex entries
in $y^{\left( L\right) }$, $M$ or $\mu $, as indicated by Eqs. (\ref{ep})
and (\ref{yN}). Furthermore, in order to successfully reproduce the neutrino
oscillation experimental data, the submatrix $\mu $, in the basis of
diagonal SM charged lepton mass matrix, should have the following form: 
\begin{equation}
\mu =M^{T}m_{\nu D}^{-1}\widetilde{\mathbf{M}}_{\nu }\left( m_{\nu
D}^{T}\right) ^{-1}M=M^{T}m_{\nu D}^{-1}U_{PMNS}\left( \widetilde{\mathbf{M}}%
_{\nu }\right) _{diag}diag\left( m_{1},m_{2},m_{3}\right) U_{PMNS}^{T}\left(
m_{\nu D}^{T}\right) ^{-1}M,
\end{equation}%
where: 
\begin{equation}
\left( \widetilde{\mathbf{M}}_{\nu }\right) _{diag}=diag\left(
m_{1},m_{2},m_{3}\right)
\end{equation}%
being $m_{1}$, $m_{2}$ and $m_{3}$ the masses of the light active neutrinos and $U_{PMNS}$ the PMNS leptonic mixing matrix.

The correlations of the baryon asymmetry and the magnitude of the Dirac neutrino Yukawa couplings $y_{11}^{(L)}$ and $y_{22}^{(L)}$ are shown in Figure \ref{etaB1} and \ref{etaB2}, respectively. Here we have set $v_R=14$ TeV, $m_{W^{\prime }}=7$ TeV, $m_{Z^{\prime }}=7.2$ TeV, $m_{N_{2}^{\pm }}=14$ TeV, $m_{N_{3}^{\pm }}=28$ TeV and $\left\vert y_{22}^{(L)}\right\vert=\left\vert y_{33}^{(L)}\right\vert=\left\vert y_{2}^{(L)}\right\vert$. As shown in Figures \ref{etaB1} and \ref{etaB2}, the measured value of the baryon asymmetry of the Universe \cite{Zyla:2020zbs}: 
\begin{equation}
\eta_{B}=\left(6.12\pm 0.04\right) \times 10^{-10}
\end{equation}
can be successfully reproduced in the simplified scenario considered in our model, provided that $\left\vert y_{1}^{(L)}\right\vert\sim\mathcal{O}\left(10^{-4}\right)$ and $\left\vert y_{2}^{(L)}\right\vert\sim\mathcal{O}\left(1\right)$. In our numerical analysis we have found that the baryon asymmetry of the Universe is generated for $z\sim\mathcal{O}\left(10\right)$, which corresponds to temperatures one order of magnitude lower than the mass $m_{N}$ of the lightest pair of pseudo-Dirac fermions $N_{1}^{\pm }=N^{\pm }$. This result is consistent with the one obtained in Ref. \cite{Blanchet:2010kw}.

The correlation of the baryon asymmetry and the mass $m_{N}$ of the lightest pair of pseudo-Dirac fermions $N_{1}^{\pm }=N^{\pm }$ is shown in Figure \ref{etaB3}. 
As shown in Figures \ref{etaB1}, \ref{etaB2} and \ref{etaB3}, our model successfully accommodates the
experimental value of the baryon asymmetry parameter $\eta_{B}$. 
\begin{figure}[tbp]
\centering
\includegraphics[width=14cm, height=10cm]{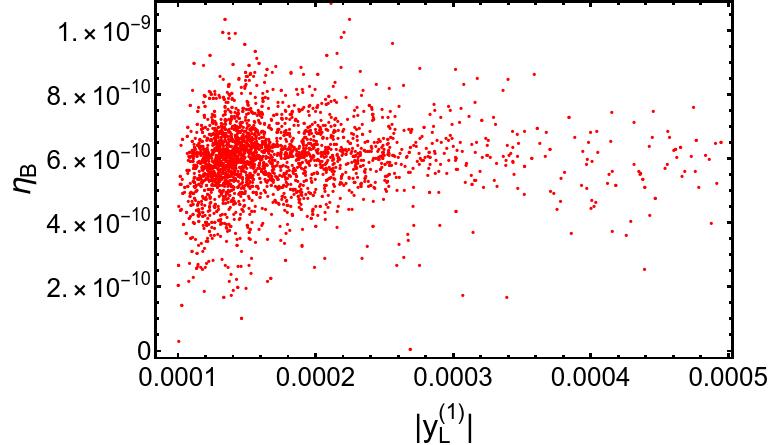}
\caption{Correlation of the baryon asymmetry and the magnitude of the Dirac neutrino Yukawa coupling $y_{11}^{(L)}$. Here we have set $v_R=14$ TeV, $m_{W^{\prime }}=7$ TeV, $m_{Z^{\prime }}=7.2$ TeV, $m_{N_{2}^{\pm }}=14$ TeV, $m_{N_{3}^{\pm }}=28$ TeV and $\left\vert y_{22}^{(L)}\right\vert=\left\vert y_{33}^{(L)}\right\vert=\left\vert y_{2}^{(L)}\right\vert$.}
\label{etaB1}
\end{figure}
\begin{figure}[tbp]
\centering
\includegraphics[width=14cm, height=10cm]{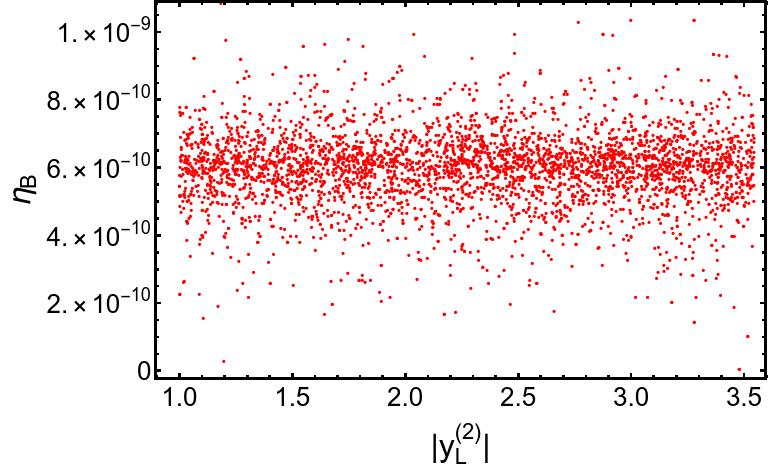}
\caption{Correlation of the baryon asymmetry and the magnitude of the Dirac neutrino Yukawa coupling $y_{22}^{(L)}$. Here we have set $v_R=14$ TeV, $m_{W^{\prime }}=7$ TeV, $m_{Z^{\prime }}=7.2$ TeV, $m_{N_{2}^{\pm }}=14$ TeV, $m_{N_{3}^{\pm }}=28$ TeV and $\left\vert y_{22}^{(L)}\right\vert=\left\vert y_{33}^{(L)}\right\vert=\left\vert y_{2}^{(L)}\right\vert$.}
\label{etaB2}
\end{figure}
\begin{figure}[tbp]
\centering
\includegraphics[width=14cm, height=10cm]{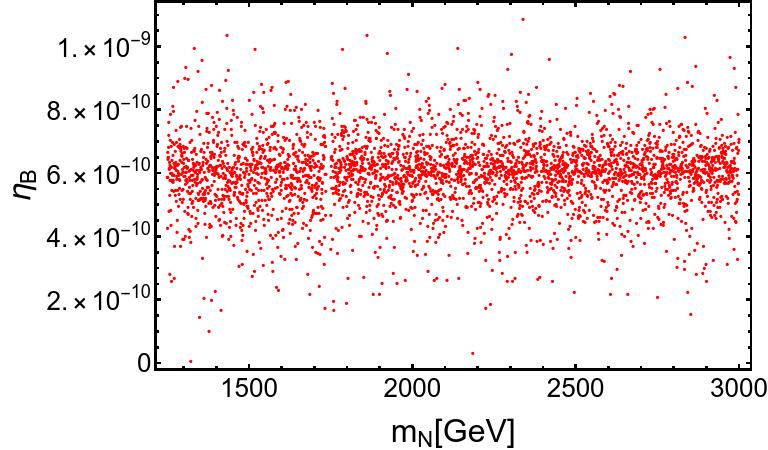}%
\caption{Correlation of the baryon asymmetry and the mass $m_{N}$ of the lightest pair of pseudo-Dirac fermions $N_{1}^{\pm }=N^{\pm }$. Here we have set $v_R=14$ TeV, $m_{W^{\prime }}=7$ TeV, $m_{Z^{\prime }}=7.2$ TeV, $m_{N_{2}^{\pm }}=14$ TeV, $m_{N_{3}^{\pm }}=28$ TeV and $\left\vert y_{22}^{(L)}\right\vert=\left\vert y_{33}^{(L)}\right\vert=\left\vert y_{2}^{(L)}\right\vert$}
\label{etaB3}
\end{figure}
\section{The simplified scalar potential}
\label{scalarpotential} In order to simplify our analysis, we will consider
a bechmark scenario where the singlet real scalar fields $\sigma $, $\eta $
and $\rho $ will not feature mixings with the neutral components of the $%
\Phi $, $\chi _{L}$ and $\chi _{R}$ scalars. Furthermore, for the sake of
simplicity, in our bechmark scenario we do not consider the trilinear terms $%
A_{1}(\chi _{R}^{\dagger }\phi _{R})\varphi $ and $A_{2}(\chi _{L}^{\dagger
}\phi _{L})\varphi $ that will give rise to mixings of the gauge singlet
scalar field $\varphi $ with the $\phi _{L}$ and $\phi _{R}$ scalars. The
justification of the benchmark scenario under consideration arises from the
fact that such gauge singlet scalars $\sigma $, $\eta $ and $\rho $ are
assumed to acquire vacuum expectation values much larger than the scale of
breaking of the left-right symmetry, thus allowing to neglect the mixings of
these fields with the $\Phi $, $\chi _{L}$ and $\chi _{R}$ scalars and to
treat their scalar potentials independently. Let us note that the mixing
angles between those fields are suppressed by the ratios of their VEVs, as
follows from the method of recursive expansion of Ref. \cite{Grimus:2000vj}.
The scalar potential for the $\Phi $, $\chi _{L}$, $\phi _{L}$, $\chi _{R}$
and $\phi _{R}$ scalars takes the form:%
\begin{eqnarray}
V &=&\mu _{1}^{2}(\chi _{L}^{\dagger }\chi _{L})+\mu _{2}^{2}(\chi
_{R}^{\dagger }\chi _{R})+\mu _{3}^{2}Tr(\Phi ^{\dagger }\Phi )+\mu
_{4}^{2}(\phi _{L}^{\dagger }\phi _{L})+\mu _{5}^{2}(\phi _{R}^{\dagger
}\phi _{R})-\mu ^{2}Tr\left[ \Phi ^{2}+\left( \Phi ^{\ast }\right) ^{2}%
\right] +\lambda _{1}(\chi _{L}^{\dagger }\chi _{L})^{2}+\lambda _{2}(\chi
_{R}^{\dagger }\chi _{R})^{2}  \notag \\
&&+\lambda _{3}(\chi _{L}^{\dagger }\chi _{L})(\chi _{R}^{\dagger }\chi
_{R})+\lambda _{4}\left[ Tr(\Phi ^{\dagger }\Phi )\right] ^{2}+\lambda _{5}Tr%
\left[ (\Phi ^{\dagger }\Phi )^{2}\right] +\lambda _{6}\left[ Tr(\widetilde{%
\Phi }\widetilde{\Phi }^{\dagger })\right] ^{2}+\lambda _{7}Tr\left[ (%
\widetilde{\Phi }\widetilde{\Phi }^{\dagger })^{2}\right] +\lambda _{8}(\chi
_{L}^{\dagger }\chi _{L})Tr(\Phi ^{\dagger }\Phi )  \notag \\
&&+\lambda _{9}(\chi _{R}^{\dagger }\chi _{R})Tr(\Phi ^{\dagger }\Phi
)+\lambda _{10}(\chi _{L}^{\dagger }\chi _{L})Tr(\widetilde{\Phi }\widetilde{%
\Phi }^{\dagger })+\lambda _{11}(\chi _{R}^{\dagger }\chi _{R})Tr(\widetilde{%
\Phi }\widetilde{\Phi }^{\dagger })+\lambda _{12}(\phi _{L}^{\dagger }\phi
_{L})^{2}+\lambda _{13}(\phi _{R}^{\dagger }\phi _{R})^{2}  \notag \\
&&+\lambda _{14}(\phi _{L}^{\dagger }\phi _{L})(\phi _{R}^{\dagger }\phi
_{R})+\lambda _{15}(\phi _{L}^{\dagger }\phi _{L})Tr(\Phi ^{\dagger }\Phi
)+\lambda _{16}(\phi _{R}^{\dagger }\phi _{R})Tr(\Phi ^{\dagger }\Phi
)+\lambda _{17}(\phi _{L}^{\dagger }\phi _{L})Tr(\widetilde{\Phi }\widetilde{%
\Phi }^{\dagger })+\lambda _{18}(\phi _{R}^{\dagger }\phi _{R})Tr(\widetilde{%
\Phi }\widetilde{\Phi }^{\dagger })  \notag \\
&&+\lambda _{19}\left[ (\phi _{L}^{\dagger }\chi _{L})(\phi _{R}^{\dagger
}\chi _{R})+(\chi _{L}^{\dagger }\phi _{L})(\chi _{R}^{\dagger }\phi _{R})%
\right]
\end{eqnarray}%
where the term $-\mu ^{2}Tr\left[ \Phi ^{2}+\left( \Phi ^{\ast }\right) ^{2}%
\right] $ softly breaks the $Z_{4}^{\left( 1\right) }$ symmetry. Such term
arises from the trilinear scalar interaction $ATr(\widetilde{\Phi }\Phi
^{\dagger }+\widetilde{\Phi }^{\dagger }\Phi )\eta $ after the $\eta $
singlet scalar field acquires a VEV.

The minimization conditions of the scalar potential yields the following
relations: 
\begin{eqnarray}
\mu _{1}^{2} &=&\frac{1}{2}\left( -2\lambda _{1}v_{L}^{2}-\lambda
_{3}v_{R}^{2}-\left( \lambda _{8}+\lambda _{10}\right) v_{1}^{2}\right) , \\
\mu _{2}^{2} &=&\frac{1}{2}\left( -\lambda _{3}v_{L}^{2}-2\lambda
_{2}v_{R}^{2}-\left( \lambda _{9}+\lambda _{11}\right) v_{1}^{2}\right) , \\
\mu _{3}^{2} &=&2\mu ^{2}+\frac{1}{2}\left( -\left( \lambda _{8}+\lambda
_{10}\right) v_{L}^{2}-\left( \lambda _{9}+\lambda _{11}\right)
v_{R}^{2}-2\left( \lambda _{4}+\lambda _{5}+\lambda _{6}+\lambda _{7}\right)
v_{1}^{2}\right) .
\end{eqnarray}

The squared mass matrix for the electrically charged scalars even under the
remnant $Z_{2}$ symmetry, in the basis $\left( \chi _{L}^{+},\chi
_{R}^{+},\phi _{1I}^{+},\phi _{2I}^{+}\right) -\left( \chi _{L}^{-},\chi
_{R}^{-},\phi _{1I}^{-},\phi _{2I}^{-}\right) $ takes the form: 
\begin{equation}
\mathbf{M}_{\text{charged}}^{2}=\left( 
\begin{array}{cccc}
0 & 0 & 0 & 0 \\ 
0 & 0 & 0 & 0 \\ 
0 & 0 & 2\mu ^{2}-\lambda _{7}v_{1}^{2} & -2\mu ^{2} \\ 
0 & 0 & -2\mu ^{2} & 2\mu ^{2}-\lambda _{5}v_{1}^{2}%
\end{array}%
\right)
\end{equation}%
where the massless scalar eigenstates $\chi _{L}^{\pm }$ and $\chi _{R}^{\pm
}$ correspond to the Goldstone bosons associated with the longitudinal
components of the $W^{\pm }$ and $W^{\prime \pm }$ gauge bosons. Besides
that, there are physical electrically charged scalars $H_{1}^{\pm }$ and $%
H_{2}^{\pm }$ , whose squared masses are given by: 
\begin{eqnarray}
m_{H_{1}^{\pm }}^{2} &=&\frac{1}{2}\left[ 4\mu ^{2}-\left( \lambda
_{5}+\lambda _{7}\right) v_{1}^{2}-\sqrt{\left( \lambda _{5}-\lambda
_{7}\right) ^{2}v_{1}^{4}+16\mu ^{4}}\right] , \\
m_{H_{2}^{\pm }}^{2} &=&\frac{1}{2}\left[ 4\mu ^{2}-\left( \lambda
_{5}+\lambda _{7}\right) v_{1}^{2}+\sqrt{\left( \lambda _{5}-\lambda
_{7}\right) ^{2}v_{1}^{4}+16\mu ^{4}}\right] .
\end{eqnarray}

Furthermore, the electrically charged scalar fields $S_{1}^{\pm }=\phi
_{L}^{\pm }$ and $S_{2}^{\pm }=\phi _{R}^{\pm }$ having non trivial charges
under the remnant $Z_{2}$ symmetry have squared masses given by:

\begin{eqnarray}
m_{S_{1}^{\pm }}^{2} &=&\mu _{4}^{2}+\left( \lambda _{15}+\lambda
_{17}\right) v_{1}^{2}, \\
m_{S_{2}^{\pm }}^{2} &=&\mu _{5}^{2}+\left( \lambda _{16}+\lambda
_{18}\right) v_{1}^{2}.
\end{eqnarray}

The squared mass matrix for the CP-odd neutral scalar sector, even under the
remnant $Z_{2}$ symmetry in the basis $\left( \func{Im}\chi _{L}^{0},\func{Im%
}\chi _{R}^{0},\phi _{1I}^{0},\phi _{2I}^{0}\right) $ has the form: 
\begin{equation}
\mathbf{M}_{CP-\text{odd}}^{2}=\left( 
\begin{array}{cccc}
0 & 0 & 0 & 0 \\ 
0 & 0 & 0 & 0 \\ 
0 & 0 & 4\mu ^{2} & 0 \\ 
0 & 0 & 0 & 4\mu ^{2}-\left( \lambda _{5}+\lambda _{7}\right) v_{1}^{2}%
\end{array}%
\right)
\end{equation}%
The massless scalar eigenstates $\func{Im}\chi _{L}^{0}$ and $\func{Im}\chi
_{R}^{0}$ are associated with the Goldstone bosons associated with the
longitudinal components of the $Z$ and $Z^{\prime }$ gauge bosons.
Furthermore, the $Z_{2}$ even CP-odd neutral scalar sector contains two
massive CP odd scalars whose squared masses are given by: 
\begin{eqnarray}
m_{A_{1}^{0}}^{2} &=&4\mu ^{2}, \\
m_{A_{2}^{0}}^{2} &=&4\mu ^{2}-\left( \lambda _{5}+\lambda _{7}\right)
v_{1}^{2}.
\end{eqnarray}

Moreover, the squared mass matrix for the CP-odd neutral scalar sector, odd
under the remnant $Z_{2}$ symmetry in the basis $\left( \func{Im}\phi
_{L}^{0},\func{Im}\phi _{R}^{0}\right) $ has the form:

\begin{equation}
\widetilde{\mathbf{M}}_{CP-\text{odd}}^{2}=\left( 
\begin{array}{cc}
\frac{1}{2}\left[ \mu _{4}^{2}+\left( \lambda _{15}+\lambda _{17}\right)
v_{1}^{2}\right] & -\lambda _{19}v_{L}v_{R} \\ 
-\lambda _{19}v_{L}v_{R} & \frac{1}{2}\left[ \mu _{5}^{2}+\left( \lambda
_{16}+\lambda _{18}\right) v_{1}^{2}\right]%
\end{array}%
\right)
\end{equation}

This matrix can be diagonalized as follows:

\begin{eqnarray}
R_{P}^{T}\widetilde{\mathbf{M}}_{CP-\text{odd}}^{2}R_{P} &=&\left( 
\begin{array}{cc}
\frac{A_{P}+B_{P}}{2}+\frac{1}{2}\sqrt{\left( A_{P}-B_{P}\right)
^{2}+4C_{P}^{2}} & 0 \\ 
0 & \frac{A_{P}+B_{P}}{2}-\frac{1}{2}\sqrt{\left( A_{P}-B_{P}\right)
^{2}+4C_{P}^{2}}%
\end{array}%
\right) ,  \notag  \label{eq:Theta-P} \\
R_{P} &=&\left( 
\begin{array}{cc}
\cos \theta _{P} & -\sin \theta _{P} \\ 
\sin \theta _{P} & \cos \theta _{P}%
\end{array}%
\right) ,  \notag \\
A_{P} &=&\frac{1}{2}\left[ \mu _{4}^{2}+\left( \lambda _{15}+\lambda
_{17}\right) v_{1}^{2}\right] ,\hspace{0.5cm}\hspace{0.7cm}B_{P}=\frac{1}{2}%
\left[ \mu _{5}^{2}+\left( \lambda _{16}+\lambda _{18}\right) v_{1}^{2}%
\right] ,  \notag \\
C_{P} &=&-\lambda _{19}v_{L}v_{R},\hspace{0.7cm}\hspace{0.7cm}\tan 2\theta
_{P}=\frac{2C_{P}}{A_{P}-B_{P}}.
\end{eqnarray}%
Consequently, the physical scalar mass eigenstates $P_{1,2}$ are given by: 
\begin{equation}
\left( 
\begin{array}{c}
P_{1} \\ 
P_{2}%
\end{array}%
\right) =\left( 
\begin{array}{cc}
\cos \theta _{P} & \sin \theta _{P} \\ 
-\sin \theta _{P} & \cos \theta _{P}%
\end{array}%
\right) \left( 
\begin{array}{c}
\func{Im}\phi _{L}^{0} \\ 
\func{Im}\phi _{R}^{0}%
\end{array}%
\right) .
\end{equation}%
Their squared masses are: 
\begin{equation}
m_{P_{1}}^{2}=\frac{A_{P}+B_{P}}{2}+\frac{1}{2}\sqrt{\left(
A_{P}-B_{P}\right) ^{2}+4C_{P}^{2}},\hspace{0.7cm}\hspace{0.7cm}%
m_{P_{2}}^{2}=\frac{A_{P}+B_{P}}{2}-\frac{1}{2}\sqrt{\left(
A_{P}-B_{P}\right) ^{2}+4C_{P}^{2}}.
\end{equation}

The squared mass matrix for the CP-even neutral scalar sector in the basis $%
\left( \phi _{1R}^{0},\func{Re}\chi _{L}^{0},\phi _{2R}^{0},\func{Re}\chi
_{R}^{0}\right) $ 
\begin{equation}
\mathbf{M}_{CP-\text{even}}^{2}=\left( 
\begin{array}{cccc}
2\left( \lambda _{4}+\lambda _{5}+\lambda _{6}+\lambda _{7}\right) v_{1}^{2}
& \left( \lambda _{8}+\lambda _{10}\right) v_{1}v_{L} & 0 & \left( \lambda
_{9}+\lambda _{11}\right) v_{1}v_{R} \\ 
\left( \lambda _{8}+\lambda _{10}\right) v_{1}v_{L} & 2\lambda _{1}v_{L}^{2}
& 0 & \lambda _{3}v_{L}v_{R} \\ 
0 & 0 & \left( \lambda _{5}+\lambda _{7}\right) \left( -v_{1}^{2}\right) & 0
\\ 
\left( \lambda _{9}+\lambda _{11}\right) v_{1}v_{R} & \lambda _{3}v_{L}v_{R}
& 0 & 2\lambda _{2}v_{R}^{2}%
\end{array}%
\right)
\end{equation}

On the other hand, the squared mass matrix for the CP-even neutral scalar
sector, odd under the remnant $Z_{2}$ symmetry in the basis $\left( \func{Re}%
\phi _{L}^{0},\func{Re}\phi _{R}^{0}\right) $ has the form:

\begin{equation}
\widetilde{\mathbf{M}}_{CP-\text{even}}^{2}=\left( 
\begin{array}{cc}
\frac{1}{2}\left[ \mu _{4}^{2}+\left( \lambda _{15}+\lambda _{17}\right)
v_{1}^{2}\right] & \lambda _{19}v_{L}v_{R} \\ 
\lambda _{19}v_{L}v_{R} & \frac{1}{2}\left[ \mu _{5}^{2}+\left( \lambda
_{16}+\lambda _{18}\right) v_{1}^{2}\right]%
\end{array}%
\right)
\end{equation}

This matrix can be diagonalized as follows:

\begin{eqnarray}
R_{S}^{T}\widetilde{\mathbf{M}}_{CP-\text{even}}^{2}R_{S} &=&\left( 
\begin{array}{cc}
\frac{A_{S}+B_{s}}{2}-\frac{1}{2}\sqrt{\left( A_{S}-B_{S}\right)
^{2}+4C_{S}^{2}} & 0 \\ 
0 & \frac{A_{S}+B_{S}}{2}+\frac{1}{2}\sqrt{\left( A_{S}-B_{S}\right)
^{2}+4C_{S}^{2}}%
\end{array}%
\right) ,  \notag  \label{eq:Theta-S} \\
R_{S} &=&\left( 
\begin{array}{cc}
\cos \theta _{S} & -\sin \theta _{S} \\ 
\sin \theta _{S} & \cos \theta _{S}%
\end{array}%
\right) ,  \notag \\
A_{S} &=&\frac{1}{2}\left[ \mu _{4}^{2}+\left( \lambda _{15}+\lambda
_{17}\right) v_{1}^{2}\right] ,\hspace{0.5cm}\hspace{0.7cm}B_{S}=\frac{1}{2}%
\left[ \mu _{5}^{2}+\left( \lambda _{16}+\lambda _{18}\right) v_{1}^{2}%
\right] ,  \notag \\
C_{S} &=&\lambda _{19}v_{L}v_{R},\hspace{0.7cm}\hspace{0.7cm}\tan 2\theta
_{S}=\frac{2C_{S}}{A_{S}-B_{S}}.
\end{eqnarray}

Consequently, the physical scalar mass eigenstates states of the matrix $%
\widetilde{\mathbf{M}}_{CP-\text{even}}^{2}$ are given by: 
\begin{equation}
\left( 
\begin{array}{c}
S_{1} \\ 
S_{2}%
\end{array}%
\right) =\left( 
\begin{array}{cc}
\cos \theta _{S} & \sin \theta _{S} \\ 
-\sin \theta _{S} & \cos \theta _{S}%
\end{array}%
\right) \left( 
\begin{array}{c}
\func{Re}\phi _{L}^{0} \\ 
\func{Re}\phi _{R}^{0}%
\end{array}%
\right) .
\end{equation}

Their squared masses are:%
\begin{equation}
m_{S_{1/2}}^{2}=\frac{A_{S}+B_{S}}{2}\pm \frac{1}{2}\sqrt{\left(
A_{S}-B_{S}\right) ^{2}+4C_{S}^{2}}\;.
\end{equation}

Correlations between the masses of the non SM scalars are shown in Figure %
\ref{scalarcorrelations} and indicates that there are a large number of
solutions for the scalar masses consistent with experimental bounds. 
\begin{figure}[tbp]
\centering
\includegraphics[width=0.5\textwidth]{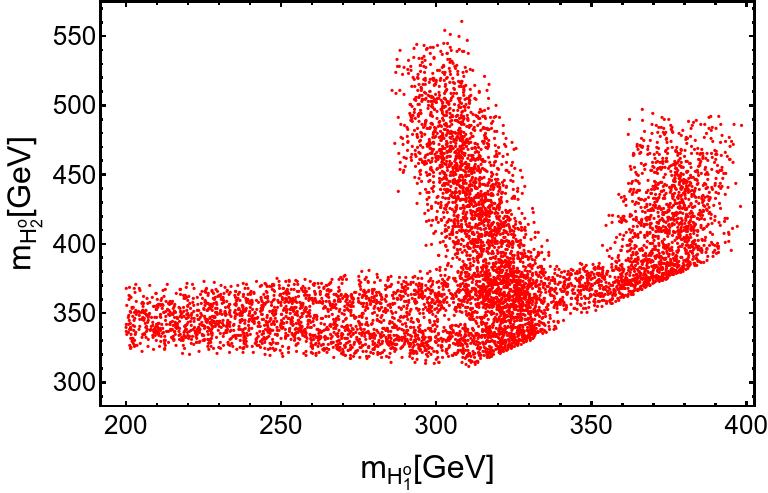}%
\includegraphics[width=0.5\textwidth]{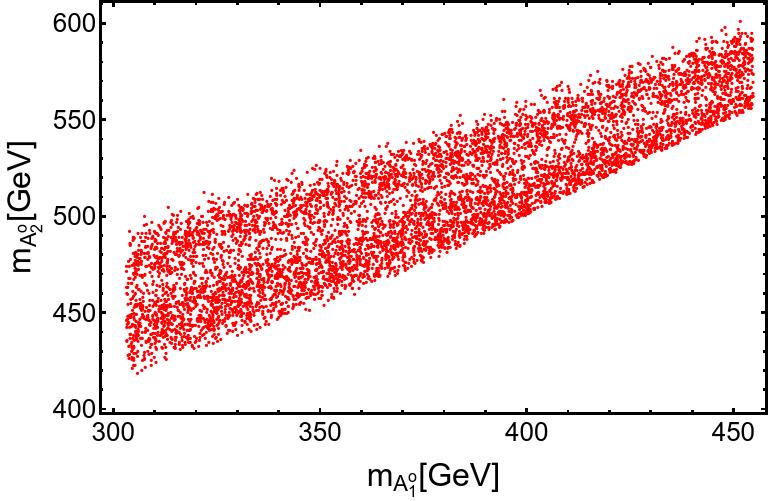}\newline
\includegraphics[width=0.5\textwidth]{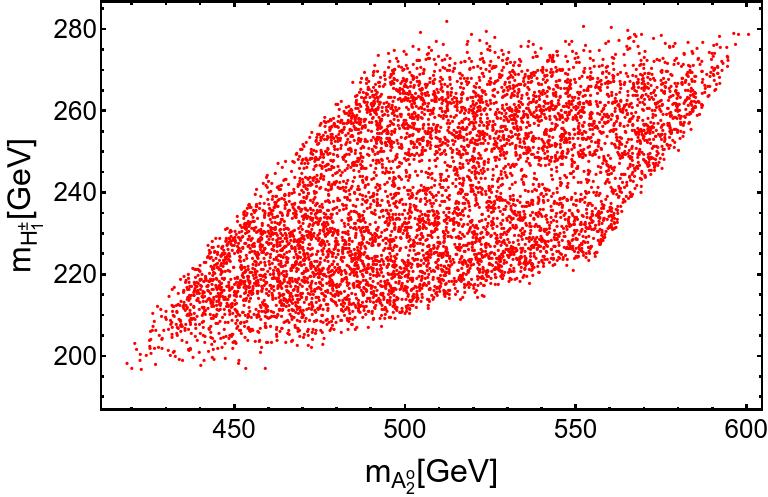}%
\includegraphics[width=0.5\textwidth]{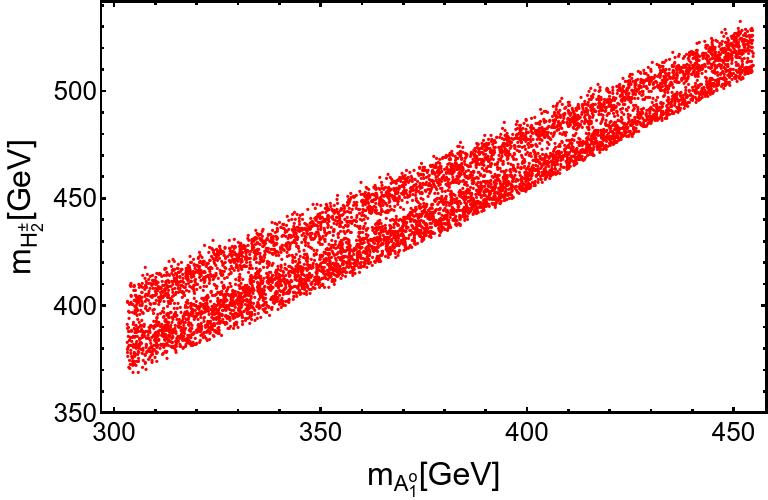}\newline
\includegraphics[width=0.5\textwidth]{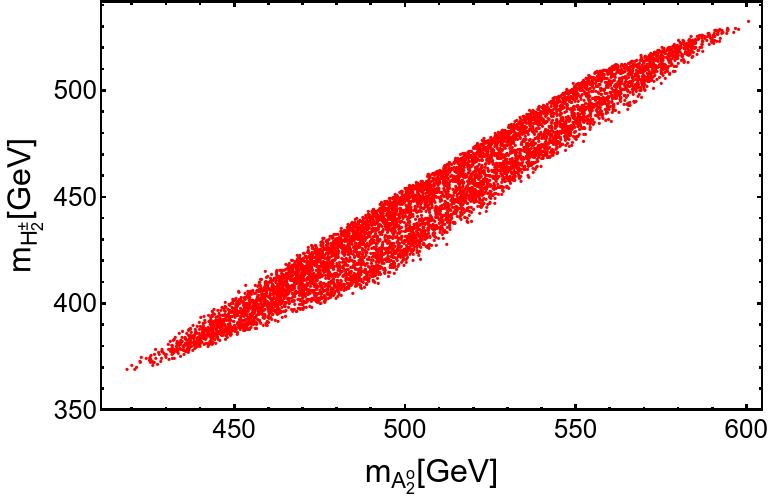}%
\includegraphics[width=0.5\textwidth]{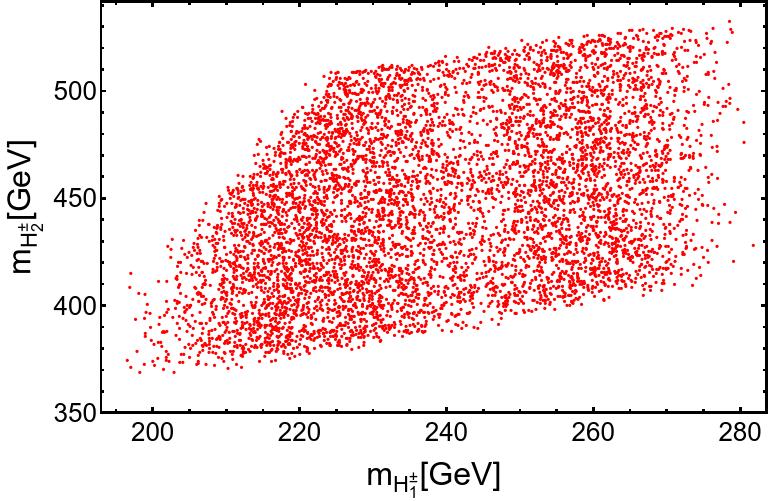}\newline
\includegraphics[width=0.5\textwidth]{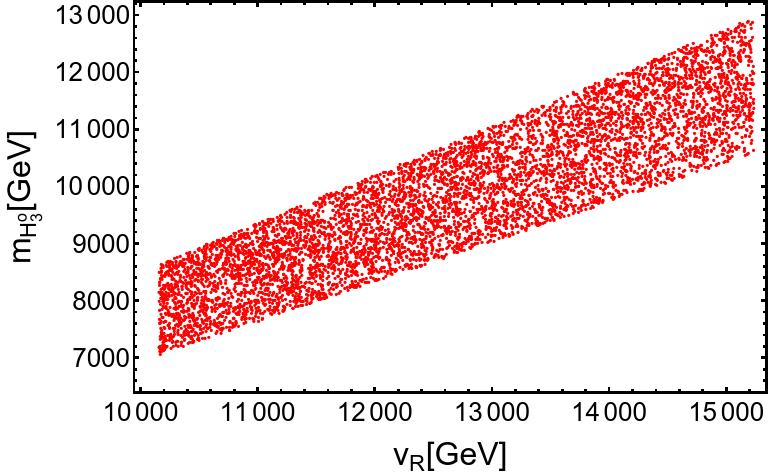}%
\caption{Correlations between the non SM scalar masses (top plots).
Correlation between the mass of the CP even neutral scalar $H_{3}^{0}$ and
the scale $v_R$ of breaking of the left-right symmetry.}
\label{scalarcorrelations}
\end{figure}

\section{Higgs diphoton decay rate}

\label{sec.Higgsdiphoton} 
The decay rate for the $h\rightarrow \gamma \gamma $ process takes the form: 
\begin{equation}
\Gamma (h\rightarrow \gamma \gamma )=\dfrac{\alpha _{em}^{2}m_{h}^{3}}{%
256\pi ^{3}v^{2}}\left\vert \sum_{f}a_{hff}N_{C}Q_{f}^{2}F_{1/2}(\rho
_{f})+a_{hWW}F_{1}(\rho _{W})+\sum_{k=1,2}\frac{C_{hH_{k}^{\pm }H_{k}^{\mp
}}v}{2m_{H_{k}^{\pm }}^{2}}F_{0}(\rho _{H_{k}^{\pm }})\right\vert ^{2},
\end{equation}%
where $\rho _{i}$ are the mass ratios $\rho _{i}=\frac{m_{h}^{2}}{4M_{i}^{2}}
$ with $M_{i}=m_{f},M_{W}$; $\alpha _{em}$ is the fine structure constant; $%
N_{C}$ is the color factor ($N_{C}=1$ for leptons and $N_{C}=3$ for quarks)
and $Q_{f}$ is the electric charge of the fermion in the loop. From the
fermion-loop contributions we only consider the dominant top quark term.
Furthermore, $C_{hH_{k}^{\pm }H_{k}^{\mp }}$ is the trilinear coupling
between the SM-like Higgs and a pair of charged Higges, whereas $a_{htt}$
and $a_{hWW}$ are the deviation factors from the SM Higgs-top quark coupling
and the SM Higgs-W gauge boson coupling, respectively (in the SM these
factors are unity). Such deviation factors are close to unity in our model,
which is a consequence of the numerical analysis of its scalar, Yukawa and
gauge sectors.

Furthermore, $F_{1/2}(z)$ and $F_{1}(z)$ are the dimensionless loop factors
for spin-$1/2$ and spin-$1$ particles running in the internal lines of the
loops. They are given by: 
\begin{align}
F_{1/2}(z)& =2(z+(z-1)f(z))z^{-2}, \\
F_{1}(z)& =-2(2z^{2}+3z+3(2z-1)f(z))z^{-2}, \\
F_{0}(z)& =-(z-f(z))z^{-2},
\end{align}%
with 
\begin{equation}
f(z)=\left\{ 
\begin{array}{lcc}
\arcsin ^{2}\sqrt{2} & \text{for} & z\leq 1 \\ 
&  &  \\ 
-\frac{1}{4}\left( \ln \left( \frac{1+\sqrt{1-z^{-1}}}{1-\sqrt{1-z^{-1}}%
-i\pi }\right) ^{2}\right) & \text{for} & z>1 \\ 
&  & 
\end{array}%
\right.
\end{equation}
In order to study the implications of our model in the decay of the $126$
GeV Higgs into a photon pair, one introduces the Higgs diphoton signal
strength $R_{\gamma \gamma }$, which is defined as: 
\begin{equation}
R_{\gamma \gamma }=\frac{\sigma (pp\rightarrow h)\Gamma (h\rightarrow \gamma
\gamma )}{\sigma (pp\rightarrow h)_{SM}\Gamma (h\rightarrow \gamma \gamma
)_{SM}}\simeq a_{htt}^{2}\frac{\Gamma (h\rightarrow \gamma \gamma )}{\Gamma
(h\rightarrow \gamma \gamma )_{SM}}.  \label{eqn:hgg}
\end{equation}%
That Higgs diphoton signal strength, normalizes the $\gamma \gamma $ signal
predicted by our model in relation to the one given by the SM. Here we have
used the fact that in our model, single Higgs production is also dominated
by gluon fusion as in the Standard Model.

The ratio $R_{\gamma \gamma }$ has been measured by CMS and ATLAS
collaborations with the best fit signals \cite{Sirunyan:2018ouh,Aad:2019mbh}%
: 
\begin{equation}
R_{\gamma \gamma }^{CMS}=1.18_{-0.14}^{+0.17}\quad \text{and}\quad R_{\gamma
\gamma }^{ATLAS}=0.96\pm 0.14.  \label{eqn:rgg}
\end{equation}%
The correlation of the Higgs diphoton signal strength with the charged
scalar mass $m_{H_{1}^{\pm }}$ is shown in Figure \ref{Higgsdiphoton}, which
indicates that our model successfully accommodates the current Higgs
diphoton decay rate constraints. Furthermore, as indicated by Figure \ref%
{Higgsdiphoton}, our model favours a Higgs diphoton decay rate lower than
the SM expectation but inside the $3\sigma$ experimentally allowed range. 
\begin{figure}[tbp]
\centering
\includegraphics[width=9.0cm, height=6.5cm]{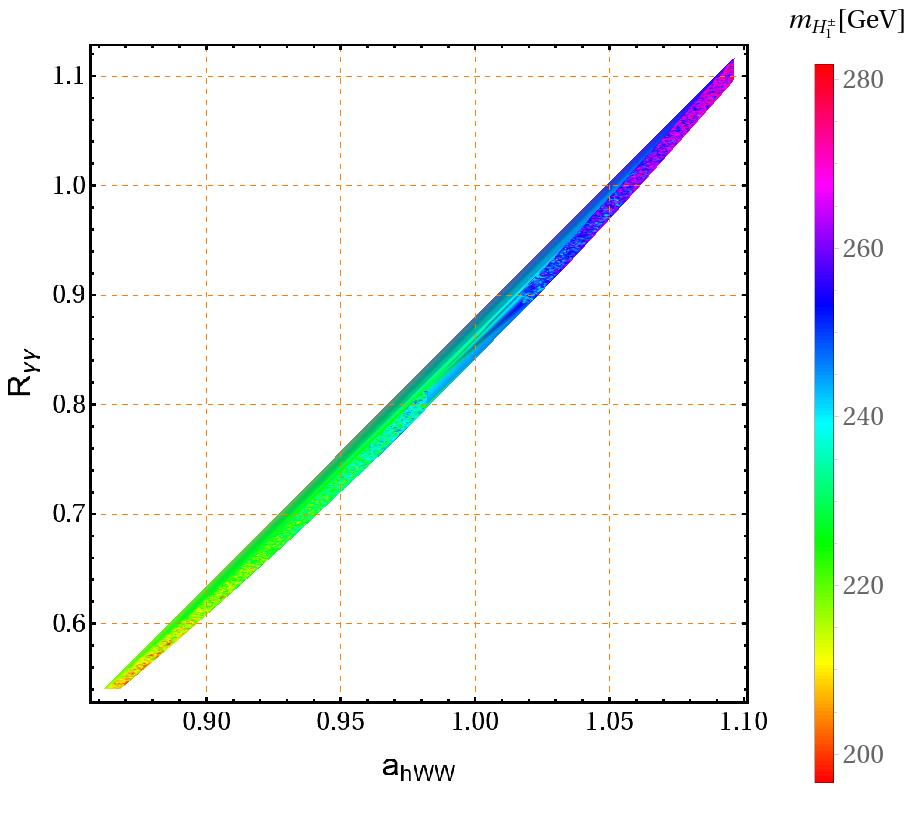}%
\includegraphics[width=9.0cm, height=6.5cm]{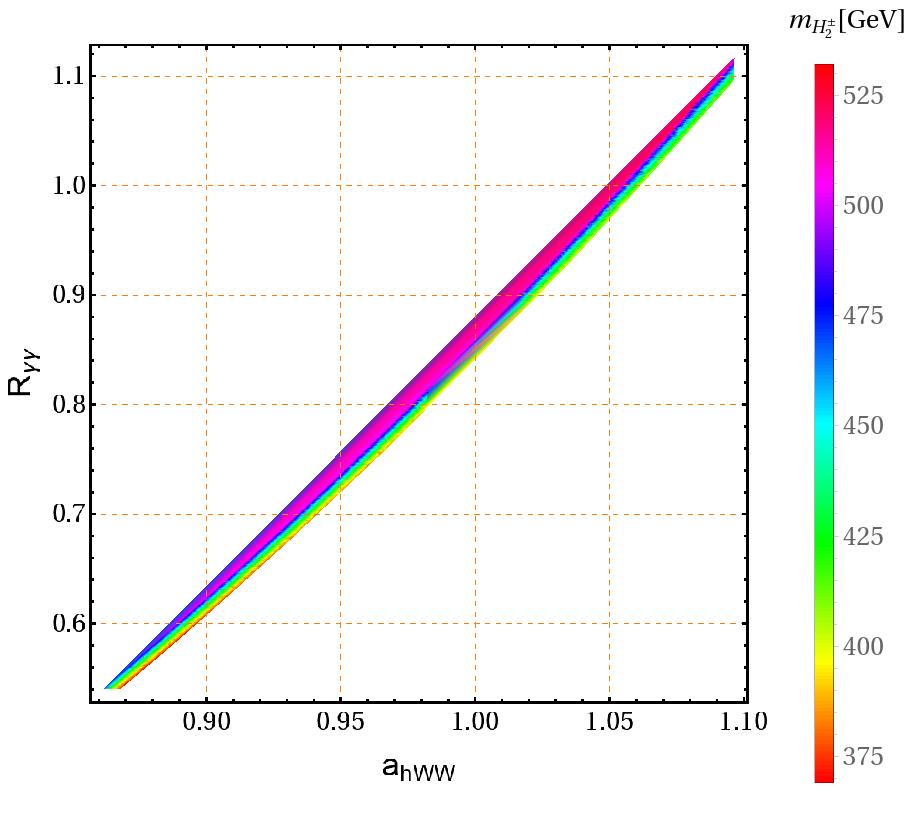} 
\caption{Correlation of the Higgs diphoton signal strength with the $a_{hWW}$
deviation factor from the SM Higgs-W gauge boson coupling.}
\label{Higgsdiphoton}
\end{figure}

\newpage

\section{Muon and electron anomalous magnetic moments}

\label{sec.gminus2} 
In this section we will analyze the implications of our model in the muon
and electron anomalous magnetic moments. 
The muon and electron anomalous magnetic moments receive contributions
arising from vertex diagrams involving the exchange of neutral scalars and
charged leptons running in the internal lines of the loop. \aech{The Feynman diagramas corresponding to these contributions are shown in Figure \ref{Diagramsgminus2}.}
\begin{figure}[tbp]
\centering
\includegraphics[width = 0.9\textwidth]{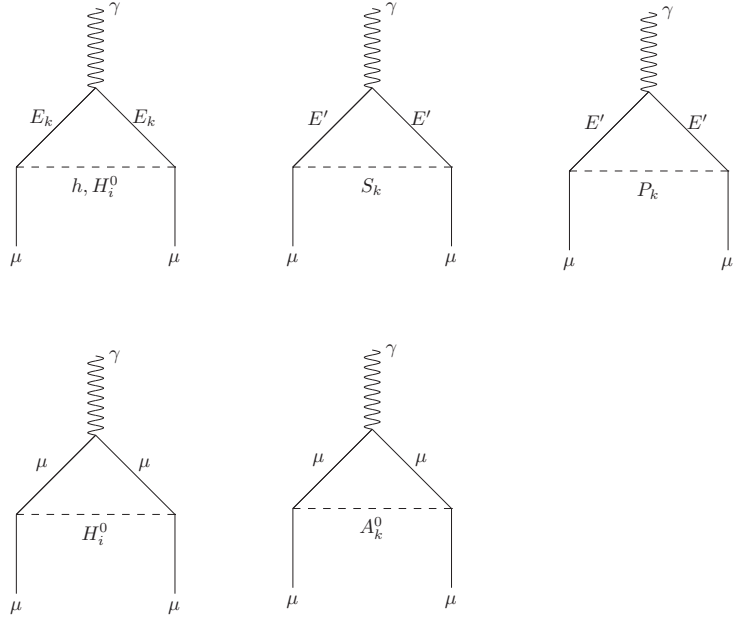}\vspace{-11cm}
\caption{One-loop Feynman diagrams contributing to the muon and electron
anomalous magnetic moments. Here $i=1,2,3$, $k=1,2$.}
\label{Diagramsgminus2}
\end{figure}
Then, in our model the contributions to the muon and electron anomalous
magnetic moments take the form: 
\begin{eqnarray}
\Delta a_{\mu } &=&\dsum\limits_{k=1}^{2}\frac{\func{Re}\left( \beta
_{2k}\gamma _{k2}^{\ast }\right) m_{\mu }^{2}}{8\pi ^{2}}\left( R_{CP-\text{%
even}}^{T}\right) _{21}\left( R_{CP-\text{even}}^{T}\right)
_{41}I_{S}^{\left( \mu \right) }\left( m_{E_{k}},m_{h^{0}}\right)  \notag \\
&&+\dsum\limits_{k=1}^{2}\frac{\func{Re}\left( \beta _{2k}\gamma _{k2}^{\ast
}\right) m_{\mu }^{2}}{8\pi ^{2}}\dsum\limits_{i=1}^{3}\left( R_{CP-\text{%
even}}^{T}\right) _{2,i+1}\left( R_{CP-\text{even}}^{T}\right)
_{4,i+1}I_{S}^{\left( \mu \right) }\left( m_{E_{k}},m_{H_{i}^{0}}\right) 
\notag \\
&&+\frac{m_{\mu }^{2}\func{Re}\left( \kappa _{2}\vartheta _{2}^{\ast
}\right) }{8\pi ^{2}}\left[ I_{S}^{\left( \mu \right) }\left( m_{E^{\prime
}},m_{S_{1}}\right) -I_{P}^{\left( \mu \right) }\left( m_{E^{\prime
}},m_{P_{1}}\right) -I_{S}^{\left( \mu \right) }\left( m_{E^{\prime
}},m_{S_{2}}\right) +I_{P}^{\left( \mu \right) }\left( m_{E^{\prime
}},m_{P_{2}}\right) \right] \sin \theta \cos \theta  \notag \\
&&+\frac{\left\vert y_{22}^{\left( L\right) }\right\vert ^{2}m_{\mu }^{2}}{%
8\pi ^{2}}\left[ \dsum\limits_{i=1}^{3}\left\vert \left( R_{CP-\text{even}%
}^{T}\right) _{3,i+1}\right\vert ^{2}I_{S}^{\left( \mu \right) }\left(
m_{\mu },m_{H_{i}^{0}}\right) +\dsum\limits_{i=1}^{2}\left\vert \left( R_{CP-%
\text{odd}}^{T}\right) _{4,i+2}\right\vert ^{2}I_{P}^{\left( \mu \right)
}\left( m_{\mu },m_{A_{i}^{0}}\right) \right]  \notag \\
\Delta a_{e} &=&\dsum\limits_{k=1}^{2}\frac{\func{Re}\left( \beta
_{1k}\gamma _{k1}^{\ast }\right) m_{e}^{2}}{8\pi ^{2}}\left( R_{CP-\text{even%
}}^{T}\right) _{21}\left( R_{CP-\text{even}}^{T}\right) _{41}I_{S}^{\left(
e\right) }\left( m_{E_{k}},m_{h^{0}}\right) \\
&&+\dsum\limits_{k=1}^{2}\frac{\func{Re}\left( \beta _{1k}\gamma _{k1}^{\ast
}\right) m_{e}^{2}}{8\pi ^{2}}\dsum\limits_{i=1}^{3}\left( R_{CP-\text{even}%
}^{T}\right) _{2,i+1}\left( R_{CP-\text{even}}^{T}\right)
_{4,i+1}I_{S}^{\left( e\right) }\left( m_{E_{k}},m_{H_{i}^{0}}\right)  \notag
\\
&&+\frac{m_{e}^{2}\func{Re}\left( \kappa _{1}\vartheta _{1}^{\ast }\right) }{%
8\pi ^{2}}\left[ I_{S}^{\left( e\right) }\left( m_{E^{\prime
}},m_{S_{1}}\right) -I_{P}^{\left( e\right) }\left( m_{E^{\prime
}},m_{P_{1}}\right) -I_{S}^{\left( e\right) }\left( m_{E^{\prime
}},m_{S_{2}}\right) +I_{P}^{\left( e\right) }\left( m_{E^{\prime
}},m_{P_{2}}\right) \right] \sin \theta \cos \theta  \notag \\
&&+\frac{\left\vert y_{11}^{\left( L\right) }\right\vert ^{2}m_{e}^{2}}{8\pi
^{2}}\left[ \dsum\limits_{i=1}^{3}\left\vert \left( R_{CP-\text{even}%
}^{T}\right) _{3,i+1}\right\vert ^{2}I_{S}^{\left( \mu \right) }\left(
m_{e},m_{H_{i}^{0}}\right) +\dsum\limits_{i=1}^{2}\left\vert \left( R_{CP-%
\text{odd}}^{T}\right) _{4,i+2}\right\vert ^{2}I_{P}^{\left( \mu \right)
}\left( m_{e},m_{A_{i}^{0}}\right) \right]  \notag
\end{eqnarray}%
where $\theta =\theta _{S}=-\theta _{P}$, being $\theta _{S}$ and $\theta
_{P}$ the $\func{Re}\phi _{L}^{0}-\func{Re}\phi _{R}^{0}$ and $\func{Im}\phi
_{L}^{0}-\func{Im}\phi _{R}^{0}$ mixing angles, respectively. Furthermore,
the loop $I_{S\left( P\right) }\left( m_{E},m\right) $ has the form \cite%
{Diaz:2002uk,Jegerlehner:2009ry,Kelso:2014qka,Lindner:2016bgg,Kowalska:2017iqv}%
: 
\begin{equation}
I_{S\left( P\right) }^{\left( e,\mu \right) }\left( m_{E},m_{S}\right)
=\int_{0}^{1}\frac{x^{2}\left( 1-x\pm \frac{m_{E}}{m_{e,\mu }}\right) }{%
m_{\mu }^{2}x^{2}+\left( m_{E}^{2}-m_{e,\mu }^{2}\right) x+m_{S,P}^{2}\left(
1-x\right) }dx
\end{equation}%
and the dimensionless parameters $\beta _{1k}$, $\beta _{2k}$, $\gamma _{k1}$%
, $\gamma _{k2}$, $\kappa _{1}$, $\kappa _{2}$, $\vartheta _{1}$, $\vartheta
_{2}$ are given by:

\begin{eqnarray}
\beta _{1k} &=&\dsum\limits_{i=1}^{3}x_{ik}^{\left( E\right) }\left(
V_{lL}^{\dagger }\right) _{1i},\hspace{0.7cm}\hspace{0.7cm}\gamma
_{k1}=\dsum\limits_{j=1}^{3}z_{kj}^{\left( E\right) }\left( V_{lR}\right)
_{j1}, \\
\beta _{2k} &=&\dsum\limits_{i=1}^{3}x_{ik}^{\left( E\right) }\left(
V_{lL}^{\dagger }\right) _{2i},\hspace{0.7cm}\hspace{0.7cm}\gamma
_{k2}=\dsum\limits_{j=1}^{3}z_{kj}^{\left( E\right) }\left( V_{lR}\right)
_{j2}, \\
\kappa _{1} &=&\dsum\limits_{i=1}^{3}w_{i}^{\left( E^{\prime }\right)
}\left( V_{lL}^{\dagger }\right) _{1i},\hspace{0.7cm}\hspace{0.7cm}\vartheta
_{1}=\dsum\limits_{j=1}^{3}r_{j}^{\left( E^{\prime }\right) }\left(
V_{lR}\right) _{j1}, \\
\kappa _{2} &=&\dsum\limits_{i=1}^{3}w_{i}^{\left( E^{\prime }\right)
}\left( V_{lL}^{\dagger }\right) _{2i},\hspace{0.7cm}\hspace{0.7cm}\vartheta
_{2}=\dsum\limits_{j=1}^{3}r_{j}^{\left( E^{\prime }\right) }\left(
V_{lR}\right) _{j2},
\end{eqnarray}

where $V_{lL}$ and $V_{lR}$ are the rotation matrices that diagonalize $%
\widetilde{M}_{E}$ according to the relation: 
\begin{equation}
V_{lL}^{\dagger }\widetilde{M}_{E}V_{lR}=diag\left( m_{e},m_{\mu },m_{\tau
}\right)
\end{equation}

Considering that the muon and electron anomalous magnetic moments are
constrained to be in the ranges \cite{Abi:2021gix,Morel:2020dww}: 
\begin{eqnarray}
\left( \Delta a_{\mu }\right) _{\exp } &=&\left( 2.51\pm 0.59\right) \times
10^{-9}  \notag \\
(\Delta a_{e})_{\text{exp}} &=&(4.8\pm 3.0)\times 10^{-13}.
\end{eqnarray}
We plot in Figure \ref{gminus2} the correlations of the muon and electron
anomalous magnetic moments with the masses $m_{A^0_{1}}$ and $m_{A^0_{2}}$
of the CP odd neutral scalar (top plots) as well as the correlation between
the electron and muon anomalous magnetic moments (bottom plot). We find that
our model can successfully accommodates the experimental values of the muon
and electron anomalous magnetic moments. 
\begin{figure}[tbp]
\centering
\includegraphics[width=8.9cm, height=6.0cm]{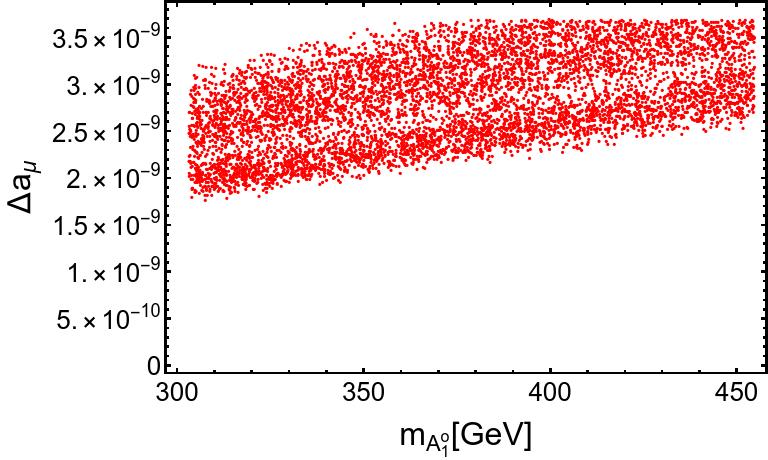} 
\includegraphics[width=8.9cm, height=6.0cm]{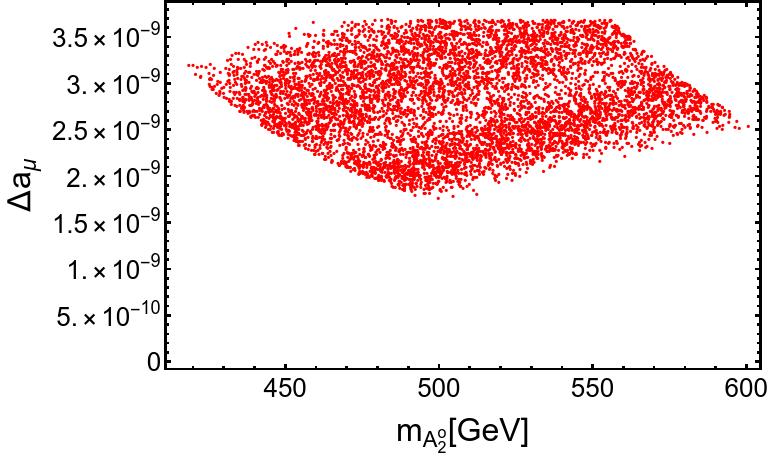}\newline
\includegraphics[width=8.9cm, height=6.0cm]{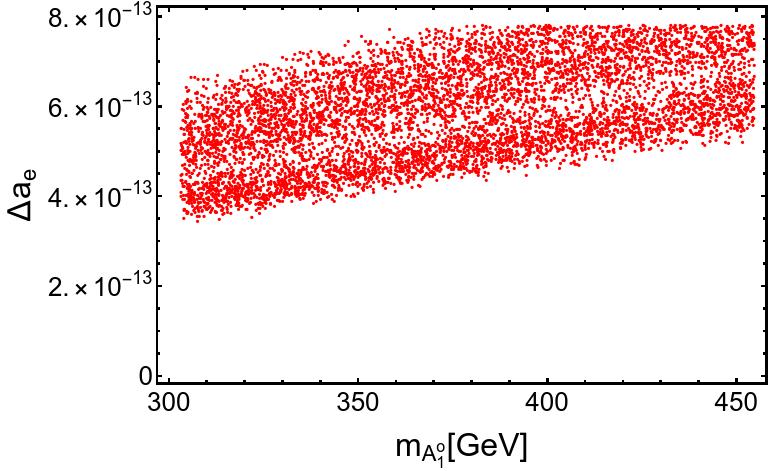} 
\includegraphics[width=8.9cm, height=6.0cm]{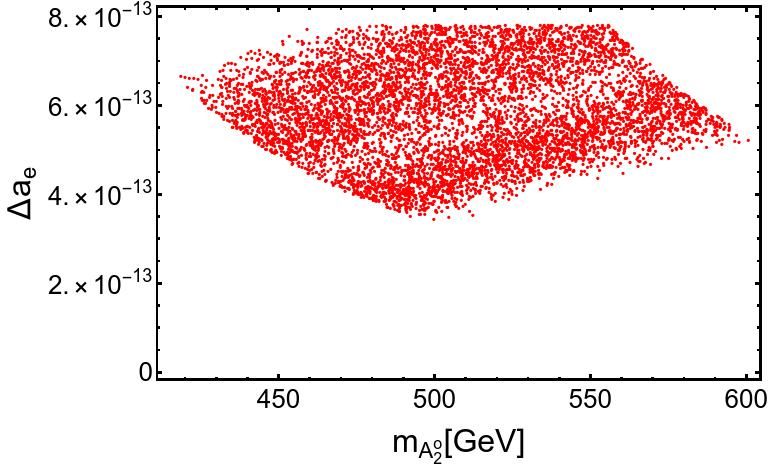}\newline
\includegraphics[width=8.9cm, height=6.0cm]{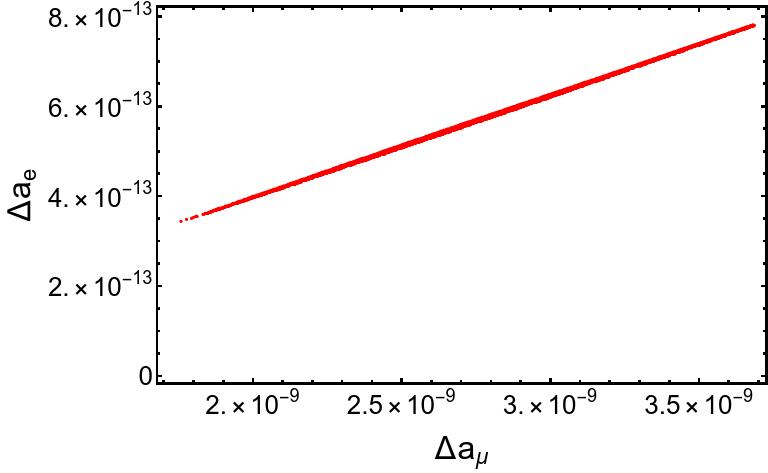}
\caption{Correlations of the muon and electron anomalous magnetic moments
with the masses $m_{A^0_{1}}$ and $m_{A^0_{2}}$ of the CP odd neutral
scalars (top plots). Correlation between the electron and muon anomalous
magnetic moments (bottom plot).}
\label{gminus2}
\end{figure}

\section{Heavy scalar production at the LHC}

\label{HeavyScalar} In this section we discuss the singly heavy scalar $%
H_{1}^{0}$ production at a proton-proton collider. Such production mechanism
at the LHC is dominated by the gluon fusion mechanism, which is a one-loop
process mediated by the top quark. Thus, the total $H_{1}^{0}$ production
cross section in proton-proton collisions with center of mass energy $\sqrt{S%
}$ takes the form: 
\begin{equation}
\sigma _{pp\rightarrow gg\rightarrow H_{1}^{0}}\left( S\right) =\frac{\alpha
_{S}^{2}a_{H_{1}^{0}t\bar{t}}^{2}m_{H_{1}^{0}}^{2}}{64\pi v^{2}S}\left[
I\left( \frac{m_{H_{1}^{0}}^{2}}{m_{t}^{2}}\right) \right] ^{2}\int_{\ln 
\sqrt{\frac{m_{H_{1}^{0}}^{2}}{S}}}^{-\ln \sqrt{\frac{m_{H_{1}^{0}}^{2}}{S}}%
}f_{p/g}\left( \sqrt{\frac{m_{H_{1}^{0}}^{2}}{S}}e^{y},\mu ^{2}\right)
f_{p/g}\left( \sqrt{\frac{m_{H_{1}^{0}}^{2}}{S}}e^{-y},\mu ^{2}\right) dy,
\end{equation}%
where $f_{p/g}\left( x_{1},\mu ^{2}\right) $ and $f_{p/g}\left( x_{2},\mu
^{2}\right) $ are the distributions of gluons in the proton which carry
momentum fractions $x_{1}$ and $x_{2}$ of the proton, respectively.
Furthermore $\mu =m_{H_{1}}$ is the factorization scale, whereas $I(z)$ has
the form: 
\begin{equation}
I(z)=\int_{0}^{1}dx\int_{0}^{1-x}dy\frac{1-4xy}{1-zxy}.  \label{g1a}
\end{equation}%
\begin{figure}[tbh]
\resizebox{8.5cm}{8cm}{\includegraphics{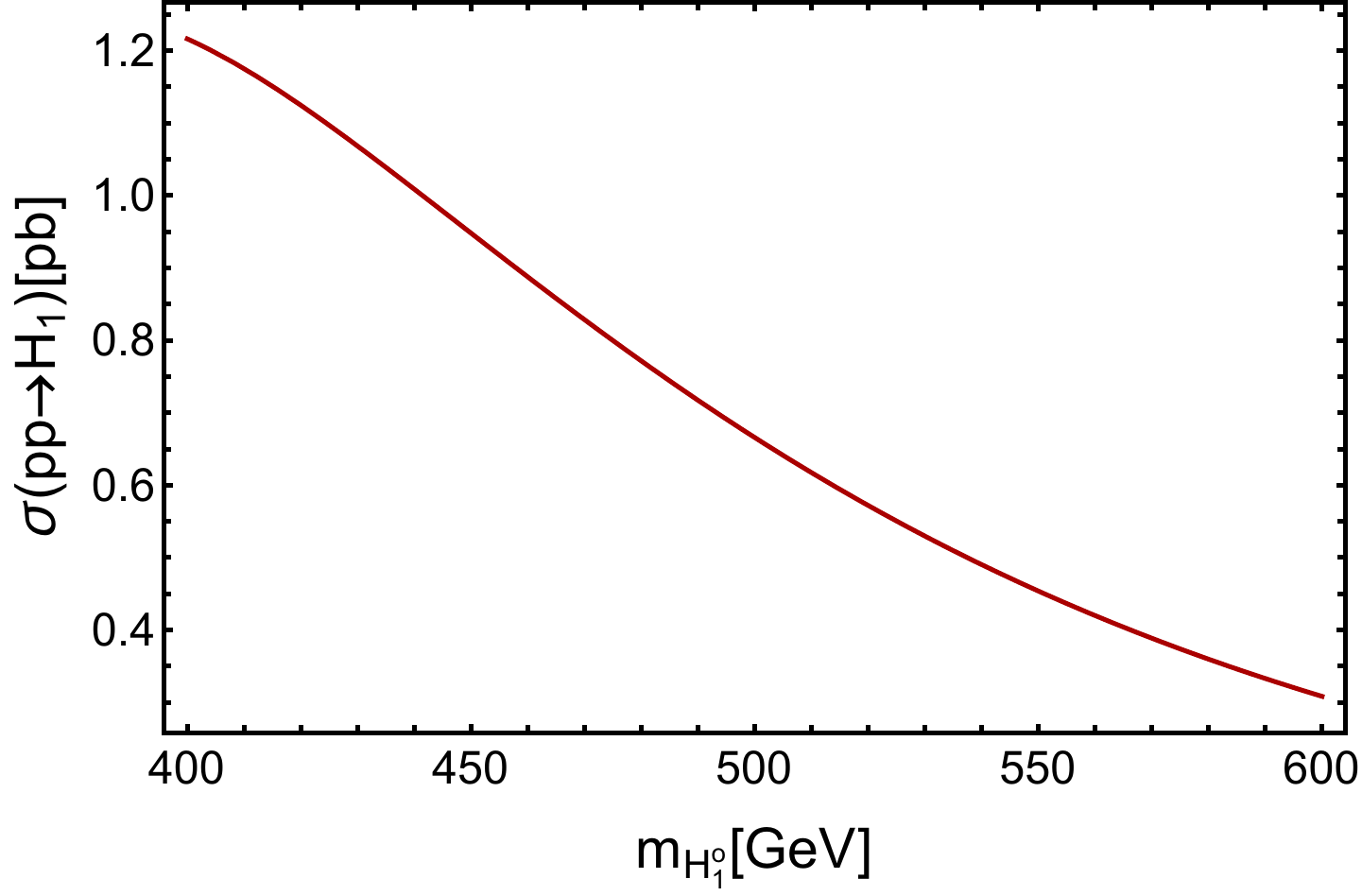}}%
\resizebox{8.5cm}{8cm}{\includegraphics{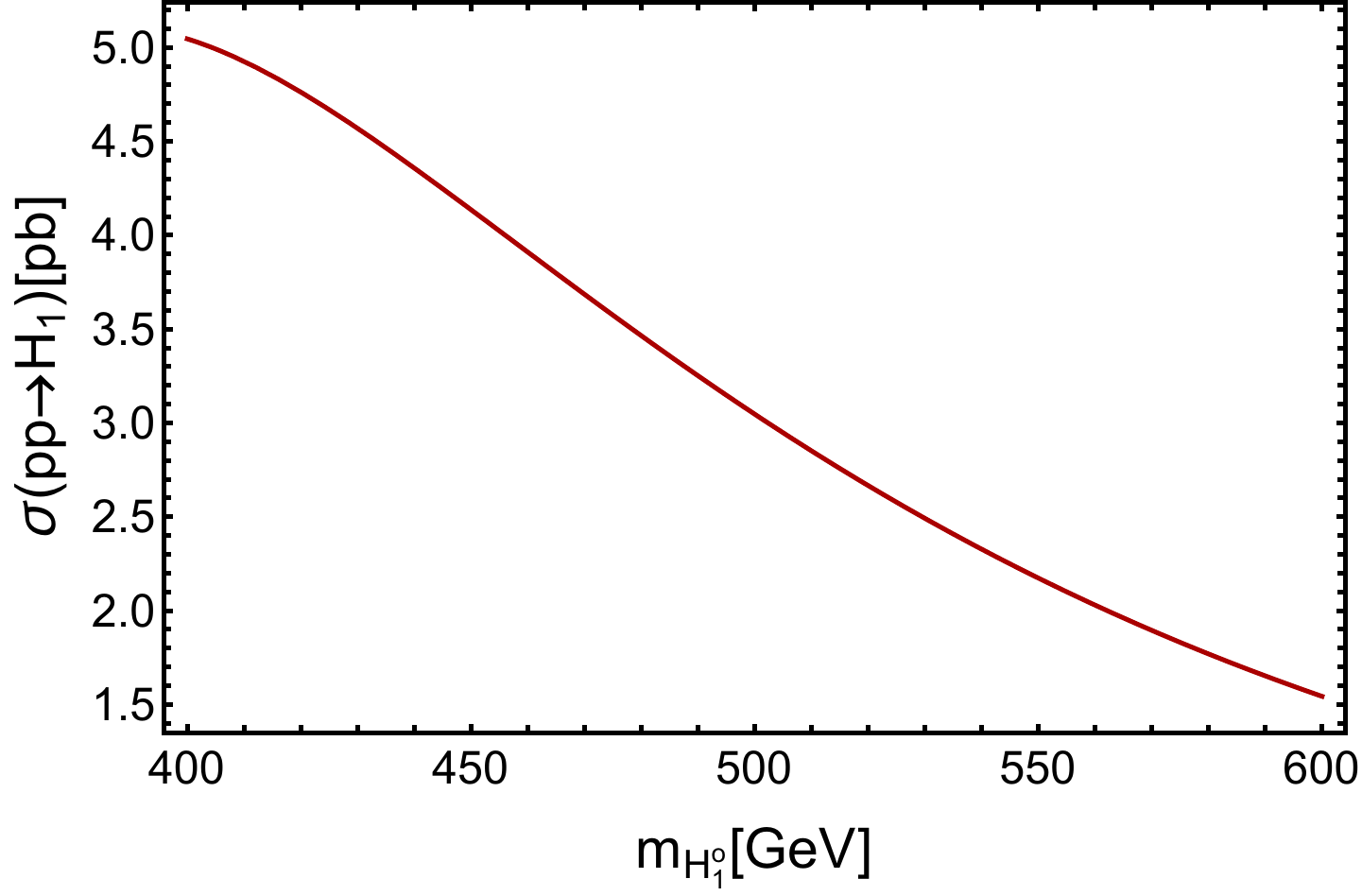}} 
\caption{Total cross section for the $H_{1}^{0}$ production via gluon fusion
mechanism at the LHC for $\protect\sqrt{s}=14$ TeV (left-panel) and $\protect%
\sqrt{S}=28$ (right-panel) TeV as a function of the heavy scalar mass $%
m_{H_{1}^{0}}$.}
\label{pptoH1}
\end{figure}
Figure~\ref{pptoH1} shows the $H_{1}^{0}$ total production cross section at
the LHC via gluon fusion mechanism for $\sqrt{S}=14$ TeV (left-plot) and $%
\sqrt{S}=28$ TeV (right-plot), as a function of the scalar mass $%
m_{H_{1}^{0}}$, which is taken to range from $400$ GeV up to $600$ GeV.
Furthermore, the coupling $a_{H_{1}^{0}t\bar{t}}$ of the heavy scalar $%
H_{1}^{0}$ with the top-antitop quark pair has been set to be equal to $0.4$%
, which is consistent with our numerical analysis of the scalar potential.
In the aforementioned region of masses for the heavy $H_{1}$ scalar, we find
that the total production cross section ranges from $1.2$ pb up to $0.3$ pb.
However, at the proposed energy upgrade of the LHC with $\sqrt{S}=28$ TeV,
the total cross section for the $H_{1}^{0}$ is enhanced reaching values
between $5$ pb and $1.5$ pb in the aforementioned mass range as indicated in
the right panel of Figure~\ref{pptoH1}. The heavy neutral $H_{1}^{0}$
scalar, after being produced, will have dominant decay modes into
top-antitop quark pairs, SM Higgs boson pairs as well as into a pair of SM
gauge bosons, thus implying that the observation of an excess of events in
the multileptons or multijet final states over the SM background can be a
smoking gun signature of this model, whose observation will be crucial to
assess its viability.

\section{$Z^\prime$ gauge boson production at the LHC}

\label{Zprime} In this section we discuss the single heavy $Z^{\prime }$
gauge boson via Drell-Yan mechanism at proton-proton collider. We consider
the dominant contributions due to the parton distribution functions of the
light up, down and strange quarks, so that the total cross section for the
production of a $Z^{\prime }$ via quark antiquark annihilation in
proton-proton collisions with center of mass energy $\sqrt{S}$ takes the
form: 
\begin{equation}
\sigma _{pp\rightarrow Z^{\prime }}^{\left( DrellYan\right) }(S)=\frac{%
g_{R}^{2}\pi }{24S}\int_{\ln \sqrt{\frac{m_{Z^{\prime }}^{2}}{S}}}^{-\ln 
\sqrt{\frac{m_{Z^{\prime }}^{2}}{S}}}\dsum\limits_{q=u,d,s}f_{p/q}\left( 
\sqrt{\frac{m_{Z^{\prime }}^{2}}{S}}e^{y},\mu ^{2}\right) f_{p/\overline{q}%
}\left( \sqrt{\frac{m_{Z^{\prime }}^{2}}{S}}e^{-y},\mu ^{2}\right) dy
\end{equation}%
where $f_{p/u}\left( x_{1},\mu ^{2}\right) $ ($f_{p/\overline{u}}\left(
x_{2},\mu ^{2}\right) $), $f_{p/d}\left( x_{1},\mu ^{2}\right) $ ($f_{p/%
\overline{d}}\left( x_{2},\mu ^{2}\right) $) and $f_{p/s}\left( x_{1},\mu
^{2}\right) $ ($f_{p/\overline{s}}\left( x_{2},\mu ^{2}\right) $) are the
distributions of the light up, down and strange quarks (antiquarks),
respectively, in the proton which carry momentum fractions $x_{1}$ ($x_{2}$)
of the proton. The factorization scale is taken to be $\mu =m_{Z^{\prime }}$%
. 
\begin{figure}[tbh]
\resizebox{8.5cm}{8cm}{\includegraphics{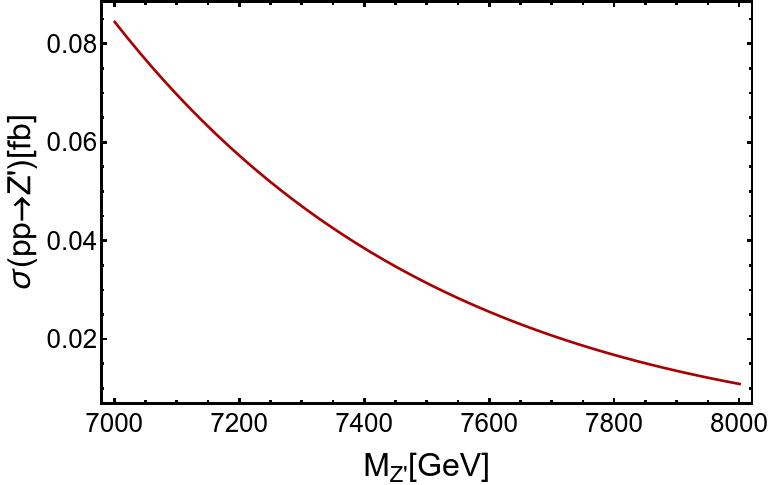}}%
\resizebox{8.5cm}{8cm}{\includegraphics{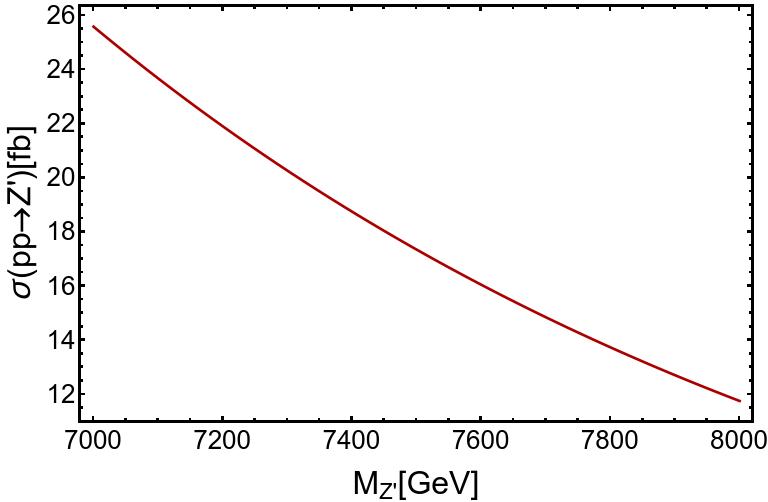}} 
\caption{Total cross section for the $Z^{\prime }$ production via Drell-Yan
mechanism at a proton-proton collider for $\protect\sqrt{S}=14$ TeV
(left-panel) and $\protect\sqrt{S}=28$ (right-panel) TeV as a function of
the $Z^{\prime }$ mass.}
\label{qqtoZprime}
\end{figure}
Fig.~\ref{qqtoZprime} displays the $Z^{\prime }$ total production cross
section at the LHC via the Drell-Yan mechanism for $\sqrt{S}=14$ TeV (left
panel) and $\sqrt{S}=28$ TeV (right panel) as a function of the $Z^{\prime }$
mass $M_{Z^{\prime }}$ in the range from $7$ TeV up to $8$ TeV. We consider $%
Z^{\prime }$ gauge boson masses larger than $7$ TeV and we set $g_R=1$, which is consistent with the constraint $\frac{M_{Z^{\prime }}}{g_R}>7$ TeV arising from LEP I and II measurements of $e^{+}e^{-}\rightarrow l^{+}l^{-}$ \cite{LEP:2004xhf,Carena:2004xs,Das:2021esm} as well
as with the ones resulting from LHC searches \cite{ATLAS:2019erb,CMS:2021ctt}. Limits on the ratio $\frac{M_{Z^{\prime }}}{g_R}$ are derived in Ref. \cite{Das:2021esm}, both for LEP II as well as for different values of the center of mass energy $\sqrt{s}$ of the future International Linear (ILC) $e^{+}e^{-}$ Collider. In this work we use the LEP II bound $\frac{M_{Z^{\prime }}}{g_R}>7$ TeV, since the other bounds correspond to future projective limits related to experiments which have not been started yet. With respect to the bounds of the $W^{\prime }$ gauge boson mass, CMS and ATLAS experiments at CERN have found that the $W^{\prime }$ gauge boson should be heavier than $6$ TeV \cite{CMS:2021qef} and $5$ TeV \cite{ATLAS:2018dcj}, respectively. }  

 For this region of $Z^{\prime }$ masses we find that the total production
cross section ranges from \ac{$0.85$ fb up to $0.01$ fb}. The heavy neutral $%
Z^{\prime }$ gauge boson, after being produced, will subsequently decay into
the pair of the SM fermion-antifermion pairs, thus implying that the
observation of an excess of events in the dileptons or dijet final states
over the SM background can be a signal of support of this model at the LHC.
On the other hand, at the proposed energy upgrade of the LHC at 28 TeV
center of mass energy, the total cross section for the Drell-Yan production
of a heavy $Z^{\prime }$ neutral gauge boson gets significantly enhanced
reaching values ranging from \ac{$26$ fb up to $12$ fb}, as indicated in the
right panel of Fig.~\ref{qqtoZprime}.

\newpage

\section{Meson oscillations}

\label{FCNC} In this section, we discuss the implications of our model in
the Flavour Changing Neutral Current (FCNC) interactions in the down type
quark sector. The FCNC Yukawa interactions in the down type quark sector
give rise to meson oscillations. The following effective Hamiltonians
describe $K^{0}-\bar{K}^{0}$, $B_{d}^{0}-\bar{B}_{d}^{0}$ and $B_{s}^{0}-%
\bar{B}_{s}^{0}$ mixings: 
\begin{equation}
\mathcal{H}_{eff}^{\left( K^{0}-\bar{K}^{0}\right) }\mathcal{=}\frac{%
G_{F}^{2}m_{W}^{2}}{16\pi ^{2}}\sum_{i=1}^{3}C_{i}^{\left( K^{0}-\bar{K}%
^{0}\right) }\left( \mu \right) O_{i}^{\left( K^{0}-\bar{K}^{0}\right)
}\left( \mu \right) ,
\end{equation}

\begin{equation}
\mathcal{H}_{eff}^{\left( B_{d}^{0}-\bar{B}_{d}^{0}\right) }\mathcal{=}\frac{%
G_{F}^{2}m_{W}^{2}}{16\pi ^{2}}\sum_{i=1}^{3}C_{i}^{\left( B_{d}^{0}-\bar{B}%
_{d}^{0}\right) }\left( \mu \right) O_{i}^{\left( B_{d}^{0}-\bar{B}%
_{d}^{0}\right) }\left( \mu \right) ,
\end{equation}

\begin{equation}
\mathcal{H}_{eff}^{\left( B_{s}^{0}-\bar{B}_{s}^{0}\right) }\mathcal{=}\frac{%
G_{F}^{2}m_{W}^{2}}{16\pi ^{2}}\sum_{i=1}^{3}C_{i}^{\left( B_{s}^{0}-\bar{B}%
_{s}^{0}\right) }\left( \mu \right) O_{i}^{\left( B_{s}^{0}-\bar{B}%
_{s}^{0}\right) }\left( \mu \right) ,
\end{equation}

In our analysis of meson oscillations we follow the approach of \cite%
{Dedes:2002er,Aranda:2012bv}. The $K^{0}-\bar{K}^{0}$, $B_{d}^{0}-\bar{B}%
_{d}^{0}$ and $B_{s}^{0}-\bar{B}_{s}^{0}$ meson mixings receive tree level
contributions corresponding to the exchange of neutral CP even and CP odd
scalars, thus giving rise to the following operators: 
\begin{eqnarray}
O_{1}^{\left( K^{0}-\bar{K}^{0}\right) } &=&\left( \overline{s}P_{L}d\right)
\left( \overline{s}P_{L}d\right) ,\hspace{1cm}O_{2}^{\left( K^{0}-\bar{K}%
^{0}\right) }=\left( \overline{s}P_{R}d\right) \left( \overline{s}%
P_{R}d\right) ,\hspace{1cm}O_{3}^{\left( K^{0}-\bar{K}^{0}\right) }=\left( 
\overline{s}P_{L}d\right) \left( \overline{s}P_{R}d\right) ,  \label{op3f} \\
O_{1}^{\left( B_{d}^{0}-\bar{B}_{d}^{0}\right) } &=&\left( \overline{d}%
P_{L}b\right) \left( \overline{d}P_{L}b\right) ,\hspace{1cm}O_{2}^{\left(
B_{d}^{0}-\bar{B}_{d}^{0}\right) }=\left( \overline{d}P_{R}b\right) \left( 
\overline{d}P_{R}b\right) ,\hspace{1cm}O_{3}^{\left( B_{d}^{0}-\bar{B}%
_{d}^{0}\right) }=\left( \overline{d}P_{L}b\right) \left( \overline{d}%
P_{R}b\right) ,\hspace{0.7cm} \\
O_{1}^{\left( B_{s}^{0}-\bar{B}_{s}^{0}\right) } &=&\left( \overline{s}%
P_{L}b\right) \left( \overline{s}P_{L}b\right) ,\hspace{1cm}O_{2}^{\left(
B_{s}^{0}-\bar{B}_{s}^{0}\right) }=\left( \overline{s}P_{R}b\right) \left( 
\overline{s}P_{R}b\right) ,\hspace{1cm}O_{3}^{\left( B_{s}^{0}-\bar{B}%
_{s}^{0}\right) }=\left( \overline{s}P_{L}b\right) \left( \overline{s}%
P_{R}b\right) ,
\end{eqnarray}%
where the corresponding Wilson coefficients are given by: 
\begin{eqnarray}
C_{1}^{\left( K^{0}-\bar{K}^{0}\right) } &=&\frac{16\pi ^{2}}{%
G_{F}^{2}m_{W}^{2}}\widetilde{C}_{1}^{\left( K^{0}-\bar{K}^{0}\right) },%
\hspace{0.7cm}\hspace{0.7cm}\widetilde{C}_{1}^{\left( K^{0}-\bar{K}%
^{0}\right) }=\frac{y_{h\overline{s}_{R}d_{L}}^{2}}{m_{h}^{2}}+\sum_{i=1}^{3}%
\frac{y_{H_{i}^{0}\overline{s}_{R}d_{L}}^{2}}{m_{H_{i}^{0}}^{2}}%
-\sum_{i=1}^{2}\frac{y_{A_{i}^{0}\overline{s}_{R}d_{L}}^{2}}{%
m_{A_{i}^{0}}^{2}}, \\
C_{2}^{\left( K^{0}-\bar{K}^{0}\right) } &=&\frac{16\pi ^{2}}{%
G_{F}^{2}m_{W}^{2}}\widetilde{C}_{2}^{\left( K^{0}-\bar{K}^{0}\right) },%
\hspace{0.7cm}\hspace{0.7cm}\widetilde{C}_{2}^{\left( K^{0}-\bar{K}%
^{0}\right) }=\frac{y_{h\overline{s}_{L}d_{R}}^{2}}{m_{h}^{2}}+\sum_{i=1}^{3}%
\frac{y_{H_{i}^{0}\overline{s}_{L}d_{R}}^{2}}{m_{H_{i}^{0}}^{2}}%
-\sum_{i=1}^{2}\frac{y_{A_{i}^{0}\overline{s}_{L}d_{R}}^{2}}{%
m_{A_{i}^{0}}^{2}}, \\
C_{3}^{\left( K^{0}-\bar{K}^{0}\right) } &=&\frac{16\pi ^{2}}{%
G_{F}^{2}m_{W}^{2}}\widetilde{C}_{3}^{\left( K^{0}-\bar{K}^{0}\right) },%
\hspace{0.3cm}\widetilde{C}_{3}^{\left( K^{0}-\bar{K}^{0}\right) }=\frac{y_{h%
\overline{s}_{R}d_{L}}y_{h\overline{s}_{L}d_{R}}}{m_{h}^{2}}+\sum_{i=1}^{3}%
\frac{y_{H_{i}^{0}\overline{s}_{R}d_{L}}y_{H_{i}^{0}\overline{s}_{L}d_{R}}}{%
m_{H_{i}^{0}}^{2}}-\sum_{i=1}^{2}\frac{y_{A_{i}^{0}\overline{s}%
_{R}d_{L}}y_{A_{i}^{0}\overline{s}_{L}d_{R}}}{m_{A_{i}^{0}}^{2}},
\end{eqnarray}%
\begin{eqnarray}
C_{1}^{\left( B_{d}^{0}-\bar{B}_{d}^{0}\right) } &=&\frac{16\pi ^{2}}{%
G_{F}^{2}m_{W}^{2}}\widetilde{C}_{1}^{\left( B_{d}^{0}-\bar{B}%
_{d}^{0}\right) },\hspace{0.7cm}\hspace{0.7cm}\widetilde{C}_{1}^{\left(
B_{d}^{0}-\bar{B}_{d}^{0}\right) }=\frac{y_{h\overline{d}_{R}b_{L}}^{2}}{%
m_{h}^{2}}+\sum_{i=1}^{3}\frac{y_{H_{i}^{0}\overline{d}_{R}b_{L}}^{2}}{%
m_{H_{i}^{0}}^{2}}-\sum_{i=1}^{2}\frac{y_{A_{i}^{0}\overline{d}_{R}b_{L}}^{2}%
}{m_{A_{i}^{0}}^{2}}, \\
C_{2}^{\left( B_{d}^{0}-\bar{B}_{d}^{0}\right) } &=&\frac{16\pi ^{2}}{%
G_{F}^{2}m_{W}^{2}}\widetilde{C}_{2}^{\left( B_{d}^{0}-\bar{B}%
_{d}^{0}\right) },\hspace{0.7cm}\hspace{0.7cm}\widetilde{C}_{2}^{\left(
B_{d}^{0}-\bar{B}_{d}^{0}\right) }=\frac{y_{h\overline{d}_{L}b_{R}}^{2}}{%
m_{h}^{2}}+\sum_{i=1}^{3}\frac{y_{H_{i}^{0}\overline{d}_{L}b_{R}}^{2}}{%
m_{H_{i}^{0}}^{2}}-\sum_{i=1}^{2}\frac{y_{A_{i}^{0}\overline{d}_{L}b_{R}}^{2}%
}{m_{A_{i}^{0}}^{2}}, \\
C_{3}^{\left( B_{d}^{0}-\bar{B}_{d}^{0}\right) } &=&\frac{16\pi ^{2}}{%
G_{F}^{2}m_{W}^{2}}\widetilde{C}_{3}^{\left( B_{d}^{0}-\bar{B}%
_{d}^{0}\right) },\hspace{0.3cm}\widetilde{C}_{3}^{\left( B_{d}^{0}-\bar{B}%
_{d}^{0}\right) }=\frac{y_{h\overline{d}_{R}b_{L}}y_{h\overline{d}_{L}b_{R}}%
}{m_{h}^{2}}+\sum_{i=1}^{3}\frac{y_{H_{i}^{0}\overline{d}%
_{R}b_{L}}y_{H_{i}^{0}\overline{d}_{L}b_{R}}}{m_{H_{i}^{0}}^{2}}%
-\sum_{i=1}^{2}\frac{y_{A_{i}^{0}\overline{d}_{R}b_{L}}y_{A_{i}^{0}\overline{%
d}_{L}b_{R}}}{m_{A_{i}^{0}}^{2}},
\end{eqnarray}%
\begin{eqnarray}
C_{1}^{\left( B_{s}^{0}-\bar{B}_{s}^{0}\right) } &=&\frac{16\pi ^{2}}{%
G_{F}^{2}m_{W}^{2}}\widetilde{C}_{1}^{\left( B_{s}^{0}-\bar{B}%
_{s}^{0}\right) },\hspace{0.7cm}\hspace{0.7cm}\widetilde{C}_{1}^{\left(
B_{s}^{0}-\bar{B}_{s}^{0}\right) }=\frac{y_{h\overline{s}_{R}b_{L}}^{2}}{%
m_{h}^{2}}+\sum_{i=1}^{3}\frac{y_{H_{i}^{0}\overline{s}_{R}b_{L}}^{2}}{%
m_{H_{i}^{0}}^{2}}-\sum_{i=1}^{2}\frac{y_{A_{i}^{0}\overline{s}_{R}b_{L}}^{2}%
}{m_{A_{i}^{0}}^{2}}, \\
C_{2}^{\left( B_{s}^{0}-\bar{B}_{s}^{0}\right) } &=&\frac{16\pi ^{2}}{%
G_{F}^{2}m_{W}^{2}}\widetilde{C}_{2}^{\left( B_{s}^{0}-\bar{B}%
_{s}^{0}\right) },\hspace{0.7cm}\hspace{0.7cm}\widetilde{C}_{2}^{\left(
B_{s}^{0}-\bar{B}_{s}^{0}\right) }=\frac{y_{h\overline{s}_{L}b_{R}}^{2}}{%
m_{h}^{2}}+\sum_{i=1}^{3}\frac{y_{H_{i}^{0}\overline{s}_{L}b_{R}}^{2}}{%
m_{H_{i}^{0}}^{2}}-\sum_{i=1}^{2}\frac{y_{A_{i}^{0}\overline{s}_{L}b_{R}}^{2}%
}{m_{A_{i}^{0}}^{2}}, \\
C_{3}^{\left( B_{s}^{0}-\bar{B}_{s}^{0}\right) } &=&\frac{16\pi ^{2}}{%
G_{F}^{2}m_{W}^{2}}\widetilde{C}_{3}^{\left( B_{s}^{0}-\bar{B}%
_{s}^{0}\right) },\hspace{0.3cm}\widetilde{C}_{3}^{\left( B_{s}^{0}-\bar{B}%
_{s}^{0}\right) }=\frac{y_{h\overline{s}_{R}b_{L}}y_{h\overline{s}_{L}b_{R}}%
}{m_{h}^{2}}+\sum_{i=1}^{3}\frac{y_{H_{i}^{0}\overline{s}%
_{R}b_{L}}y_{H_{i}^{0}\overline{s}_{L}b_{R}}}{m_{H_{i}^{0}}^{2}}%
-\sum_{i=1}^{2}\frac{y_{A_{i}^{0}\overline{s}_{R}b_{L}}y_{A_{i}^{0}\overline{%
s}_{L}b_{R}}}{m_{A_{i}^{0}}^{2}},
\end{eqnarray}%
Furthermore, the $K-\bar{K}$, $B_{d}^{0}-\bar{B}_{d}^{0}$ and $B_{s}^{0}-%
\bar{B}_{s}^{0}$\ mass splittings can be written as: 
\begin{equation}
\Delta m_{K}=\left( \Delta m_{K}\right) _{SM}+\Delta m_{K}^{\left( NP\right)
},\hspace{1cm}\Delta m_{B_{d}}=\left( \Delta m_{B_{d}}\right) _{SM}+\Delta
m_{B_{d}}^{\left( NP\right) },\hspace{1cm}\Delta m_{B_{s}}=\left( \Delta
m_{B_{s}}\right) _{SM}+\Delta m_{B_{s}}^{\left( NP\right) },  \label{Deltam}
\end{equation}%
where $\left( \Delta m_{K}\right) _{SM}$, $\left( \Delta m_{B_{d}}\right)
_{SM}$ and $\left( \Delta m_{B_{s}}\right) _{SM}$ are the SM contributions,
whereas $\Delta m_{K}^{\left( NP\right) }$ , $\Delta m_{B_{d}}^{\left(
NP\right) }$ and $\left( \Delta m_{B_{s}}\right) _{SM}$ are new physics
contributions.

\begin{figure}[h]
\includegraphics[width=0.7\textwidth]{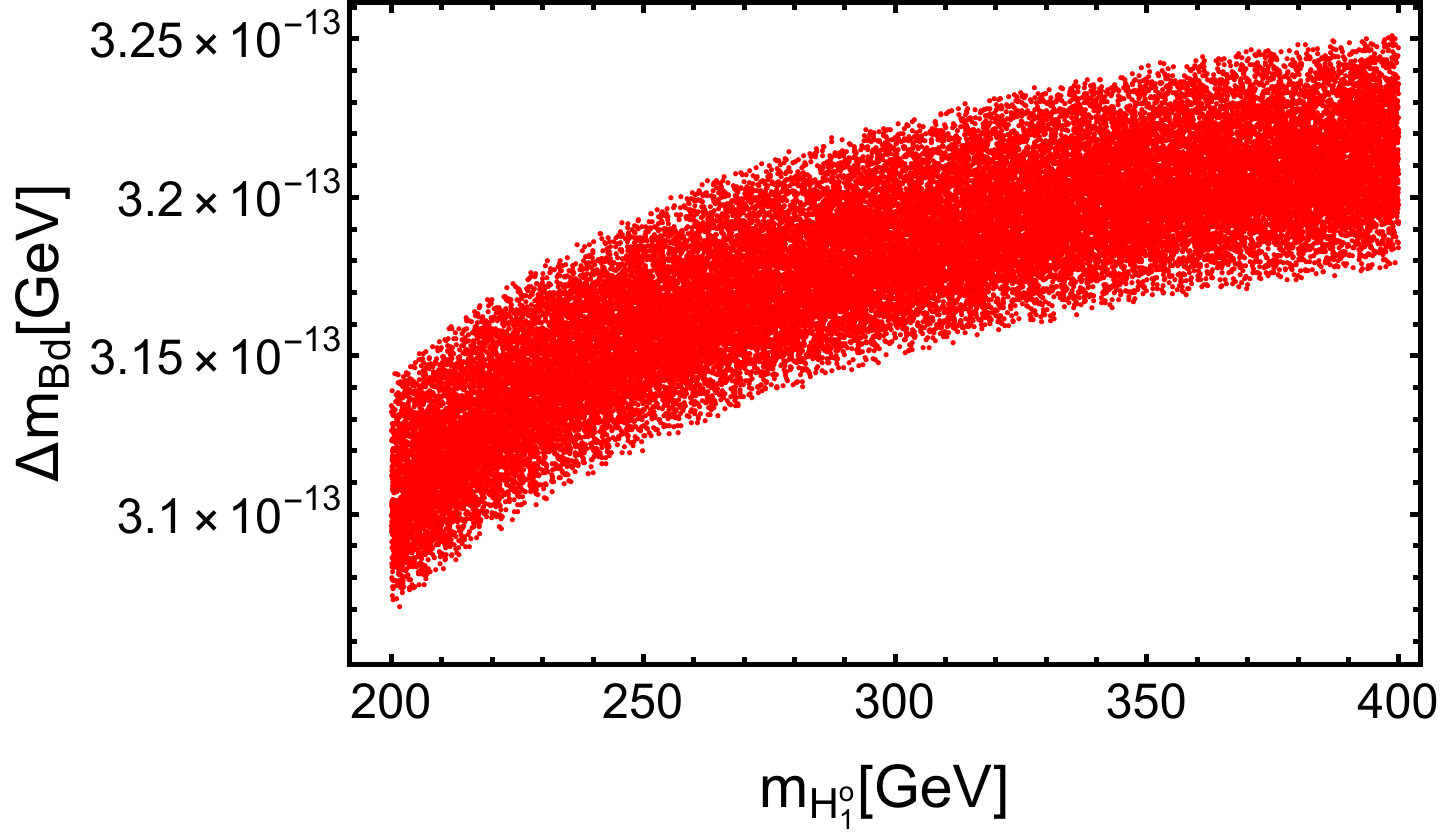}
\caption{Correlation between the $\Delta m_{B_{d}}$ mass splitting and the
heavy CP even scalar mass $m_{H_{1}^{0}}$. The couplings of the flavor
changing neutral Yukawa interactions have been set to be equal to $10^{-4}$.}
\label{BBbar}
\end{figure}

In our model, the new physics contributions to the meson differences are
given by: 
\begin{eqnarray}
\Delta m_{K}^{\left( NP\right) } &=&\frac{G_{F}^{2}m_{W}^{2}}{6\pi ^{2}}%
m_{K}f_{K}^{2}\eta _{K}B_{K}\left[ P_{2}^{\left( K^{0}-\bar{K}^{0}\right)
}C_{3}^{\left( K^{0}-\bar{K}^{0}\right) }+P_{1}^{\left( K^{0}-\bar{K}%
^{0}\right) }\left( C_{1}^{\left( K^{0}-\bar{K}^{0}\right) }+C_{2}^{\left(
K^{0}-\bar{K}^{0}\right) }\right) \right]  \notag \\
&=&\frac{8}{3}m_{K}f_{K}^{2}\eta _{K}B_{K}\left[ P_{2}^{\left( K^{0}-\bar{K}%
^{0}\right) }\widetilde{C}_{3}^{\left( K^{0}-\bar{K}^{0}\right)
}+P_{1}^{\left( K^{0}-\bar{K}^{0}\right) }\left( \widetilde{C}_{1}^{\left(
K^{0}-\bar{K}^{0}\right) }+\widetilde{C}_{2}^{\left( K^{0}-\bar{K}%
^{0}\right) }\right) \right]
\end{eqnarray}%
\begin{eqnarray}
\Delta m_{B_{d}}^{\left( NP\right) } &=&\frac{G_{F}^{2}m_{W}^{2}}{6\pi ^{2}}%
m_{B_{d}}f_{B_{d}}^{2}\eta _{B_{d}}B_{B_{d}}\left[ P_{2}^{\left( B_{d}^{0}-%
\bar{B}_{d}^{0}\right) }C_{3}^{\left( B_{d}^{0}-\bar{B}_{d}^{0}\right)
}+P_{1}^{\left( B_{d}^{0}-\bar{B}_{d}^{0}\right) }\left( C_{1}^{\left(
B_{d}^{0}-\bar{B}_{d}^{0}\right) }+C_{2}^{\left( B_{d}^{0}-\bar{B}%
_{d}^{0}\right) }\right) \right]  \notag \\
&=&\frac{8}{3}m_{B_{d}}f_{B_{d}}^{2}\eta _{B_{d}}B_{B_{d}}\left[
P_{2}^{\left( B_{d}^{0}-\bar{B}_{d}^{0}\right) }\widetilde{C}_{3}^{\left(
B_{d}^{0}-\bar{B}_{d}^{0}\right) }+P_{1}^{\left( B_{d}^{0}-\bar{B}%
_{d}^{0}\right) }\left( \widetilde{C}_{1}^{\left( B_{d}^{0}-\bar{B}%
_{d}^{0}\right) }+\widetilde{C}_{2}^{\left( B_{d}^{0}-\bar{B}_{d}^{0}\right)
}\right) \right]
\end{eqnarray}%
\begin{eqnarray}
\Delta m_{B_{s}}^{\left( NP\right) } &=&\frac{G_{F}^{2}m_{W}^{2}}{6\pi ^{2}}%
m_{B_{s}}f_{B_{s}}^{2}\eta _{B_{s}}B_{B_{s}}\left[ P_{2}^{\left( B_{s}^{0}-%
\bar{B}_{s}^{0}\right) }C_{3}^{\left( B_{s}^{0}-\bar{B}_{s}^{0}\right)
}+P_{1}^{\left( B_{s}^{0}-\bar{B}_{s}^{0}\right) }\left( C_{1}^{\left(
B_{s}^{0}-\bar{B}_{s}^{0}\right) }+C_{2}^{\left( B_{s}^{0}-\bar{B}%
_{s}^{0}\right) }\right) \right]  \notag \\
&=&\frac{8}{3}m_{B_{s}}f_{B_{s}}^{2}\eta _{B_{s}}B_{B_{s}}\left[
P_{2}^{\left( B_{s}^{0}-\bar{B}_{s}^{0}\right) }\widetilde{C}_{3}^{\left(
B_{s}^{0}-\bar{B}_{s}^{0}\right) }+P_{1}^{\left( B_{s}^{0}-\bar{B}%
_{s}^{0}\right) }\left( \widetilde{C}_{1}^{\left( B_{s}^{0}-\bar{B}%
_{s}^{0}\right) }+\widetilde{C}_{2}^{\left( B_{s}^{0}-\bar{B}_{s}^{0}\right)
}\right) \right]
\end{eqnarray}%
Using the following parameters \cite%
{Dedes:2002er,Aranda:2012bv,Khalil:2013ixa,Queiroz:2016gif,Buras:2016dxz,Ferreira:2017tvy,Duy:2020hhk,Branco:2021vhs,Zyla:2020zbs}%
: 
\begin{eqnarray}
\Delta m_{K} &=&\left( 3.484\pm 0.006\right) \times 10^{-12}MeV,\hspace{1.5cm%
}\left( \Delta m_{K}\right) _{SM}=3.483\times 10^{-12}MeV  \notag \\
f_{K} &=&\left( 155.7\pm 0.3\right) MeV,\hspace{1.5cm}B_{K}=0.717\pm 0.024,%
\hspace{1.5cm}\eta _{K}=0.57,  \notag \\
P_{1}^{\left( K^{0}-\bar{K}^{0}\right) } &=&-9.3,\hspace{1.5cm}P_{2}^{\left(
K^{0}-\bar{K}^{0}\right) }=30.6,\hspace{1.5cm}m_{K}=\left( 497.611\pm
0.013\right) MeV,\hspace{1.5cm}
\end{eqnarray}%
\begin{eqnarray}
\left( \Delta m_{B_{d}}\right) _{\exp } &=&\left( 3.334\pm 0.013\right)
\times 10^{-10}MeV,\hspace{1.5cm}\left( \Delta m_{B_{d}}\right)
_{SM}=3.582\times 10^{-10}MeV,  \notag \\
f_{B_{d}} &=&\left( 190.0\pm 1.3\right) MeV,\hspace{1.5cm}B_{B_{d}}=1.30\pm
0.10,\hspace{1.5cm}\eta _{B_{d}}=0.55,  \notag \\
P_{1}^{\left( B_{d}^{0}-\bar{B}_{d}^{0}\right) } &=&-0.52,\hspace{1.5cm}%
P_{2}^{\left( B_{d}^{0}-\bar{B}_{d}^{0}\right) }=0.88,\hspace{1.5cm}%
m_{B_{d}}=\left( 5279.65\pm 0.12\right) MeV,\hspace{1.5cm}
\end{eqnarray}%
\begin{eqnarray}
\left( \Delta m_{B_{s}}\right) _{\exp } &=&\left( 1.1683\pm 0.0013\right)
\times 10^{-8}MeV,\hspace{1.5cm}\left( \Delta m_{B_{s}}\right)
_{SM}=1.21103\times 10^{-8}MeV,  \notag \\
f_{B_{s}} &=&\left( 230.3\pm 1.3\right) MeV,\hspace{1.5cm}B_{B_{s}}=1.35\pm
0.06,\hspace{1.5cm}\eta _{B_{s}}=0.55,  \notag \\
P_{1}^{\left( B_{s}^{0}-\bar{B}_{s}^{0}\right) } &=&-0.52,\hspace{1.5cm}%
P_{2}^{\left( B_{s}^{0}-\bar{B}_{s}^{0}\right) }=0.88,\hspace{1.5cm}%
m_{B_{s}}=\left( 5366.88\pm 0.14\right) MeV,\hspace{1.5cm}
\end{eqnarray}%
Figure \ref{BBbar} displays the correlation between the $\Delta m_{B_{d}}$
mass splitting and the heavy CP even scalar mass $m_{H_{1}^{0}}$. In our
numerical analysis, for the sake of simplicity, we have set the couplings of
the flavor changing neutral Yukawa interactions that produce the $B_{d}^{0}-%
\bar{B}_{d}^{0}$ oscillations to be equal to $10^{-4}$. Furthermore, we have
fixed $m_{H_{3}^{0}}=10$ TeV and we have varied the masses of $H_{1}^{0}$, $%
H_{2}^{0}$ and $A_{1}^{0}$ in the ranges $200$ GeV$\leqslant
m_{H_{1}^{0}}\leqslant $ $400$ GeV, $350$ GeV$\leqslant
m_{H_{2}^{0}}\leqslant $ $550$ GeV and $300$ GeV$\leqslant
m_{A_{1}^{0}}\leqslant $ $450$ GeV, whereas we have also set $%
m_{A_{2}^{0}}=m_{A_{1}^{0}}+150GeV$. It is worth mentioning that the above
described ranges of scalar masses is consistent with the ones described in
the correlation plots of heavy scalar masses shown in Figure \ref%
{scalarcorrelations}. As indicated in Figure \ref{BBbar}, the experimental
constraints arising from $B_{d}^{0}-\bar{B}_{d}^{0}$ meson oscillations are
successfully fullfilled for the aforementioned range of parameter space. We
have numerically checked that in the above described range of masses, the
obtained values for the $\Delta m_{B_{s}}$ and $\Delta m_{K}$ mass
splittings are consistent with the experimental data on meson oscillations
for flavor violating Yukawa couplings equal to $2.5\times 10^{-4}$ and $%
10^{-6}$ for the $B_{s}^{0}-\bar{B}_{s}^{0}$ and $K^{0}-\bar{K}^{0}$
mixings, respectively. 

\section{Conclusions}

\label{conclusions} We have built a renormalizable left-right symmetric
theory with additional symmetry $Z_{4}^{\left( 1\right) }\times
Z_{4}^{\left( 2\right) }$ consistent with the observed SM fermion mass
hierarchy, the tiny values for the light active neutrino masses, the lepton
and baryon asymmetries of the Universe, the constraints arising from meson
oscillations, from charged lepton flavor violation, as well as the muon and electron anomalous magnetic moments. As
the main appealing feature of the proposed model, the top and exotic
fermions get their masses at tree level whereas the masses of the bottom,
charm and strange quarks, tau and muon leptons are generated from a tree
level Universal Seesaw mechanism thanks to their mixings with charged exotic
vector like fermions. The first generation SM charged fermions masses are
produced from a radiative seesaw mechanism at one loop level mediated by
charged vector like fermions and electrically neutral scalars. The tiny
masses of the light active neutrinos arise from an inverse seesaw mechanism
at one-loop level. Furthermore, we have also shown that the proposed model
successfully accommodates the current Higgs diphoton decay rate constraints,
yielding a Higgs diphoton decay rate lower than the SM expectation but
inside the $3\sigma $ experimentally allowed range. We also studied the
heavy $H_{1}^{0}$ scalar and $Z^{\prime }$ gauge boson production in a
proton-proton collider at $\sqrt{S}=14$ TeV and $\sqrt{S}=28$ TeV, via the
gluon fusion and Drell-Yan mechanisms, respectively. We found that the
singly $H_{1}^{0}$ scalar production cross section reach values of $1.2$ and 
$5$ pb at $\sqrt{S}=14$ TeV and $\sqrt{S}=28$ TeV, respectively, for a $400$
GeV heavy scalar mass. On the other hand, we found that the total cross
section for the $Z^{\prime }$ gauge boson production takes the values of $0.85$
fb and $26$ fb at $\sqrt{S}=14$ TeV and $\sqrt{S}=28$ TeV, respectively,
for a $7$ TeV $Z^{\prime }$ gauge boson mass.

\section*{Acknowledgments}

A.E.C.H and I.S. are supported by ANID-Chile FONDECYT 1210378, ANID-Chile
FONDECYT 1180232, ANID-Chile FONDECYT 3150472, ANID PIA/APOYO AFB180002 and
Milenio-ANID-ICN2019\_044

\appendix

\section{Analytical argument of the minimal number of seesaw mediators}

\label{M} In this appendix we provide an analytical argument of the minimal
number of fermionic seesaw mediators required to generate the masses of SM
fermions via a seesaw-like mechanism. We start by considering the case of
two heavy seesaw mediators which mix the three fermion families, thus giving
rise to the following general structure of the low energy fermionic mass
matrix: 
\begin{equation}
M=\left( 
\begin{array}{ccc}
F_{1}G_{1}+X_{1}Y_{1} & F_{1}G_{2}+X_{1}Y_{2} & F_{1}G_{3}+X_{1}Y_{3} \\ 
F_{2}G_{1}+X_{2}Y_{1} & F_{2}G_{2}+X_{2}Y_{2} & F_{2}G_{3}+X_{2}Y_{3} \\ 
F_{3}G_{1}+X_{3}Y_{1} & F_{3}G_{2}+X_{3}Y_{2} & F_{3}G_{3}+X_{3}Y_{3}%
\end{array}%
\right) ,  \label{Mwith2}
\end{equation}

Then, the $\left( i,j\right) $ matrix element of $M$ can be written as: 
\begin{equation}
M_{ij}=M_{j}^{i}=F_{i}G_{j}+X_{i}Y_{j}=F^{i}G_{j}+X^{i}Y_{j}  \label{Mij}
\end{equation}

where 
\begin{equation}
F^{i}=F_{i},\hspace{1.5cm}X^{i}=X_{i}
\end{equation}

being $F_{i}$, $G_{j}$, $X_{i}$ and $Y_{j}$ ($i=1,2,3$)\ functions including
the Yukawa couplings corresponding the interactions generating the vertices
of the loop as well as the loop integrals depending on the masses of the
heavy fermions and scalars running in the internal lines of the loop.

As it will be shown, below, the structure of the mass matrix $M$ of Eq. {(%
\ref{Mwith2}}) is so that $\det \left( M\right) =0$, thus implying the
existence of one massless fermion. To prove that, we start by the
considering the general expresion of the determinant for the $n\times n$
matrix: 
\begin{eqnarray}
\det \left( M_{\mu }^{\nu }\right) &=&\frac{1}{n!}\sum_{\mu _{1},\mu
_{2},\cdots \mu _{k},\nu _{1},\nu _{2},\cdots \nu _{k}=1}^{n}\bigskip
\varepsilon _{\mu _{1}\mu _{2}\ldots \mu _{k}}\varepsilon ^{\nu _{1}\nu
_{2}\ldots \nu _{k}}M_{\nu _{1}}^{\mu _{1}}M_{\nu _{2}}^{\mu _{2}}\cdots
M_{\nu _{k}}^{\mu _{k}}  \notag \\
&=&\frac{1}{n!}\sum_{\mu _{1},\mu _{2},\cdots \mu _{k},\nu _{1},\nu
_{2},\cdots \nu _{k}=1}^{n}\delta _{\mu _{1}\ldots \mu _{k}}^{\nu _{1}\ldots
\nu _{k}}M_{\nu _{1}}^{\mu _{1}}M_{\nu _{2}}^{\mu _{2}}\cdots M_{\nu
_{k}}^{\mu _{k}}  \label{detMgen}
\end{eqnarray}

where $\delta _{\mu _{1}\ldots \mu _{j}}^{\nu _{1}\ldots \nu _{j}}$ is the
generalized Kronecker delta defined by the following determinant:

\begin{eqnarray}
\delta _{\mu _{1}\ldots \mu _{j}}^{\nu _{1}\ldots \nu _{j}} &=&\left\vert 
\begin{array}{cccc}
\delta _{\mu _{1}}^{\nu _{1}} & \delta _{\mu _{2}}^{\nu _{1}} & \ldots & 
\delta _{\mu _{j}}^{\nu _{1}} \\ 
\delta _{\mu _{1}}^{\nu _{2}} & \delta _{\mu _{2}}^{\nu _{2}} & \ldots & 
\delta _{\mu _{j}}^{\nu _{2}} \\ 
\ldots & \ldots & \ldots & \ldots \\ 
\delta _{\mu _{1}}^{\nu _{j}} & \delta _{\mu _{2}}^{\nu _{j}} & \ldots & 
\delta _{\mu _{j}}^{\nu _{j}}%
\end{array}%
\right\vert =\delta _{\mu _{1}}^{\nu _{1}}\delta _{\mu _{2}\mu _{3}\ldots
\mu _{j}}^{\nu _{2}\nu _{3}\ldots \nu _{j}}-\delta _{\mu _{2}}^{\nu
_{1}}\delta _{\mu _{1}\mu _{3}\ldots \mu _{j}}^{\nu _{2}\nu _{3}\ldots \nu
_{j}}+\delta _{\mu _{3}}^{\nu _{1}}\delta _{\mu _{1}\mu _{2}\mu _{4}\ldots
\mu _{j}}^{\nu _{2}\nu _{3}\nu _{4}\ldots \nu _{j}}+\cdots +\left( -1\right)
^{j+1}\delta _{\mu _{j}}^{\nu _{1}}\delta _{\mu _{1}\mu _{2}\ldots \mu
_{j-1}}^{\nu _{2}\nu _{3}\ldots \nu _{j}}  \notag \\
&=&\delta _{\mu _{1}}^{\nu _{1}}\left( \delta _{\mu _{2}}^{\nu _{2}}\delta
_{\mu _{3}\mu _{4}\ldots \mu _{j}}^{\nu _{3}\nu _{4}\ldots \nu _{j}}-\delta
_{\mu _{3}}^{\nu _{2}}\delta _{\mu _{2}\mu _{4}\ldots \mu _{j}}^{\nu _{3}\nu
_{4}\ldots \nu _{j}}+\delta _{\mu _{4}}^{\nu _{2}}\delta _{\mu _{2}\mu
_{3}\mu _{5}\ldots \mu _{j}}^{\nu _{3}\nu _{4}\nu _{5}\ldots \nu
_{j}}+\cdots +\left( -1\right) ^{j}\delta _{\mu _{j}}^{\nu _{2}}\delta _{\mu
_{2}\mu _{3}\ldots \mu _{j-1}}^{\nu _{3}\nu _{4}\ldots \nu _{j}}\right) 
\notag \\
&&-\delta _{\mu _{2}}^{\nu _{1}}\left( \delta _{\mu _{1}}^{\nu _{2}}\delta
_{\mu _{3}\mu _{4}\ldots \mu _{j}}^{\nu _{3}\nu _{4}\ldots \nu _{j}}-\delta
_{\mu _{3}}^{\nu _{2}}\delta _{\mu _{1}\mu _{4}\ldots \mu _{j}}^{\nu _{3}\nu
_{4}\ldots \nu _{j}}+\delta _{\mu _{4}}^{\nu _{2}}\delta _{\mu _{1}\mu
_{3}\mu _{5}\ldots \mu _{j}}^{\nu _{3}\nu _{4}\nu _{5}\ldots \nu
_{j}}+\cdots +\left( -1\right) ^{j}\delta _{\mu _{j}}^{\nu _{2}}\delta _{\mu
_{1}\mu _{3}\ldots \mu _{j-1}}^{\nu _{3}\nu _{4}\ldots \nu _{j}}\right) 
\notag \\
&&+\delta _{\mu _{3}}^{\nu _{1}}\left( \delta _{\mu _{1}}^{\nu _{2}}\delta
_{\mu _{2}\mu _{4}\ldots \mu _{j}}^{\nu _{3}\nu _{4}\ldots \nu _{j}}-\delta
_{\mu _{2}}^{\nu _{2}}\delta _{\mu _{1}\mu _{4}\ldots \mu _{j}}^{\nu _{3}\nu
_{4}\ldots \nu _{j}}+\delta _{\mu _{4}}^{\nu _{2}}\delta _{\mu _{1}\mu
_{2}\mu _{5}\ldots \mu _{j}}^{\nu _{3}\nu _{4}\nu _{5}\ldots \nu
_{j}}+\cdots +\left( -1\right) ^{j}\delta _{\mu _{j}}^{\nu _{2}}\delta _{\mu
_{1}\mu _{2}\ldots \mu _{j-1}}^{\nu _{3}\nu _{4}\ldots \nu _{j}}\right) 
\notag \\
&&+\cdots +\left( -1\right) ^{j+1}\delta _{\mu _{j}}^{\nu _{1}}\left( \delta
_{\mu _{1}}^{\nu _{2}}\delta _{\mu _{2}\mu _{3}\ldots \mu _{j-1}}^{\nu
_{3}\nu _{4}\ldots \nu _{j}}-\delta _{\mu _{2}}^{\nu _{2}}\delta _{\mu
_{1}\mu _{3}\ldots \mu _{j-1}}^{\nu _{3}\nu _{4}\ldots \nu _{j}}+\delta
_{\mu _{3}}^{\nu _{2}}\delta _{\mu _{1}\mu _{2}\mu _{4}\ldots \mu
_{j-1}}^{\nu _{3}\nu _{4}\nu _{5}\ldots \nu _{j}}+\cdots +\left( -1\right)
^{j}\delta _{\mu _{j-1}}^{\nu _{2}}\delta _{\mu _{1}\mu _{2}\ldots \mu
_{j-2}}^{\nu _{3}\nu _{4}\ldots \nu _{j}}\right)  \notag \\
&=&\sum_{\sigma =1}^{j!}sign\left( \sigma \right) \prod_{k=1}^{j}\delta
_{\mu _{\sigma \left( k\right) }}^{\nu _{\sigma \left( k\right) }}=j!\delta
_{\mu _{1}}^{[\nu _{1}}\delta _{\mu _{2}}^{\nu _{2}}\cdots \delta _{\mu
_{j}}^{\nu _{j}]}  \label{1c15}
\end{eqnarray}

and the $\left[ \cdots \right] $ denotes antisymmetrization on the enclosed
indices as usual. This antisymmetrization for a tensor $A_{\mu _{1}\ldots
\mu _{j}}$ is defined as:\ 
\begin{equation}
A_{\left[ \mu _{1}\ldots \mu _{j}\right] }=\frac{1}{j!}\sum_{\sigma
=1}^{j!}sign\left( \sigma \right) A_{\sigma \left( \mu _{1}\right) \cdots
\sigma \left( \mu _{j}\right) }  \label{1c16}
\end{equation}

Then the following relations are fullfilled:

\begin{equation}
\delta _{ik}^{jl}=\left\vert 
\begin{array}{cc}
\delta _{i}^{j} & \delta _{k}^{j} \\ 
\delta _{i}^{l} & \delta _{k}^{l}%
\end{array}%
\right\vert =\delta _{i}^{j}\delta _{k}^{l}-\delta _{k}^{j}\delta _{i}^{l}
\end{equation}

\begin{equation}
\delta _{ikh}^{jrs}=\left\vert 
\begin{array}{ccc}
\delta _{i}^{j} & \delta _{k}^{j} & \delta _{h}^{j} \\ 
\delta _{i}^{r} & \delta _{k}^{r} & \delta _{h}^{r} \\ 
\delta _{i}^{s} & \delta _{k}^{s} & \delta _{h}^{s}%
\end{array}%
\right\vert =\sum_{\sigma =1}^{3!}sign\left( \sigma \right)
\prod_{k=1}^{3}\delta _{\mu _{\sigma \left( k\right) }}^{\nu _{\sigma \left(
k\right) }}=3!\delta _{\mu _{1}}^{[\nu _{1}}\delta _{\mu _{2}}^{\nu
_{2}}\delta _{\mu _{3}}^{\nu _{3}]}
\end{equation}

As a consistency check of Eq. {(\ref{detMgen}}), we show the result obtained
for the case of a $2\times 2$ matrix:

\begin{eqnarray}
\det \left( M_{j}^{i}\right)
&=&M_{1}^{1}M_{2}^{2}-M_{2}^{1}M_{1}^{2}=\sum_{\mu _{1},\mu
_{2}=1}^{2}\left( \delta _{1}^{\mu _{1}}\delta _{2}^{\mu _{2}}-\delta
_{2}^{\mu _{1}}\delta _{1}^{\mu _{2}}\right) M_{\mu _{1}}^{1}M_{\mu
_{2}}^{2}=\frac{1}{2}\sum_{\mu _{1},\mu _{2},\nu _{1},\nu _{2}=1}^{2}\left(
\delta _{\nu _{1}}^{\mu _{1}}\delta _{\nu _{2}}^{\mu _{2}}-\delta _{\nu
_{2}}^{\mu _{1}}\delta _{\nu _{1}}^{\mu _{2}}\right) M_{\mu _{1}}^{\nu
_{1}}M_{\mu _{2}}^{\nu _{2}}  \notag \\
&=&\frac{1}{2}\sum_{\mu _{1},\mu _{2},\nu _{1},\nu _{2}=1}^{2}\varepsilon
_{\nu _{1}\nu _{2}}\varepsilon ^{\mu _{1}\mu _{2}}M_{\mu _{1}}^{\nu
_{1}}M_{\mu _{2}}^{\nu _{2}},\hspace{1.5cm}\mu _{1},\mu _{2},\nu _{1},\nu
_{2}=1,2
\end{eqnarray}

Now considering the case of a $3\times 3$ matrix:

\begin{equation}
M=\left( 
\begin{array}{ccc}
M_{11} & M_{12} & M_{13} \\ 
M_{21} & M_{22} & M_{23} \\ 
M_{31} & M_{32} & M_{33}%
\end{array}%
\right)
\end{equation}

It follows that its determinant can be written as follows:

\begin{eqnarray}
\det \left( M_{j}^{i}\right) &=&M_{1}^{1}\left(
M_{2}^{2}M_{3}^{3}-M_{3}^{2}M_{2}^{3}\right) -M_{2}^{1}\left(
M_{1}^{2}M_{3}^{3}-M_{3}^{2}M_{1}^{3}\right) +M_{3}^{1}\left(
M_{1}^{2}M_{2}^{3}-M_{2}^{2}M_{1}^{3}\right)  \notag \\
&=&\frac{1}{3!}\delta
_{j_{1}j_{2}j_{3}}^{i_{1}i_{2}i_{3}}M_{i_{1}}^{j_{1}}M_{i_{2}}^{j_{2}}M_{i_{3}}^{j_{3}}=%
\frac{1}{3!}\varepsilon _{j_{1}j_{2}j_{3}}\varepsilon
^{i_{1}i_{2}i_{3}}M_{i_{1}}^{j_{1}}M_{i_{2}}^{j_{2}}M_{i_{3}}^{j_{3}}
\end{eqnarray}

where:

\begin{equation}
M_{j}^{i}=M_{ij}
\end{equation}

Then, coming back to the case of the matrix $M_{j}^{i}$ arising from a
seesaw mechanism involving two seesaw mediators and given in Eq. {(\ref{Mij}}%
), it follows that:

\begin{equation}
\det \left( M_{j}^{i}\right) =\frac{1}{3!}\delta
_{j_{1}j_{2}j_{3}}^{i_{1}i_{2}i_{3}}M_{j}^{i}=\frac{1}{3!}\varepsilon
_{j_{1}j_{2}j_{3}}\varepsilon ^{i_{1}i_{2}i_{3}}\left(
F^{j_{1}}G_{i_{1}}+X^{j_{1}}Y_{i_{1}}\right) \left(
F^{j_{2}}G_{i_{2}}+X^{j_{2}}Y_{i_{2}}\right) \left(
F^{j_{3}}G_{i_{3}}+X^{j_{3}}Y_{i_{3}}\right) =0  \label{detM}
\end{equation}

which is due to the fact that every term in Eq. {(\ref{detM}}) involves the
contraction of symmetric and antisymmetric tensors which always yields a
vanishing result. We see that, as a result of the linear dependance of the
rows and columns of the matrix $M_{j}^{i}$ of Eq. {(\ref{Mij}}), the
existence of one vanishing eigenvalue.

Finally, let's consider the case of three fermionic seesaw mediators. Then,
the the $\left( i,j\right) $ element of the low energy fermionic mass matrix
arising from the seesaw mechanism has the form:

\begin{equation}
M_{ij}=M_{j}^{i}=F_{i}G_{j}+X_{i}Y_{j}+R_{i}S_{j}=F^{i}G_{j}+X^{i}Y_{j}+R^{i}S_{j}
\end{equation}

Then, it follows that:

\begin{equation}
\det \left( M_{j}^{i}\right) =\varepsilon _{j_{1}j_{2}j_{3}}\varepsilon
^{i_{1}i_{2}i_{3}}F^{j_{1}}G_{i_{1}}X^{j_{2}}Y_{i_{2}}R^{j_{3}}S_{i_{3}}\neq
0
\end{equation}

provided that:

\begin{equation}
G_{i_{1}}\neq Y_{i_{2}}\neq S_{i_{3}},\hspace{1.5cm}F^{j_{1}}\neq
X^{j_{2}}\neq R^{j_{3}}
\end{equation}

Therefore, we have shown that in order to generate the masses of three
fermion families via a seesaw mechanism, there should be at least three
fermionic seesaw mediators. Furthermore, the number of the massless states
obtained in a mass matrix resulting from a seesaw mechanism is $3-n$, where $%
n$ is the number of fermionic seesaw mediators.\newpage

\centerline{\bf{REFERENCES}}\vspace{-0.4cm} 
\bibliographystyle{utphys}
\bibliography{BiblioLR6thFebruary2022}

\providecommand{\href}[2]{#2}\begingroup\raggedright\begin{thebibliography}{10}

\bibitem{Pati:1974yy}
J.~C. Pati and A.~Salam, ``{Lepton Number as the Fourth Color},''
  \href{http://dx.doi.org/10.1103/PhysRevD.10.275,
  10.1103/PhysRevD.11.703.2}{{\em Phys. Rev.} {\bfseries D10} (1974) 275--289}.
[Erratum: Phys. Rev.D11,703(1975)].

\bibitem{Mohapatra:1974gc}
R.~N. Mohapatra and J.~C. Pati, ``{A Natural Left-Right Symmetry},''
\href{http://dx.doi.org/10.1103/PhysRevD.11.2558}{{\em Phys. Rev.} {\bfseries
  D11} (1975) 2558}.

\bibitem{Davidson:1987mh}
A.~Davidson and K.~C. Wali, ``{Universal Seesaw Mechanism?},''
\href{http://dx.doi.org/10.1103/PhysRevLett.59.393}{{\em Phys. Rev. Lett.}
  {\bfseries 59} (1987) 393}.

\bibitem{Davidson:1987mi}
A.~Davidson and K.~C. Wali, ``{SU(5)-L x SU(5)-R HYBRID UNIFICATION},''
\href{http://dx.doi.org/10.1103/PhysRevLett.58.2623}{{\em Phys. Rev. Lett.}
  {\bfseries 58} (1987) 2623}.

\bibitem{CarcamoHernandez:2018hst}
A.~E. Cárcamo~Hernández, S.~Kovalenko, J.~W.~F. Valle, and C.~A.
  Vaquera-Araujo, ``{Neutrino predictions from a left-right symmetric flavored
  extension of the standard model},''
  \href{http://dx.doi.org/10.1007/JHEP02(2019)065}{{\em JHEP} {\bfseries 02}
  (2019) 065},
\href{http://arxiv.org/abs/1811.03018}{{\ttfamily arXiv:1811.03018 [hep-ph]}}.

\bibitem{Dekens:2014ina}
W.~Dekens and D.~Boer, ``{Viability of minimal left–right models with
  discrete symmetries},''
  \href{http://dx.doi.org/10.1016/j.nuclphysb.2014.10.025}{{\em Nucl. Phys.}
  {\bfseries B889} (2014) 727--756},
\href{http://arxiv.org/abs/1409.4052}{{\ttfamily arXiv:1409.4052 [hep-ph]}}.

\bibitem{Nomura:2016run}
T.~Nomura, H.~Okada, and Y.~Orikasa, ``{Radiative neutrino mass in alternative
  left–right model},''
  \href{http://dx.doi.org/10.1140/epjc/s10052-017-4657-4}{{\em Eur. Phys. J.}
  {\bfseries C77} no.~2, (2017) 103},
\href{http://arxiv.org/abs/1602.08302}{{\ttfamily arXiv:1602.08302 [hep-ph]}}.

\bibitem{Brdar:2018sbk}
V.~Brdar and A.~Y. Smirnov, ``{Low Scale Left-Right Symmetry and Naturally
  Small Neutrino Mass},'' \href{http://dx.doi.org/10.1007/JHEP02(2019)045}{{\em
  JHEP} {\bfseries 02} (2019) 045},
\href{http://arxiv.org/abs/1809.09115}{{\ttfamily arXiv:1809.09115 [hep-ph]}}.

\bibitem{Ma:2020lnm}
E.~Ma, ``{Universal Scotogenic Fermion Masses in Left-Right Gauge Model},''
  \href{http://dx.doi.org/10.1016/j.nuclphysb.2021.115406}{{\em Nucl. Phys.}
  {\bfseries B967} (2021) 115406},
\href{http://arxiv.org/abs/2012.03128}{{\ttfamily arXiv:2012.03128 [hep-ph]}}.

\bibitem{Babu:2020bgz}
K.~S. Babu and A.~Thapa, ``{Left-Right Symmetric Model without Higgs
  Triplets},''
\href{http://arxiv.org/abs/2012.13420}{{\ttfamily arXiv:2012.13420 [hep-ph]}}.

\bibitem{Escudero:2016gzx}
M.~Escudero, A.~Berlin, D.~Hooper, and M.-X. Lin, ``{Toward (Finally!) Ruling
  Out Z and Higgs Mediated Dark Matter Models},''
  \href{http://dx.doi.org/10.1088/1475-7516/2016/12/029}{{\em JCAP} {\bfseries
  1612} (2016) 029},
\href{http://arxiv.org/abs/1609.09079}{{\ttfamily arXiv:1609.09079 [hep-ph]}}.

\bibitem{Bernal:2017xat}
N.~Bernal, A.~E. Cárcamo~Hernández, I.~de~Medeiros~Varzielas, and
  S.~Kovalenko, ``{Fermion masses and mixings and dark matter constraints in a
  model with radiative seesaw mechanism},''
  \href{http://dx.doi.org/10.1007/JHEP05(2018)053}{{\em JHEP} {\bfseries 05}
  (2018) 053},
\href{http://arxiv.org/abs/1712.02792}{{\ttfamily arXiv:1712.02792 [hep-ph]}}.

\bibitem{CarcamoHernandez:2020ehn}
A.~E. Cárcamo~Hernández, J.~W.~F. Valle, and C.~A. Vaquera-Araujo, ``{Simple
  theory for scotogenic dark matter with residual matter-parity},''
  \href{http://dx.doi.org/10.1016/j.physletb.2020.135757}{{\em Phys. Lett.}
  {\bfseries B809} (2020) 135757},
\href{http://arxiv.org/abs/2006.06009}{{\ttfamily arXiv:2006.06009 [hep-ph]}}.

\bibitem{Han:2019lux}
Z.-L. Han and W.~Wang, ``{Predictive Scotogenic Model with Flavor Dependent
  Symmetry},'' \href{http://dx.doi.org/10.1140/epjc/s10052-019-7033-8}{{\em
  Eur. Phys. J.} {\bfseries C79} no.~6, (2019) 522},
\href{http://arxiv.org/abs/1901.07798}{{\ttfamily arXiv:1901.07798 [hep-ph]}}.

\bibitem{Cabrera:2020lmg}
M.~E. Cabrera, J.~A. Casas, A.~Delgado, and S.~Robles, ``{2HDM singlet portal
  to dark matter},'' \href{http://dx.doi.org/10.1007/JHEP01(2021)123}{{\em
  JHEP} {\bfseries 01} (2021) 123},
\href{http://arxiv.org/abs/2011.09101}{{\ttfamily arXiv:2011.09101 [hep-ph]}}.

\bibitem{CarcamoHernandez:2021iat}
A.~E. C\'arcamo~Hern\'andez, C.~Espinoza, J.~Carlos G\'omez-Izquierdo, and
  M.~Mondrag\'on, ``{Fermion masses and mixings, dark matter, leptogenesis and
  $g-2$ muon anomaly in an extended 2HDM with inverse seesaw},''
  \href{http://arxiv.org/abs/2104.02730}{{\ttfamily arXiv:2104.02730
  [hep-ph]}}.

\bibitem{Abada:2021yot}
A.~Abada, N.~Bernal, A.~E.~C. Hern\'andez, X.~Marcano, and G.~Piazza, ``{Gauged
  inverse seesaw from dark matter},''
  \href{http://dx.doi.org/10.1140/epjc/s10052-021-09535-5}{{\em Eur. Phys. J.
  C} {\bfseries 81} no.~8, (2021) 758},
  \href{http://arxiv.org/abs/2107.02803}{{\ttfamily arXiv:2107.02803
  [hep-ph]}}.

\bibitem{Xing:2020ijf}
Z.-z. Xing, ``{Flavor structures of charged fermions and massive neutrinos},''
  \href{http://dx.doi.org/10.1016/j.physrep.2020.02.001}{{\em Phys. Rept.}
  {\bfseries 854} (2020) 1--147},
  \href{http://arxiv.org/abs/1909.09610}{{\ttfamily arXiv:1909.09610
  [hep-ph]}}.

\bibitem{Zyla:2020zbs}
{\bfseries Particle Data Group} Collaboration, P.~Zyla {\em et~al.}, ``{Review
  of Particle Physics},'' \href{http://dx.doi.org/10.1093/ptep/ptaa104}{{\em
  PTEP} {\bfseries 2020} no.~8, (2020) 083C01}.

\bibitem{Pilaftsis:1991ug}
A.~Pilaftsis, ``{Radiatively induced neutrino masses and large Higgs neutrino
  couplings in the standard model with Majorana fields},''
  \href{http://dx.doi.org/10.1007/BF01482590}{{\em Z. Phys. C} {\bfseries 55}
  (1992) 275--282}, \href{http://arxiv.org/abs/hep-ph/9901206}{{\ttfamily
  arXiv:hep-ph/9901206}}.

\bibitem{Catano:2012kw}
M.~E. Catano, R.~Martinez, and F.~Ochoa, ``{Neutrino masses in a 331 model with
  right-handed neutrinos without doubly charged Higgs bosons via inverse and
  double seesaw mechanisms},''
  \href{http://dx.doi.org/10.1103/PhysRevD.86.073015}{{\em Phys. Rev.}
  {\bfseries D86} (2012) 073015},
\href{http://arxiv.org/abs/1206.1966}{{\ttfamily arXiv:1206.1966 [hep-ph]}}.

\bibitem{Abada:2018nio}
A.~Abada and A.~M. Teixeira, ``{Heavy neutral leptons and high-intensity
  observables},'' \href{http://dx.doi.org/10.3389/fphy.2018.00142}{{\em Front.
  in Phys.} {\bfseries 6} (2018) 142},
  \href{http://arxiv.org/abs/1812.08062}{{\ttfamily arXiv:1812.08062
  [hep-ph]}}.

\bibitem{Fernandez-Martinez:2016lgt}
E.~Fernandez-Martinez, J.~Hernandez-Garcia, and J.~Lopez-Pavon, ``{Global
  constraints on heavy neutrino mixing},''
  \href{http://dx.doi.org/10.1007/JHEP08(2016)033}{{\em JHEP} {\bfseries 08}
  (2016) 033}, \href{http://arxiv.org/abs/1605.08774}{{\ttfamily
  arXiv:1605.08774 [hep-ph]}}.

\bibitem{AguilarSaavedra:2012fu}
J.~A. Aguilar-Saavedra, F.~Deppisch, O.~Kittel, and J.~W.~F. Valle, ``{Flavour
  in heavy neutrino searches at the LHC},''
  \href{http://dx.doi.org/10.1103/PhysRevD.85.091301}{{\em Phys. Rev.}
  {\bfseries D85} (2012) 091301},
\href{http://arxiv.org/abs/1203.5998}{{\ttfamily arXiv:1203.5998 [hep-ph]}}.

\bibitem{Das:2012ii}
S.~P. Das, F.~F. Deppisch, O.~Kittel, and J.~W.~F. Valle, ``{Heavy Neutrinos
  and Lepton Flavour Violation in Left-Right Symmetric Models at the LHC},''
  \href{http://dx.doi.org/10.1103/PhysRevD.86.055006}{{\em Phys. Rev.}
  {\bfseries D86} (2012) 055006},
\href{http://arxiv.org/abs/1206.0256}{{\ttfamily arXiv:1206.0256 [hep-ph]}}.

\bibitem{Dev:2009aw}
P.~S.~B. Dev and R.~N. Mohapatra, ``{TeV Scale Inverse Seesaw in SO(10) and
  Leptonic Non-Unitarity Effects},''
  \href{http://dx.doi.org/10.1103/PhysRevD.81.013001}{{\em Phys. Rev.}
  {\bfseries D81} (2010) 013001},
\href{http://arxiv.org/abs/0910.3924}{{\ttfamily arXiv:0910.3924 [hep-ph]}}.

\bibitem{BhupalDev:2012zg}
P.~S. Bhupal~Dev, R.~Franceschini, and R.~N. Mohapatra, ``{Bounds on TeV Seesaw
  Models from LHC Higgs Data},''
  \href{http://dx.doi.org/10.1103/PhysRevD.86.093010}{{\em Phys. Rev.}
  {\bfseries D86} (2012) 093010},
\href{http://arxiv.org/abs/1207.2756}{{\ttfamily arXiv:1207.2756 [hep-ph]}}.

\bibitem{Das:2012ze}
A.~Das and N.~Okada, ``{Inverse seesaw neutrino signatures at the LHC and
  ILC},'' \href{http://dx.doi.org/10.1103/PhysRevD.88.113001}{{\em Phys. Rev.}
  {\bfseries D88} (2013) 113001},
\href{http://arxiv.org/abs/1207.3734}{{\ttfamily arXiv:1207.3734 [hep-ph]}}.

\bibitem{Dev:2013oxa}
C.-H. Lee, P.~S. Bhupal~Dev, and R.~N. Mohapatra, ``{Natural TeV-scale
  left-right seesaw mechanism for neutrinos and experimental tests},''
  \href{http://dx.doi.org/10.1103/PhysRevD.88.093010}{{\em Phys. Rev.}
  {\bfseries D88} no.~9, (2013) 093010},
\href{http://arxiv.org/abs/1309.0774}{{\ttfamily arXiv:1309.0774 [hep-ph]}}.

\bibitem{Das:2014jxa}
A.~Das, P.~S. Bhupal~Dev, and N.~Okada, ``{Direct bounds on electroweak scale
  pseudo-Dirac neutrinos from $\sqrt s=8$ TeV LHC data},''
  \href{http://dx.doi.org/10.1016/j.physletb.2014.06.058}{{\em Phys. Lett.}
  {\bfseries B735} (2014) 364--370},
\href{http://arxiv.org/abs/1405.0177}{{\ttfamily arXiv:1405.0177 [hep-ph]}}.

\bibitem{Das:2016hof}
A.~Das, P.~Konar, and S.~Majhi, ``{Production of Heavy neutrino in
  next-to-leading order QCD at the LHC and beyond},''
  \href{http://dx.doi.org/10.1007/JHEP06(2016)019}{{\em JHEP} {\bfseries 06}
  (2016) 019},
\href{http://arxiv.org/abs/1604.00608}{{\ttfamily arXiv:1604.00608 [hep-ph]}}.

\bibitem{Das:2017gke}
A.~Das, P.~Konar, and A.~Thalapillil, ``{Jet substructure shedding light on
  heavy Majorana neutrinos at the LHC},''
  \href{http://dx.doi.org/10.1007/JHEP02(2018)083}{{\em JHEP} {\bfseries 02}
  (2018) 083},
\href{http://arxiv.org/abs/1709.09712}{{\ttfamily arXiv:1709.09712 [hep-ph]}}.

\bibitem{Das:2017nvm}
A.~Das and N.~Okada, ``{Bounds on heavy Majorana neutrinos in type-I seesaw and
  implications for collider searches},''
  \href{http://dx.doi.org/10.1016/j.physletb.2017.09.042}{{\em Phys. Lett.}
  {\bfseries B774} (2017) 32--40},
\href{http://arxiv.org/abs/1702.04668}{{\ttfamily arXiv:1702.04668 [hep-ph]}}.

\bibitem{Das:2017zjc}
A.~Das, P.~S.~B. Dev, and C.~S. Kim, ``{Constraining Sterile Neutrinos from
  Precision Higgs Data},''
  \href{http://dx.doi.org/10.1103/PhysRevD.95.115013}{{\em Phys. Rev.}
  {\bfseries D95} no.~11, (2017) 115013},
\href{http://arxiv.org/abs/1704.00880}{{\ttfamily arXiv:1704.00880 [hep-ph]}}.

\bibitem{Das:2017rsu}
A.~Das, Y.~Gao, and T.~Kamon, ``{Heavy neutrino search via semileptonic Higgs
  decay at the LHC},''
  \href{http://dx.doi.org/10.1140/epjc/s10052-019-6937-7}{{\em Eur. Phys. J.}
  {\bfseries C79} no.~5, (2019) 424},
\href{http://arxiv.org/abs/1704.00881}{{\ttfamily arXiv:1704.00881 [hep-ph]}}.

\bibitem{Das:2018usr}
A.~Das, S.~Jana, S.~Mandal, and S.~Nandi, ``{Probing right handed neutrinos at
  the LHeC and lepton colliders using fat jet signatures},''
  \href{http://dx.doi.org/10.1103/PhysRevD.99.055030}{{\em Phys. Rev.}
  {\bfseries D99} no.~5, (2019) 055030},
\href{http://arxiv.org/abs/1811.04291}{{\ttfamily arXiv:1811.04291 [hep-ph]}}.

\bibitem{Das:2018hph}
A.~Das, ``{Searching for the minimal Seesaw models at the LHC and beyond},''
  \href{http://dx.doi.org/10.1155/2018/9785318}{{\em Adv. High Energy Phys.}
  {\bfseries 2018} (2018) 9785318},
\href{http://arxiv.org/abs/1803.10940}{{\ttfamily arXiv:1803.10940 [hep-ph]}}.

\bibitem{Bhardwaj:2018lma}
A.~Bhardwaj, A.~Das, P.~Konar, and A.~Thalapillil, ``{Looking for Minimal
  Inverse Seesaw scenarios at the LHC with Jet Substructure Techniques},''
  \href{http://dx.doi.org/10.1088/1361-6471/ab7769}{{\em J. Phys.} {\bfseries
  G47} no.~7, (2020) 075002},
\href{http://arxiv.org/abs/1801.00797}{{\ttfamily arXiv:1801.00797 [hep-ph]}}.

\bibitem{Helo:2018rll}
J.~C. Helo, H.~Li, N.~A. Neill, M.~Ramsey-Musolf, and J.~C. Vasquez, ``{Probing
  neutrino Dirac mass in left-right symmetric models at the LHC and next
  generation colliders},''
  \href{http://dx.doi.org/10.1103/PhysRevD.99.055042}{{\em Phys. Rev.}
  {\bfseries D99} no.~5, (2019) 055042},
\href{http://arxiv.org/abs/1812.01630}{{\ttfamily arXiv:1812.01630 [hep-ph]}}.

\bibitem{Pascoli:2018heg}
S.~Pascoli, R.~Ruiz, and C.~Weiland, ``{Heavy neutrinos with dynamic jet
  vetoes: multilepton searches at $ \sqrt{s}=14 $ , 27, and 100 TeV},''
  \href{http://dx.doi.org/10.1007/JHEP06(2019)049}{{\em JHEP} {\bfseries 06}
  (2019) 049},
\href{http://arxiv.org/abs/1812.08750}{{\ttfamily arXiv:1812.08750 [hep-ph]}}.

\bibitem{Ilakovac:1994kj}
A.~Ilakovac and A.~Pilaftsis, ``{Flavor violating charged lepton decays in
  seesaw-type models},''
  \href{http://dx.doi.org/10.1016/0550-3213(94)00567-X}{{\em Nucl. Phys. B}
  {\bfseries 437} (1995) 491},
  \href{http://arxiv.org/abs/hep-ph/9403398}{{\ttfamily arXiv:hep-ph/9403398}}.

\bibitem{Deppisch:2004fa}
F.~Deppisch and J.~W.~F. Valle, ``{Enhanced lepton flavor violation in the
  supersymmetric inverse seesaw model},''
  \href{http://dx.doi.org/10.1103/PhysRevD.72.036001}{{\em Phys. Rev. D}
  {\bfseries 72} (2005) 036001},
  \href{http://arxiv.org/abs/hep-ph/0406040}{{\ttfamily arXiv:hep-ph/0406040}}.

\bibitem{Lindner:2016bgg}
M.~Lindner, M.~Platscher, and F.~S. Queiroz, ``{A Call for New Physics : The
  Muon Anomalous Magnetic Moment and Lepton Flavor Violation},''
  \href{http://dx.doi.org/10.1016/j.physrep.2017.12.001}{{\em Phys. Rept.}
  {\bfseries 731} (2018) 1--82},
  \href{http://arxiv.org/abs/1610.06587}{{\ttfamily arXiv:1610.06587
  [hep-ph]}}.

\bibitem{Deppisch:2013cya}
F.~F. Deppisch, N.~Desai, and J.~W.~F. Valle, ``{Is charged lepton flavor
  violation a high energy phenomenon?},''
  \href{http://dx.doi.org/10.1103/PhysRevD.89.051302}{{\em Phys. Rev. D}
  {\bfseries 89} no.~5, (2014) 051302},
  \href{http://arxiv.org/abs/1308.6789}{{\ttfamily arXiv:1308.6789 [hep-ph]}}.

\bibitem{Kuno:1999jp}
Y.~Kuno and Y.~Okada, ``{Muon decay and physics beyond the standard model},''
  \href{http://dx.doi.org/10.1103/RevModPhys.73.151}{{\em Rev. Mod. Phys.}
  {\bfseries 73} (2001) 151--202},
  \href{http://arxiv.org/abs/hep-ph/9909265}{{\ttfamily arXiv:hep-ph/9909265}}.

\bibitem{Blanchet:2009kk}
S.~Blanchet, T.~Hambye, and F.-X. Josse-Michaux, ``{Reconciling leptogenesis
  with observable mu ---> e gamma rates},''
  \href{http://dx.doi.org/10.1007/JHEP04(2010)023}{{\em JHEP} {\bfseries 04}
  (2010) 023},
\href{http://arxiv.org/abs/0912.3153}{{\ttfamily arXiv:0912.3153 [hep-ph]}}.

\bibitem{Buchmuller:2004nz}
W.~Buchmuller, P.~Di~Bari, and M.~Plumacher, ``{Leptogenesis for
  pedestrians},'' \href{http://dx.doi.org/10.1016/j.aop.2004.02.003}{{\em
  Annals Phys.} {\bfseries 315} (2005) 305--351},
  \href{http://arxiv.org/abs/hep-ph/0401240}{{\ttfamily arXiv:hep-ph/0401240}}.

\bibitem{Covi:1996wh}
L.~Covi, E.~Roulet, and F.~Vissani, ``{CP violating decays in leptogenesis
  scenarios},'' \href{http://dx.doi.org/10.1016/0370-2693(96)00817-9}{{\em
  Phys. Lett. B} {\bfseries 384} (1996) 169--174},
  \href{http://arxiv.org/abs/hep-ph/9605319}{{\ttfamily arXiv:hep-ph/9605319}}.

\bibitem{Rangarajan:1999kt}
R.~Rangarajan and H.~Mishra, ``{Leptogenesis with heavy Majorana neutrinos
  revisited},'' \href{http://dx.doi.org/10.1103/PhysRevD.61.043509}{{\em Phys.
  Rev. D} {\bfseries 61} (2000) 043509},
  \href{http://arxiv.org/abs/hep-ph/9908417}{{\ttfamily arXiv:hep-ph/9908417}}.

\bibitem{Gu:2010xc}
P.-H. Gu and U.~Sarkar, ``{Leptogenesis with Linear, Inverse or Double
  Seesaw},'' \href{http://dx.doi.org/10.1016/j.physletb.2010.09.062}{{\em Phys.
  Lett.} {\bfseries B694} (2011) 226--232},
\href{http://arxiv.org/abs/1007.2323}{{\ttfamily arXiv:1007.2323 [hep-ph]}}.

\bibitem{Pilaftsis:1997jf}
A.~Pilaftsis, ``{CP violation and baryogenesis due to heavy Majorana
  neutrinos},'' \href{http://dx.doi.org/10.1103/PhysRevD.56.5431}{{\em Phys.
  Rev.} {\bfseries D56} (1997) 5431--5451},
\href{http://arxiv.org/abs/hep-ph/9707235}{{\ttfamily arXiv:hep-ph/9707235
  [hep-ph]}}.

\bibitem{Plumacher:1996kc}
M.~Plumacher, ``{Baryogenesis and lepton number violation},''
  \href{http://dx.doi.org/10.1007/s002880050418}{{\em Z. Phys. C} {\bfseries
  74} (1997) 549--559}, \href{http://arxiv.org/abs/hep-ph/9604229}{{\ttfamily
  arXiv:hep-ph/9604229}}.

\bibitem{Cosme:2004xs}
N.~Cosme, ``{Leptogenesis, neutrino masses and gauge unification},''
  \href{http://dx.doi.org/10.1088/1126-6708/2004/08/027}{{\em JHEP} {\bfseries
  08} (2004) 027}, \href{http://arxiv.org/abs/hep-ph/0403209}{{\ttfamily
  arXiv:hep-ph/0403209}}.

\bibitem{Frere:2008ct}
J.-M. Frere, T.~Hambye, and G.~Vertongen, ``{Is leptogenesis falsifiable at
  LHC?},'' \href{http://dx.doi.org/10.1088/1126-6708/2009/01/051}{{\em JHEP}
  {\bfseries 01} (2009) 051}, \href{http://arxiv.org/abs/0806.0841}{{\ttfamily
  arXiv:0806.0841 [hep-ph]}}.

\bibitem{Blanchet:2010kw}
S.~Blanchet, P.~S.~B. Dev, and R.~N. Mohapatra, ``{Leptogenesis with TeV Scale
  Inverse Seesaw in SO(10)},''
  \href{http://dx.doi.org/10.1103/PhysRevD.82.115025}{{\em Phys. Rev. D}
  {\bfseries 82} (2010) 115025},
  \href{http://arxiv.org/abs/1010.1471}{{\ttfamily arXiv:1010.1471 [hep-ph]}}.

\bibitem{Dolan:2018qpy}
M.~J. Dolan, T.~P. Dutka, and R.~R. Volkas, ``{Dirac-Phase Thermal Leptogenesis
  in the extended Type-I Seesaw Model},''
  \href{http://dx.doi.org/10.1088/1475-7516/2018/06/012}{{\em JCAP} {\bfseries
  1806} (2018) 012},
\href{http://arxiv.org/abs/1802.08373}{{\ttfamily arXiv:1802.08373 [hep-ph]}}.

\bibitem{Luty:1992un}
M.~A. Luty, ``{Baryogenesis via leptogenesis},''
  \href{http://dx.doi.org/10.1103/PhysRevD.45.455}{{\em Phys. Rev. D}
  {\bfseries 45} (1992) 455--465}.

\bibitem{Grimus:2000vj}
W.~Grimus and L.~Lavoura, ``{The Seesaw mechanism at arbitrary order:
  Disentangling the small scale from the large scale},''
  \href{http://dx.doi.org/10.1088/1126-6708/2000/11/042}{{\em JHEP} {\bfseries
  11} (2000) 042},
\href{http://arxiv.org/abs/hep-ph/0008179}{{\ttfamily arXiv:hep-ph/0008179
  [hep-ph]}}.

\bibitem{Sirunyan:2018ouh}
{\bfseries CMS} Collaboration, A.~M. Sirunyan {\em et~al.}, ``{Measurements of
  Higgs boson properties in the diphoton decay channel in proton-proton
  collisions at $\sqrt{s} =$ 13 TeV},''
  \href{http://dx.doi.org/10.1007/JHEP11(2018)185}{{\em JHEP} {\bfseries 11}
  (2018) 185},
\href{http://arxiv.org/abs/1804.02716}{{\ttfamily arXiv:1804.02716 [hep-ex]}}.

\bibitem{Aad:2019mbh}
{\bfseries ATLAS} Collaboration, G.~Aad {\em et~al.}, ``{Combined measurements
  of Higgs boson production and decay using up to $80$ fb$^{-1}$ of
  proton-proton collision data at $\sqrt{s}=$ 13 TeV collected with the ATLAS
  experiment},'' \href{http://dx.doi.org/10.1103/PhysRevD.101.012002}{{\em
  Phys. Rev.} {\bfseries D101} no.~1, (2020) 012002},
\href{http://arxiv.org/abs/1909.02845}{{\ttfamily arXiv:1909.02845 [hep-ex]}}.

\bibitem{Diaz:2002uk}
R.~A. Diaz, R.~Martinez, and J.~A. Rodriguez, ``{Phenomenology of lepton flavor
  violation in 2HDM(3) from (g-2)(mu) and leptonic decays},''
  \href{http://dx.doi.org/10.1103/PhysRevD.67.075011}{{\em Phys. Rev.}
  {\bfseries D67} (2003) 075011},
\href{http://arxiv.org/abs/hep-ph/0208117}{{\ttfamily arXiv:hep-ph/0208117
  [hep-ph]}}.

\bibitem{Jegerlehner:2009ry}
F.~Jegerlehner and A.~Nyffeler, ``{The Muon g-2},''
  \href{http://dx.doi.org/10.1016/j.physrep.2009.04.003}{{\em Phys. Rept.}
  {\bfseries 477} (2009) 1--110},
\href{http://arxiv.org/abs/0902.3360}{{\ttfamily arXiv:0902.3360 [hep-ph]}}.

\bibitem{Kelso:2014qka}
C.~Kelso, H.~N. Long, R.~Martinez, and F.~S. Queiroz, ``{Connection of
  $g-2_{\mu}$, electroweak, dark matter, and collider constraints on 331
  models},'' \href{http://dx.doi.org/10.1103/PhysRevD.90.113011}{{\em Phys.
  Rev.} {\bfseries D90} no.~11, (2014) 113011},
\href{http://arxiv.org/abs/1408.6203}{{\ttfamily arXiv:1408.6203 [hep-ph]}}.

\bibitem{Kowalska:2017iqv}
K.~Kowalska and E.~M. Sessolo, ``{Expectations for the muon g-2 in simplified
  models with dark matter},''
  \href{http://dx.doi.org/10.1007/JHEP09(2017)112}{{\em JHEP} {\bfseries 09}
  (2017) 112},
\href{http://arxiv.org/abs/1707.00753}{{\ttfamily arXiv:1707.00753 [hep-ph]}}.

\bibitem{Abi:2021gix}
{\bfseries Muon g-2} Collaboration, B.~Abi {\em et~al.}, ``{Measurement of the
  Positive Muon Anomalous Magnetic Moment to 0.46 ppm},''
  \href{http://dx.doi.org/10.1103/PhysRevLett.126.141801}{{\em Phys. Rev.
  Lett.} {\bfseries 126} no.~14, (2021) 141801},
\href{http://arxiv.org/abs/2104.03281}{{\ttfamily arXiv:2104.03281 [hep-ex]}}.

\bibitem{Morel:2020dww}
L.~Morel, Z.~Yao, P.~Cladé, and S.~Guellati-Khélifa, ``{Determination of the
  fine-structure constant with an accuracy of 81 parts per trillion},''
\href{http://dx.doi.org/10.1038/s41586-020-2964-7}{{\em Nature} {\bfseries 588}
  no.~7836, (2020) 61--65}.

\bibitem{LEP:2004xhf}
{\bfseries LEP, ALEPH, DELPHI, L3, LEP Electroweak Working Group, SLD
  Electroweak Group, SLD Heavy Flavour Group, OPAL} Collaboration, ``{A
  Combination of preliminary electroweak measurements and constraints on the
  standard model},'' \href{http://arxiv.org/abs/hep-ex/0412015}{{\ttfamily
  arXiv:hep-ex/0412015}}.

\bibitem{Carena:2004xs}
M.~Carena, A.~Daleo, B.~A. Dobrescu, and T.~M.~P. Tait, ``{$Z^\prime$ gauge
  bosons at the Tevatron},''
  \href{http://dx.doi.org/10.1103/PhysRevD.70.093009}{{\em Phys. Rev. D}
  {\bfseries 70} (2004) 093009},
  \href{http://arxiv.org/abs/hep-ph/0408098}{{\ttfamily arXiv:hep-ph/0408098}}.

\bibitem{Das:2021esm}
A.~Das, P.~S.~B. Dev, Y.~Hosotani, and S.~Mandal, ``{Probing the minimal
  $U(1)_X$ model at future electron-positron colliders via the fermion
  pair-production channel},'' \href{http://arxiv.org/abs/2104.10902}{{\ttfamily
  arXiv:2104.10902 [hep-ph]}}.

\bibitem{ATLAS:2019erb}
{\bfseries ATLAS} Collaboration, G.~Aad {\em et~al.}, ``{Search for high-mass
  dilepton resonances using 139 fb$^{-1}$ of $pp$ collision data collected at
  $\sqrt{s}=$13 TeV with the ATLAS detector},''
  \href{http://dx.doi.org/10.1016/j.physletb.2019.07.016}{{\em Phys. Lett. B}
  {\bfseries 796} (2019) 68--87},
  \href{http://arxiv.org/abs/1903.06248}{{\ttfamily arXiv:1903.06248
  [hep-ex]}}.

\bibitem{CMS:2021ctt}
{\bfseries CMS} Collaboration, A.~M. Sirunyan {\em et~al.}, ``{Search for
  resonant and nonresonant new phenomena in high-mass dilepton final states at
  $ \sqrt{s} $ = 13 TeV},''
  \href{http://dx.doi.org/10.1007/JHEP07(2021)208}{{\em JHEP} {\bfseries 07}
  (2021) 208}, \href{http://arxiv.org/abs/2103.02708}{{\ttfamily
  arXiv:2103.02708 [hep-ex]}}.

\bibitem{CMS:2021qef}
{\bfseries CMS} Collaboration, ``{Search for a right-handed W boson and heavy
  neutrino in proton-proton collisions at $\sqrt{s}=13$ TeV},''.

\bibitem{ATLAS:2018dcj}
{\bfseries ATLAS} Collaboration, M.~Aaboud {\em et~al.}, ``{Search for heavy
  Majorana or Dirac neutrinos and right-handed $W$ gauge bosons in final states
  with two charged leptons and two jets at $ \sqrt{s}=13 $ TeV with the ATLAS
  detector},'' \href{http://dx.doi.org/10.1007/JHEP01(2019)016}{{\em JHEP}
  {\bfseries 01} (2019) 016}, \href{http://arxiv.org/abs/1809.11105}{{\ttfamily
  arXiv:1809.11105 [hep-ex]}}.

\bibitem{Dedes:2002er}
A.~Dedes and A.~Pilaftsis, ``{Resummed Effective Lagrangian for Higgs Mediated
  FCNC Interactions in the CP Violating MSSM},''
  \href{http://dx.doi.org/10.1103/PhysRevD.67.015012}{{\em Phys. Rev. D}
  {\bfseries 67} (2003) 015012},
  \href{http://arxiv.org/abs/hep-ph/0209306}{{\ttfamily arXiv:hep-ph/0209306}}.

\bibitem{Aranda:2012bv}
A.~Aranda, C.~Bonilla, and J.~L. Diaz-Cruz, ``{Three generations of Higgses and
  the cyclic groups},''
  \href{http://dx.doi.org/10.1016/j.physletb.2012.09.011}{{\em Phys. Lett. B}
  {\bfseries 717} (2012) 248--251},
  \href{http://arxiv.org/abs/1204.5558}{{\ttfamily arXiv:1204.5558 [hep-ph]}}.

\bibitem{Khalil:2013ixa}
S.~Khalil and S.~Salem, ``{Enhancement of $H \to \gamma\gamma$ in $SU(5)$ model
  with 45$_{H^1}$ plet},''
  \href{http://dx.doi.org/10.1016/j.nuclphysb.2013.08.016}{{\em Nucl. Phys. B}
  {\bfseries 876} (2013) 473--492},
  \href{http://arxiv.org/abs/1304.3689}{{\ttfamily arXiv:1304.3689 [hep-ph]}}.

\bibitem{Queiroz:2016gif}
F.~S. Queiroz, C.~Siqueira, and J.~W.~F. Valle, ``{Constraining Flavor Changing
  Interactions from LHC Run-2 Dilepton Bounds with Vector Mediators},''
  \href{http://dx.doi.org/10.1016/j.physletb.2016.10.057}{{\em Phys. Lett. B}
  {\bfseries 763} (2016) 269--274},
  \href{http://arxiv.org/abs/1608.07295}{{\ttfamily arXiv:1608.07295
  [hep-ph]}}.

\bibitem{Buras:2016dxz}
A.~J. Buras and F.~De~Fazio, ``{331 Models Facing the Tensions in $\Delta F=2$
  Processes with the Impact on $\varepsilon^\prime/\varepsilon$,
  $B_s\to\mu^+\mu^-$ and $B\to K^*\mu^+\mu^-$},''
  \href{http://dx.doi.org/10.1007/JHEP08(2016)115}{{\em JHEP} {\bfseries 08}
  (2016) 115}, \href{http://arxiv.org/abs/1604.02344}{{\ttfamily
  arXiv:1604.02344 [hep-ph]}}.

\bibitem{Ferreira:2017tvy}
P.~M. Ferreira, I.~P. Ivanov, E.~Jim\'enez, R.~Pasechnik, and H.~Ser\^odio,
  ``{CP4 miracle: shaping Yukawa sector with CP symmetry of order four},''
  \href{http://dx.doi.org/10.1007/JHEP01(2018)065}{{\em JHEP} {\bfseries 01}
  (2018) 065}, \href{http://arxiv.org/abs/1711.02042}{{\ttfamily
  arXiv:1711.02042 [hep-ph]}}.

\bibitem{Duy:2020hhk}
N.~T. Duy, T.~Inami, and D.~T. Huong, ``{Physical constraints derived from FCNC
  in the 3-3-1-1 model},'' \href{http://arxiv.org/abs/2009.09698}{{\ttfamily
  arXiv:2009.09698 [hep-ph]}}.

\bibitem{Branco:2021vhs}
G.~C. Branco, J.~T. Penedo, P.~M.~F. Pereira, M.~N. Rebelo, and J.~I.
  Silva-Marcos, ``{Addressing the CKM unitarity problem with a vector-like up
  quark},'' \href{http://dx.doi.org/10.1007/JHEP07(2021)099}{{\em JHEP}
  {\bfseries 07} (2021) 099}, \href{http://arxiv.org/abs/2103.13409}{{\ttfamily
  arXiv:2103.13409 [hep-ph]}}.

\end{thebibliography}\endgroup

\end{document}